\documentstyle[12pt,epsfig]{article}
 1
 1
 1

\newcommand{\beq}{\begin{equation}}
\newcommand{\eeq}{\end{equation}}

\newcommand{\beqn}{\begin{eqnarray}}
\newcommand{\eeqn}{\end{eqnarray}}
 
\newcommand{\ra}{\rightarrow}

\newcommand{\tr}{{\rm Tr}}

\newcommand{\gag}{$\gamma \gamma$ }

\newcommand{\la}{\lambda}
\newcommand{\lag}{\lambda_{\gamma}}

\newcommand{\lambdae}{\lambda_{e}}

\newcommand{\np}{{\em Nucl.\ Phys.\ }}
\newcommand{\pl}{{\em Phys.\ Lett.\ }}
\newcommand{\pr}{{\em Phys.\ Rev.\ }}
\newcommand{\prl}{{\em Phys.\ Rev.\,Lett.\ }}
\newcommand{\prep}{{\em Phys.\ Rep.\ }}
\newcommand{\zp}{{\em Z.\ Phys.\ }}
\newcommand{\sovjnp}{{\em Sov.\ J.\ Nucl.\ Phys.\ }}
\newcommand{\nuclinst}{{\em Nucl.\ Instrum.\ Meth.\ }}
\newcommand{\annp}{{\em Ann.\ Phys.\ }}
\newcommand{\intjmp}{{\em Int.\ J.\ of Mod.\  Phys.\ }}

\newcommand{\eps}{\epsilon}
\newcommand{\mhmh}{M_{H}^2}
\newcommand{\mh}{M_{H}}

\newcommand{\Mhe}{M_{H}}

\newcommand{\she}{\hat{s}}

\newcommand{\mww}{M_{W}^{2}}
\newcommand{\mwmw}{M_{W}^{2}}

\newcommand{\mzz}{M_{Z}^{2}}
\newcommand{\cl}{{{\cal L}}}
\newcommand{\cd}{{{\cal D}}}

\newcommand{\ppout}{ P^+_{34}}
\newcommand{\pmout}{ P^-_{34}}
\newcommand{\sinsq}{\sin^2\theta}
\newcommand{\cossq}{\cos^2\theta}

\newcommand{\hppll}{++;00}
\newcommand{\hpmll}{+-;00}
\newcommand{\hpplt}{++;\lambda_30}
\newcommand{\hpmlt}{+-;\lambda_30}
\newcommand{\hpptt}{++;\lambda_3\lambda_4}
\newcommand{\hpmtt}{+-;\lambda_3\lambda_4} 
\newcommand{\dk}{\Delta\kappa}
\newcommand{\klam}{\Delta\kappa \lambda_\gamma }
\newcommand{\kac}{\Delta\kappa^2 }
\newcommand{\lac}{\lambda_\gamma^2 }

\def\slashc{c\kern -.400em {/}}
\def\slashL{L\kern -.450em {/}}
\def\slashcl{\cl\kern -.600em {/}}
\def\gam{\gamma \gamma}
\def\ggzz{$\gamma \gamma \ra ZZ \;$}
\def\gag{$\gamma \gamma\;$}
\def\ggww{$\gamma \gamma \ra W^+ W^-\;$}
\def\ggwwe{\gamma \gamma \ra W^+ W^-}
\def\W{{\bf W}}
\def\B{{\bf B}}
\def\noi{\noindent}
\def\sm{${\cal{S}} {\cal{M}}\;$}
\def\smx{{\cal{S}} {\cal{M}}\;}

\def\ssb{${\cal{S}} {\cal{B}}\;$}
\def\cviol{${\cal{C}}\;$}
\def\pviol{${\cal{P}}\;$}
\def\cpviol{${\cal{C}} {\cal{P}}\;$}
\newcommand{\epm}{$e^{+} e^{-}\;$}

\newcommand{\epem}{e^{+} e^{-}\;}

\newcommand{\ggwwht}{$\gamma \gamma \ra W^+ W^- H \;$}
\newcommand{\ggwwh}{\gamma \gamma \ra W^+ W^- H \;}

\begin{document}
\bibliographystyle{unsrt}
\def\baselinestretch{1.2}

\relax

\begin{titlepage}
\rightline{ENSLAPP-A-473/94}
\rightline{hep-ph/9405359}
\rightline{May 1994}
\vspace*{\fill}
\begin{center}
{\large 
{\bf Electroweak Physics Issues at a High Energy Photon Collider $^*$}}

\vspace*{0.5cm}

\begin{tabular}[t]{c}
Marc Baillargeon$^{\#}$, Genevi\`eve B\'elanger$^{\#}$ and Fawzi 
Boudjema\\
{\it Laboratoire de Physique Th\'eorique 
EN{\large S}{\Large L}{\large A}PP} $^{\S}$\\
{\it B.P.110, 74941 Annecy-Le-Vieux Cedex, France}
\end{tabular}
\end{center}
\vspace*{\fill}

\centerline{ {\bf Abstract} }
\baselineskip=14pt
\noindent
 {\small
The main attractions of studying the bosonic sector of the electroweak 
standard model and its extensions at 
a future high energy photon collider are reviewed. 
A presentation 
of the laser scheme for obtaining such a collider is 
given where we emphasize the importance of polarised \gag spectra. 
The need for {\em measuring} the differential luminosities is stressed. 
We show that, in a large variety of processes, 
the yield of weak vector bosons 
is much higher than at the \epm mode but unfortunately, the cross 
sections are dominated by the transverse modes. Investigation 
of the physics related to the symmetry breaking sector both in $W$ and $Z$ 
pair production 
is given and contrasted with what we expect to obtain in the \epm mode and 
at the LHC. We reassess the issue of whether an intermediate-mass 
Higgs can be observed 
as a resonance when we have a broad spectrum that allows the simultaneous 
study of a host of electroweak phenomena. This investigation includes 
the important background of the so-called ``resolved" photon. We analyse 
the important issues of the mass resolution, the $b$ tagging efficiencies, 
and the polarisation of the beams at two typical \epm energies: 300GeV 
and 500GeV. New efficient cuts are found to suppress the background. 
The importance of $WWH$ production at $\sim$1TeV is emphasized and 
contrasted with other Higgs production 
mechanisms at both \epm and $e\gamma$. We take into account various 
backgrounds and look at the effect of polarisation. Finally, 
we examine the interesting problem of the longitudinal $W$ ($W_L$) 
content of the photon. A new set of polarised structure functions 
for the $W_L$ inside the photon is proposed.  We test the validity 
of the ensuing effective $W$ approximation in the \ggwwht process.}

\vspace*{\fill}

{\footnotesize $\;^*$ Based on invited talks given at the ``{\it
Two-Photon Physics from DA$\Phi$NE to LEP200 and Beyond"}, 2-4 February 
1994, Paris.}
\vspace*{0.1cm}

{\footnotesize $\;^{\#}$ On leave from {\em Laboratoire de Physique 
Nucl\'eaire, Universit\'e de Montr\'eal, C.P. 6128, Succ. A, Montr\'eal, 
Qu\'ebec, H3C 3J7, Canada.} }\\
\noindent 
{\footnotesize $\;^{\S}$ URA 14-36 du CNRS, associ\'ee \`a l'E.N.S. 
de Lyon et au LAPP 
d'Annecy-le-Vieux.}

\end{titlepage}
\baselineskip=18pt

\setcounter{section}{1}
\setcounter{subsection}{0}
\setcounter{equation}{0}
\def\thesubsection {\thesection.\arabic{subsection}}
\def\theequation{\thesection.\arabic{equation}}

\setcounter{equation}{0}
\def\thequation{\thesection.\arabic{equation}}

\setcounter{section}{0} 
\setcounter{subsection}{0}

\section{Introduction}
Photons have, since very long, proved to be 
an excellent probe of the structure of matter and the forces that govern it. 
In the area of particle physics a good example is the investigation 
of the electromagnetic properties (form-factors) of the particles. 
Analysis 
of final state photons can also be a unique way of revealing the presence 
of new physics. Another important on-going area of research is two-photon 
physics at the various \epm storage rings. These photons are quasi-real photons 
which give a predominantly soft luminosity spectrum. Here, the bulk of the 
studies 
is devoted to various aspects of hadronic interactions and tests of QCD. 
A new source of highly energetic photons will open up additional 
possibilities for the investigation of electroweak phenomena, not only 
because one reaches higher thresholds, but also because the photon 
is intimately connected to the weak vector bosons. After all, in the 
$SU(2)\times U(1)$ theory the photon emerges as a combination of 
$W^0$ and the hypercharge gauge boson.  \\
The main attractions of \gag collisions are: 
\begin{itemize}
\item
\gag is a quite ``democratic" means of producing any {\em charged}
particle. Phase space allowing, independently of their origin, elementary 
particles are produced with a predictable cross-section. Signatures of 
different models/particles reside in their decay. Usually the production 
mechanisms via \epm are more complicated and model dependent.
\item From the point of view of the electroweak physics 
and most importantly the sector of symmetry breaking, \gag allows to 
access the $J_Z=0$ directly with the most spectacular manifestation being 
the production of a scalar as a resonance. This is perhaps the most salient 
advantage over \epm collisions where chirality highly suppresses this 
$s-$channel production. 
\end{itemize}

On the other hand a clear disadvantage is that 
a $J_Z=1$ resonance can not occur in the \gag mode. This should be viewed as 
a complementarity between the \epm and the \gag modes. At this point it is 
worth adding a slight ``undertone". While the $J_Z=1$ resonance in \epm 
is not suppressed at all, the neutral scalar can not be coupled in a point-like
manner to the two photon state. This coupling can be effectively parameterized 
by a dimension-5 operator, essential for gauge invariance, 
which involves a factor of $\alpha$: 
\beqn
(S,P)\gamma \gamma \ra \propto \frac{1}{M} \frac{\alpha}{\pi} 
F_{\mu \nu} (F_{\mu \nu},\tilde{F}^{\mu \nu})
\eeqn

\noi where $S$ is for a scalar and $P$ for a pseudo-scalar and $M$ is a typical 
scale. Therefore the $S\gamma \gamma$ coupling is suppressed and the peak 
is not expected to be as prominent as what we would have with a gauge boson
in \epm. For example, the two-photon coupling of the Higgs is only induced 
at one-loop.

\noi 
Nonetheless, with the unique possibility of easily accessing the $J_Z=0$ 
state and the observation that the photon has an $SU(2)$ part, \gag
collisions seem to have all the ingredients to study $W$ physics and notably 
the mechanism of symmetry breaking. If only we had large cross-sections.....
and correspondingly very energetic and ``luminous" beams.....
 
\section{Typical sizes of electroweak cross-sections}
\renewcommand{\thequation}{\thesection-\arabic{equation}}
\setcounter{equation}{0}

\begin{figure*}[htbp]
\begin{center}
\caption{\label{allgamegam}
{\em Typical sizes of non hadronic \gag and $e\gamma$ processes.
The subscripts in Higgs processes refer to the mass of the Higgs. For 
$t \bar t$ production the top mass was set to $130$~GeV. }}
 \mbox{\epsfxsize=16.5cm\epsfysize=20.cm\epsffile{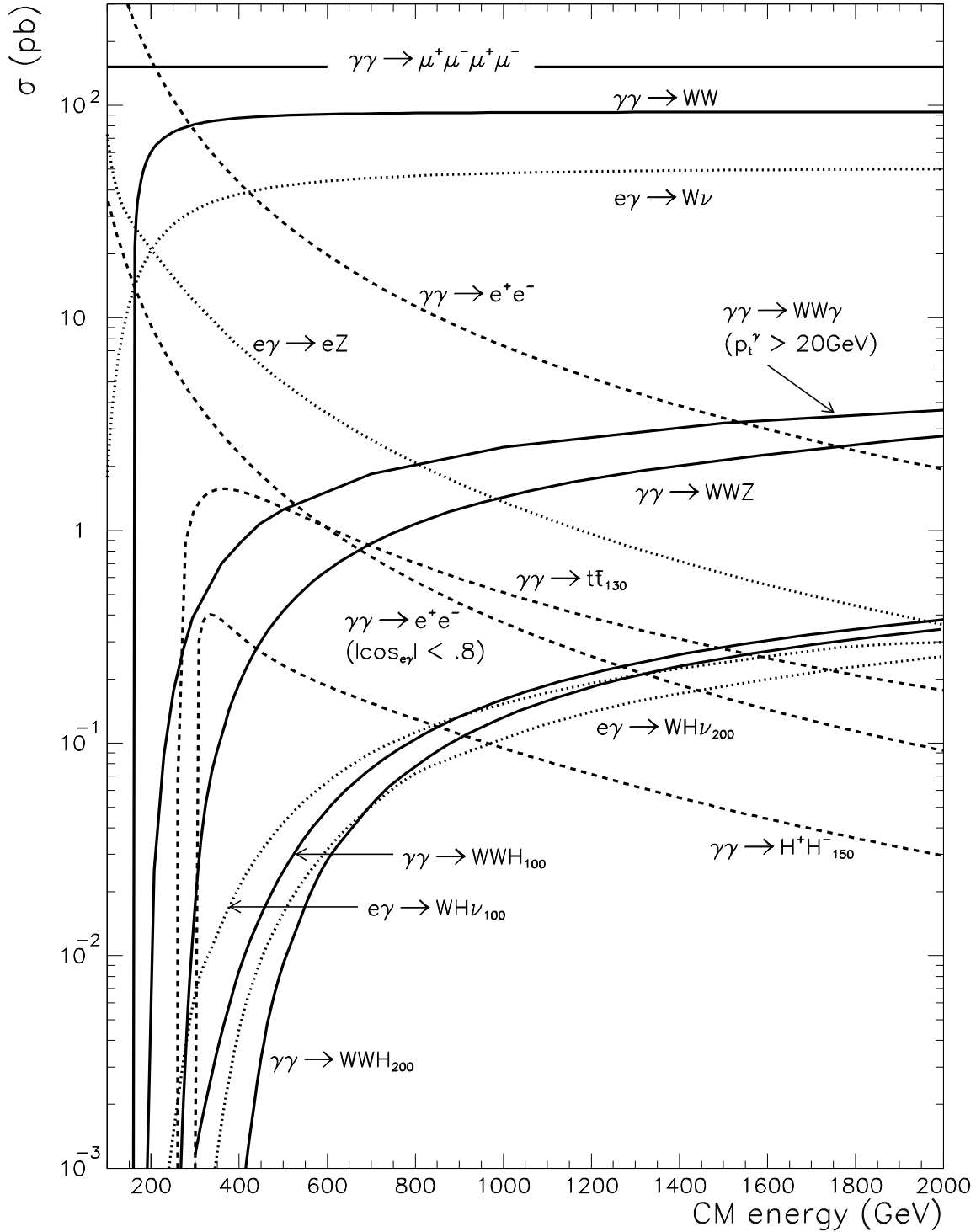}}
\vspace*{-1.cm}
\end{center}
\end{figure*}

At high enough energies, in fact soon after the opening up of the 
corresponding thresholds, 
production of $W$ bosons gives very large cross-sections. In 
Fig.~\ref{allgamegam} we show 
some typical processes that occur in \gag collisions. For the sake of 
comparison we have also included $e\gamma$ processes, although our talks 
concentrate essentially on $\gamma \gamma$ physics. Vector boson production 
dominates in \gag collisions due the t-channel spin-1 exchanges. Most 
prominent is the $W$ pair cross-section, which very quickly reaches a plateau
of about $90$pb\cite{Ginzburg83,nousggvv}. At $500$~GeV this cross-section is larger 
than the total \epm production that scales as $1/s$. At yet higher energies 
triple vector production, $WWZ$ and $WW\gamma$
(with a fixed $p_t^\gamma>20$~GeV) 
become more important than fermion pair production\cite{nousgg3v}. In fact $W$ pair production 
is so important that we can envisage to use it to trigger $H$ production. We 
see that substantial $WWH$ cross-sections are possible\cite{nousgg3v}. From the 
experimental point of view one could use the large cross-section for W pair
production as a 
luminosity monitor. The total $\mu^+\mu^-\mu^+\mu^-$\cite{4mu} that could also 
be used as a luminosity monitor is not sensibly larger, while it remains to be 
seen how well one could tag the extreme forward 4 muons. \\

\begin{figure*}[htbp]
\caption{\label{gammaee}{\em Comparison between the sizes of \gag 
and \epm cross-sections with the same final state (All bold curves are
for \gag processes). 
 The subscripts refer to Higgs and top masses.
In the case of  ZZ production 
 (from \protect\cite{Jikiazz}), 
  the dotted line  corresponds to $M_H=300GeV$ while  
the plain curve corresponds to the infinite Higgs mass.}}
\begin{center}
\mbox{\epsfxsize=16.5cm\epsfysize=20cm\epsffile{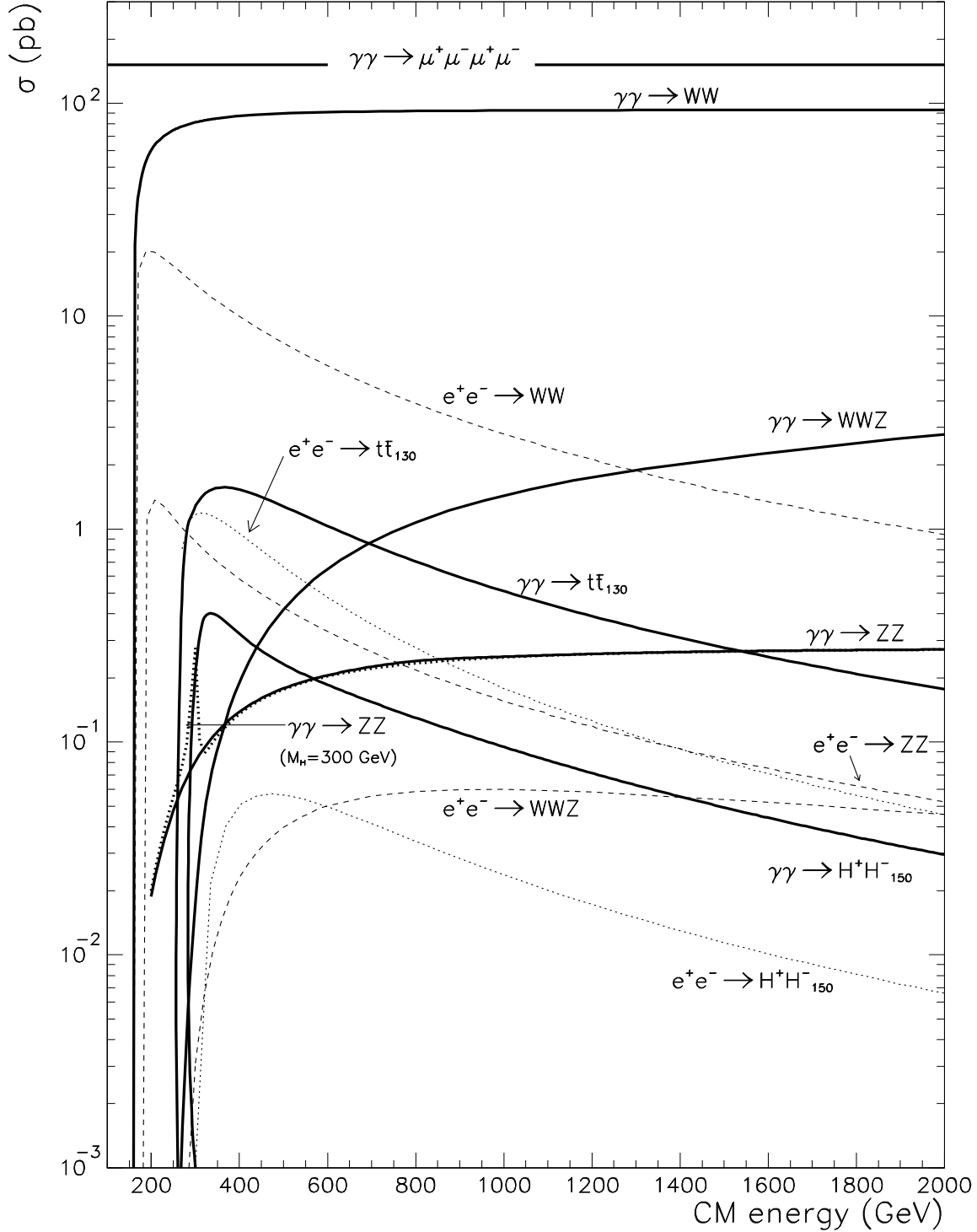}}
\vspace*{-1.cm}
\end{center}
\end{figure*}

To clearly see how important and how rich \gag reactions become 
at TeV energies it is very educating to compare a few characteristic 
cross-sections with the corresponding same-final-state processes at \epm. 
The first observation 
is that independently of the spin of the particle, at the same centre-of-mass 
energy, \gag initiated processes are, at high enough energy, about an order of 
magnitude larger than the corresponding \epm reactions
(Fig.~\ref{gammaee}).  Let us briefly contrast
the situation as regards the production of charged scalars, fermions and 
vector bosons that we will take to be the $W$. As stated earlier, for \gag 
reactions one only needs to know the electric charge assignment and for the 
weak vectors we take them to be elementary gauge bosons. We express the 
cross-section in units of $\sigma_0=4\pi \alpha^2/(3s)$ with $s$ being 
the centre-of-mass energy of either the \gag or \epm system. For pair 
production of a particle of mass $M$ 
we will also use $\tilde{\sigma}_0=\sigma_0(s=4 M^2)=\pi \alpha^2/3M^2$.\\

\noi $\bullet$\underline{Scalars}\\
\noi To compute the pair production rate of
charged scalars in \epm shown in Fig.~\ref{gammaee} 
we have chosen as an example the charged Higgses of the MSSM which 
(at tree-level) is like taking a simple two-doublet extension of the 
standard model (\sm). 
For \gag collisions only the charge assignment is needed. This is 
a particular example of what we alluded to earlier: tree-level cross-sections 
in \epm require the knowledge of a model dependent part which is interesting 
in its own. However, this argument can also be turned to the advantage of the 
\gag mode by arguing that the model dependent part necessarily will show up 
in the decay patterns leaving us with a cleaner initial state. 
The Z exchange, though, does not change the features of the comparison between 
the \gag and \epm modes, so for the sake of clarity our discussion 
only keeps the photonic exchanges in \epm both for the case of scalars and 
fermions, without any incidence on the unitarity of the cross-section. 
The first characteristic
is the behaviour of the cross-section at threshold

\beq
\sigma_{thr}(e^+ e^- \ra H^+ H^-)= \frac{1}{4} \beta^3 \tilde{\sigma}_0 
\;\;\;\;\;\;
\sigma_{thr}(\gam  \ra H^+ H^-)= \frac{3}{2} \beta \tilde{\sigma}_0 
\eeq

\noi 
Clearly the \epm cross-section suffers from the P-wave suppression 
($\propto \beta^3$). At higher energies, $s\gg 4M^2$, there is a factor 
$6$ in favour of the \gag cross-section:

\beq
\sigma(e^+ e^- \ra H^+ H^-)= \frac{1}{4} \sigma_0\;\;\;\;\;\;
\sigma(\gam  \ra H^+ H^-)= \frac{3}{2} \sigma_0
\eeq

\noi $\bullet$\underline{Fermions} \\ 
Again, only taking the photon exchange in \epm and considering a charged 
particle 
with unit charge one has 
\beq
\sigma_{thr}(e^+ e^- \ra f \bar{f})= \frac{3}{2} \beta \tilde{\sigma}_0 
\;\;\;\;\;\;
\sigma_{thr}(\gam  \ra f \bar{f})= 3 \beta \tilde{\sigma}_0 
\eeq

\noi 
The cross-section is twice as large, although this threshold enhancement 
factor can be offset for fermions with 
less-than-unity charges. Far from threshold the \gag cross-section 
has a logarithmic factor enhancement
\beq
\sigma(e^+ e^- \ra f \bar{f})= \sigma_0\;\;\;\;\;\;
\sigma(\gam  \ra f \bar{f})= 3 (\log(s/M^2) -1 ) \sigma_0
\eeq

\noi $\bullet$\underline{Vectors} \\
\noi It is with the vector bosons that we have the most interesting results. 
Here we include all of the electroweak diagrams. First, 
at threshold one has
\beqn
\sigma_{thr}(e^+ e^- \ra W^+ W^-)= \frac{3}{4 s_W^4} \beta \tilde{\sigma}_0 
\approx 12 \beta \tilde{\sigma}_0 \nonumber \\
\sigma_{thr}(\gam  \ra W^+ W^-)\approx  \frac{57}{2} \beta \tilde{\sigma}_0 
\eeqn
 \noi Again the \gag cross-section is about a factor 2 larger. 
At asymptotic energies, as advertised earlier, the \gag cross-section reaches 
a plateau while the \epm decreases with energy: 
\beq
\sigma(e^+ e^- \ra W^+ W^-)= \frac{3}{8 s_W^4} \log(s/M_W^2)\sigma_0
\;\;\;\;\;\;
\sigma(\gam  \ra W^+ W^-)= \frac{8\pi \alpha^2}{M_W^2} 
\eeq

\noi Another very important fact concerns the reaction 
$\gamma \gamma \ra ZZ$\cite{Jikiazz,Bajc,Berger,Dicus,Veltman}. While $ZZ$ production in \epm occurs at tree-level and 
decreases with energy, $\gamma \gamma \ra ZZ$ is a purely quantum 
effect very akin 
to the scattering of light-by-light. Even so, at tree-level 
the cross-section very rapidly 
picks up and takes over the corresponding \epm process! Again this is due 
to the rescattering effect, $\gamma \gamma \ra W^+W^- \ra ZZ$. The reaction 
$\gam \ra W^+W^-$ is the backbone reaction for a host of electroweak 
processes. Even triple vector boson production, $WWZ$, which can be considered 
as a $Z$ radiation off $W$ plays an important role at TeV energies in 
\gag\cite{nousgg3v}. Compared to the same final state  
in \epm\cite{nousee3v,Barger3v,BargerH2v}, at 2~TeV, there is a factor 
of  two orders of magnitude in favour of the \gag mode. 

\subsection{To be fair: $t$-channel cross-sections and transverse 
{\em versus} longitudinal vector bosons yield}
As stressed earlier, the importance of the vector boson pair production 
and the fact that the cross-section for $W$ pair becomes constant at asymptotic 
energies is due to the spin-1 $t$ channel exchange. Of course, fusion 
processes via spin-1 do occur at the \epm collider but the interesting 
processes with production of vector bosons are generally accompanied 
with missing momentum (neutrinos or electrons lost in the beam 
pipe) as in single $W$ or $Z$ production. Even when compared 
to these cases, the $WW$ in \gag has a larger cross-section. It is true 
that the $W$'s are produced quite forward and hence once a cut on the 
scattering angle of the $W$ is imposed the cross-section does decrease 
with energy. However, even with a cut on the $W$ scattering angle 
such that $|\cos \theta|<0.8$, the cross-section 
is still substantial and one is doing with $WW$ ``{\em picobarn physics}" 
all the way 
up to 1~TeV centre-of-mass. Moreover, even when the $W$'s are very forward, at 
sub-TeV energies one could in principle still recognize them through their 
decay. Therefore, even after angular cuts the $WW$ cross-section is still 
very large. \\

\begin{figure*}[hbt]
\vspace*{-.5cm}
\begin{center}
\caption{\label{eeggwwfig2}{\em Comparing the total $WW$ cross-sections 
and the longitudinal $W_L W_L$ in \epm {\it versus} \gag as well as 
the ratio of longitudinal over total. For the latter, 
the scale can be read off on the same $y$ axis. The second figure shows what 
happens when a cut on the scattering is imposed.}}
\mbox{\epsfxsize=14cm\epsfysize=9.cm\epsffile{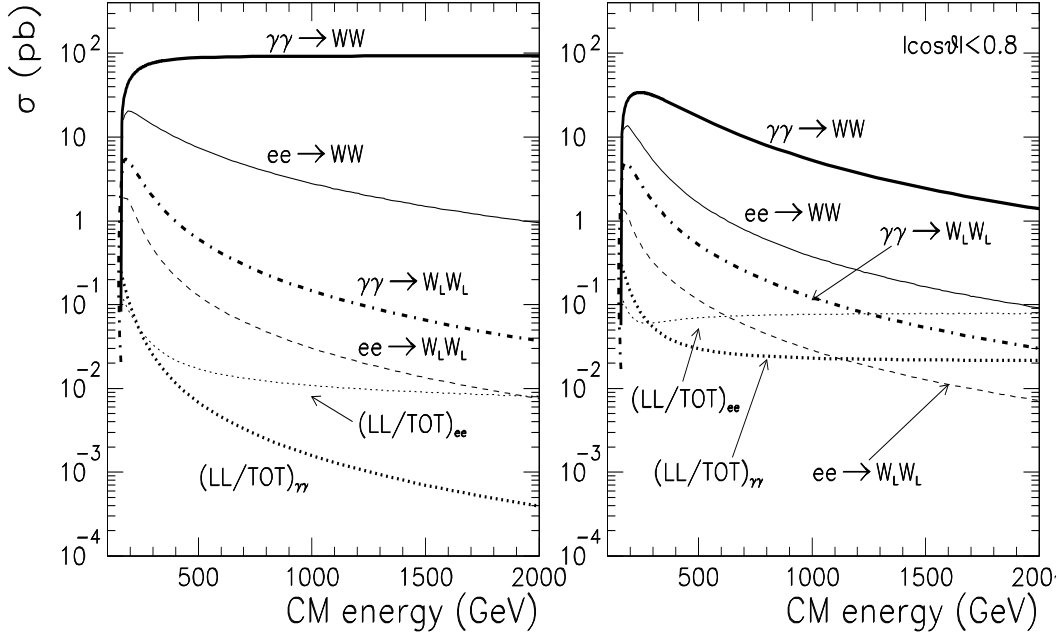}}
\vspace*{-1.cm}
\end{center}
\end{figure*}

\noi Another aspect that must be addressed is to know how much of the 
$W$ sample would be conducive to interesting electroweak tests. 
For instance, having in view the mechanism of 
symmetry breaking (\ssb), what is the amount of longitudinal $W$'s at high energy? 
The production of $W_L W_L$ in \gag is given by the same asymptotic 
formula for the production of two scalars. This is as much as 5 times 
more than what we have in \epm. This is the good news. However, one has to 
realize that the extraction of these longitudinals, if the new physics 
does not substantially increase their yield, is an incredibly tiny 
portion of all $WW$'s. From this perspective the situation in \epm 
is not so bad. To wit,
\beqn
\sigma_{s\gg M_W^2}(e^+ e^- \ra W^+_L W^-_L ) \sim \frac{1}{3} \sigma_0
\;\;\;\;\;
\sigma_{s\gg M_W^2}(\gam  \ra W^+_L W^-_L )= \frac{3}{2} \sigma_0
 \eeqn
but 
\beqn
\frac{\sigma_{s\gg M_W^2}(e^+ e^- \ra W^+_L W^-_L )}
{\sigma_{s\gg M_W^2}(e^+ e^- \ra W^+ W^-)} \sim \frac{1}{20} 
\frac{1}{\log(s/M_W^2)}
\; ; \;\;
\frac{\sigma_{s\gg M_W^2}(\gam \ra W^+_L W^-_L )}
{\sigma_{s\gg M_W^2}(\gam \ra W^+ W^-)} \sim \frac{1}{4} \frac{M_W^2}{s}
\eeqn

So, as a summary regarding the comparison between \epm and \gag processes 
and before addressing  
the subject of how so large \gag cms energies can be obtained, one could say 
that the $WW$ cross-section system provides a good example of 
the characteristics of the 
electroweak cross-sections in \gag in the sense that there would be 
plenty of electroweak events but we will 
have to fight very hard to extract the interesting samples of longitudinal
vector bosons out of the huge sample of transverse vector bosons. Angular 
cuts do, though, improve the $LL/TT$ ratio (see Fig.~\ref{eeggwwfig2}).

  \section{The laser set-up for a high energy photon collider and the luminosity 
spectra}
\renewcommand{\thequation}{\thesection-\arabic{equation}}
\setcounter{equation}{0}
\subsection{The laser set-up} 
\noi 
Until now, two-photon processes at $e^+e^-$ storage rings have exploited the 
``Weisz\"acker-Williams"
spectrum\cite{EPAWW}, which is essentially a ``soft-photons" spectrum. The $\gamma \gamma$ 
luminosity peaks for very small fractions of the invariant $\gamma \gamma$ mass 
$\sqrt{s}_{\gamma \gamma}$, i.e., for 
$\tau=s_{\gamma \gamma}/s_{e^+e^-} \ll 1$. 
Recently, with the intense activity in the physics of a linear \epm 
collider there has been a growing interest and some excitement about 
converting the \underline{single pass } electron into a very energetic photon 
through Compton backscattering of 
an intense laser light.  The seeds of the idea are some 
30 years old\cite{PhotonColold} but the most comprehensive analysis of the scheme 
has been performed by the Novosibirsk group\cite{PhotonCol}. The detailed analyses 
of this group have also 
provided the working basis\cite{Borden} to investigate new physical processes. 

\begin{figure*}[hbt]
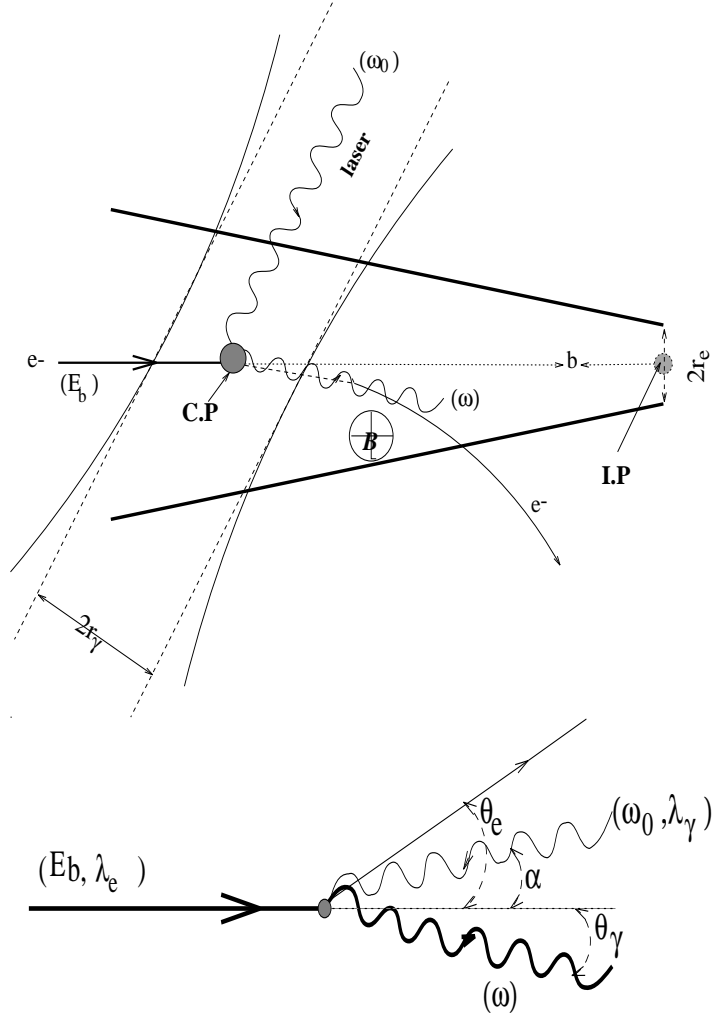

\caption{\label{lasernew}{\em The laser scheme of converting an electron of the linac into a 
highly energetic photon (see text).}}
\begin{center}
 \mbox{\epsfxsize=9.5cm\epsfysize=9.5cm\epsffile{lasernew.eps}}
\mbox{\epsfxsize=9cm\epsfysize=4.cm\epsffile{lasernew2pol.eps}}
\end{center}
\end{figure*}

\noi The set-up of such a scheme is shown in Fig.~\ref{lasernew}

\noi The principle of the scheme is quite simple. A laser beam of frequency 
$\omega_0$ (of order the $eV$) is focused at an extremely small 
angle  $(\alpha)$ on the electron 
beam of energy  $E_b$. 
Compton backscattering occurs at the conversion point (C.P.) which is 
a distance $b$ (a few $cm$) 
away from the interaction point (I.P.), where the electron would have converged. 
This gives a very energetic photon of energy $\omega$ 
emitted at a very small angle and a soft electron. 
In order that the ``soft" 
electron does not end up in the interaction region and hence would interact 
with a hard photon from the opposite arm of the collider or with another 
soft electron, it is suggested to use a very strong transversal magnetic field 
between the conversion region and the interaction point\cite{PhotonCol}.
With this suggestion, 
additional and unnecessary backgrounds due to non-\gag  initiated 
processes as well as degradation of  the overall \gag luminosity
are avoided. However, 
considering that the conversion distance $b$ is only a few 
$cm$ at best, we will have a highly complicated interaction region where a lot 
of disruptions can occur as far as placing a detector goes. The whole issue 
of the interaction requires a dedicated investigation. This particular aspect 
of the strong magnetic field so close to the interaction region 
should also be borne in mind when discussing particular processes that are 
very sensitive on detector performances. We, here, have in mind the issue 
of $b$-tagging for instance with a vertex detector which is a pivotal issue
in the detection of a Higgs with an intermediate mass. The situation is worse 
in the case of $e\gamma$ high energy collider.  The issue of the choice between
converting at some distance from the interaction point or hitting at the interaction
point still requires a detailed investigation as the merits of a finite conversion
distance are offset by the large magnetic field.

\subsection{Parameters of a Photon Collider}
A key parameter of the machine, $x_0$, is directly related to the 
maximum energy, $\omega_{max}$, 
of the ``collider" photon. It is introduced through the 
scaled invariant mass of the original $e\gamma$ system and 
for a head-on hit of the laser is given by:
\beq
x_0=\frac{M_{e\gamma_0}^2}{m_e^2}-1=\frac{4 E_b \omega_0}{m_e^2} \simeq
(15.3) \left((\frac{E_b}{TeV})\right) \left(\frac{\omega_0}{eV}\right) 
\;\;\; \mbox{{\rm so that }}\;\; 
\omega_{\max.}=\frac{x_0}{x_0+1} E_b 
\label{omemax}
\eeq
Most of the photons are emitted at extremely small angles with 
the most energetic photons scattered at zero angle. 
With the typical angle 
$\theta_0=(m_e/E_b) \sqrt{x+1}$ of order some $\mu$rd, the spread of the 
high-energy photon beam is thus of order some 10's {\em nm}. The energy spread 
is roughly given by $\omega\approx \omega_{max}/
(1+(\theta/\theta_0)^2)$.
 It is clear that the further away from the I.P. the conversion occurs, 
those photons that make it to the I.P. are those with the smallest scattering 
angle and hence with the maximum energy. These are the ones that will 
contribute most to the luminosity. 
Therefore, with a large distance of conversion 
one has a high monochromaticity at the expense of a small integrated (over 
the energy spectrum) luminosity. For some processes whose cross-section is 
largest for the highest possible energy, this particular set-up would be 
advantageous especially in reducing possible backgrounds that 
dominate at smaller invariant \gag masses. 

\noi 
From  Eq.~\ref{omemax} it is clear that in order to reach 
the highest possible 
photon energies one should aim at having as large a $x_0$ as possible. 
However, one should be careful that the produced photon and the laser photon 
do not interact so that they create a \epm pair (first threshold); the laser 
frequency should be chosen or tuned such that one is below the \epm 
threshold\cite{PhotonCol}. If we want maximum energy, it 
is by far best to choose the largest $x_0$ taking into 
account this restriction. The optimal $x_0$ is then given by 
$x_0 \leq 2 (1+\sqrt{2}) \sim 4.83$. This value 
means that the photon can take up as much as $83\%$ of the beam energy. 

\subsection{The luminosity spectra}

\noi 
Naturally, the luminosity spectrum depends directly on the differential Compton 
cross-section. 
The original electron as well as the laser 
can be polarised, resulting in quite distinctive  spectra depending 
on how one chooses the polarisations. Introducing 
 $y$, the fraction of the initial electron energy retained
  by the backscattered photon,
\beqn
y = {\omega\over E_b}\leq {\omega_{max}\over E_b}  \;\;\;  r = {y\over x_0(1-y)}\leq 1 \;\;\;
\overline{\sigma}_0=\frac{2\pi \alpha^2}{m_e^2 x_0}
\eeqn
 \noi the energy spectrum of the photons is given by \cite{PhotonCol}
\beqn
f_c(y)&=&\frac{1}{\sigma_c} \frac{{\rm d}{\sigma_c}}{{\rm d}y}= 
\frac{\overline\sigma_0}{\sigma_c} C_{00}(y)\;\;\;\mbox{{\rm with}} \; \nonumber \\
C_{00} (y)&=&{1\over 1-y} + 1-y-4r(1-r)-2\;
\mbox{\boldmath {$\lambda_e P_c$}}\;rx_0(2r-1)(2-y)
\label{c00}
\eeqn
 
\noi 
where $\lambda_e$ is the average helicity of the initial electron and $P_c$ is
the degree of circular polarisation of the initial laser beam. 
The spectrum is normalised from the Compton cross section ($\sigma_c$)
\beqn
\sigma_c& =&\sigma_c^{n.p}+2 \lambda_e P_c \sigma_1\nonumber\\
\sigma_c^{n.p}& =&\overline{\sigma}_0\biggl[
(1-\frac{4}{x_0}-\frac{8}{x_0^2})\log(1+x_0)+\frac{1}{2}+\frac{8}{x}
-\frac{1}{2(1+x_0)^2}\biggr]\nonumber\\
\sigma_1& =&\overline{\sigma}_0\biggl[
(1+\frac{2}{x_0})\log(1+x_0)-\frac{5}{2}+\frac{1}{1+x}
-\frac{1}{2(1+x_0)^2}\biggr]
\eeqn
\noi
One of the important observation is that the spectrum depends on the 
product of the helicity of the electron {\it and} the photon. 
The backscattered photons will retain 
a certain amount of the polarisation of the laser photon beam.
This polarisation, which is energy-dependent is determined by the three
Stoke's parameters 
$\xi_i$, $<\xi_i> =C_{i0}/C_{00}$. 
The functions $C_{i0}$ are written as
\begin{eqnarray}
C_{10} = 2r^2P_t \sin 2\phi,  \phantom{sssss} 
C_{30} = 2r^2P_t\cos 2\phi 
\end{eqnarray}
\begin{eqnarray}
 C_{20} = 2\;\mbox{\boldmath {$\lambda_e$}}\; rx_0(1+(1-y)(2r-1)^2)
-\mbox{\boldmath {$P_c$}} (2r-1)({1\over 1-y}+1-y)
\label{c20}
\end{eqnarray}
\noi 
where now $P_t$ is the degree of transverse polarisation of the laser photon
beam and $\phi$ specifies the direction of the maximum laser polarisation.
 The mean helicity of the high energy photon is given by $<\xi_2>$
while its degree of transverse polarisation by $<\xi_1>$ and 
$<\xi_3>$. 
 We will be dealing with processes that are \cviol and \pviol 
conserving so that there is 
no need in requiring any transverse polarisation.
Such polarisation has been shown to be very useful when looking
for \cpviol violating signals in scalar production\cite{Gunion} or in W pair
production \cite{belcou}.
 It is very crucial to 
realise that the mean helicity of the produced photon does not, in contrary 
to the energy spectrum, depend on the \underline{product} 
of the mean initial helicities (eq.~\ref{c20}).
Therefore, one can get a dominant helicity configuration for the colliding 
photons by having only the laser polarised, which is very easily obtained. 

\noi The \gag luminosity spectrum is a convolution involving the 
differential Compton cross-sections of the two photons  
 as well as a 
conversion function that depends very sensitively on the conversion distances 
and the characteristics of the linac beams. The energy dependence of the 
former function is only through the energy 
fraction $\sqrt{\tau}$, while the conversion function involves the \epm cm 
energy explicitly. 
Realistically other 
considerations should be taken into account. These have to do with the 
laser power. In the following it is assumed that the density of the laser 
photons is such that all the electrons are converted (this assumes 
a conversion coefficient, $k=1$) and that multiple 
scattering is negligible. A compact analytical form for the conversion function 
is obtained in the case of a Gaussian profile for the electron beam with 
an azimuthal symmetry. In this case 
the electron energy density, for a beam with  spotsize 
$\sigma_e$ can be written as \cite{PhotonCol}

\beq
F_e(r)=\frac{1}{2\pi\sigma_e^2} e^{-\frac{r^2}{2 \sigma_e^2}}
\eeq

\noi 
Taking the same conversion distances for both arms of the collider and with the 
same initial electron beams characteristics one gets for the double 
differential cross-section

\beqn
  \frac{{\rm d}^2 L}{{\rm d} \tau{\rm d}\eta}&=&
I_0\left(\rho_0^2 \frac{(x_0+1)}{\sqrt{\tau}} X \right) 
\exp\left(-\rho_0^2 \frac{(x_0+1)}{\sqrt{\tau}} Y \right) f_c(y_1)
f_c(y_2) \nonumber \\
L&=&\frac{{{\cal L}}_{\gamma \gamma}}{k^2 {{\cal L}}_{ee}} \;\;\;
k \; \mbox{{\rm is  the conversion coefficient}}
\eeqn

\noi 
where $y_{1,2}$ are the energy fractions of the two collider-photons which can 
be re-expressed in terms of the reduced invariant \gag mass and the ``rapidity" 
$\eta$ through $y_{1,2}=\sqrt{\tau} e^{\pm \eta}$. $I_0$ 
is the modified Bessel function of zeroth-order 
($I_0(z)=\frac{1}{\pi} \int_{0}^{\pi} {\rm d}\theta e^{-z \cos \theta}$). 
The effect of a non-zero conversion distance is all contained in 
$\rho_0$ whose value is a measure of the (improved) monochromaticity of 
the spectrum due to the conversion distance:
\beq
\rho_0=\frac{b}{\sqrt{2} \sigma_e} \frac{m_e}{E_b}
\eeq
while $X$ and $Y$ are given by:
\beqn
X=\sqrt{\left(\frac{x_0}{x_0+1} e^{-\eta} -\sqrt{\tau}\right)
\left(\frac{x_0}{x_0+1} e^{\eta} -\sqrt{\tau}\right)}, \;\; ; \;\; 
Y=\frac{x_0}{x_0+1}\cosh(\eta)-\sqrt{\tau} 
\eeqn

\noi Note that $\rho_0$ is inversely proportional to the electron 
spotsize and to the beam energy. Hence increasing the spotsize and the energy 
degrades the monochromaticity due to the conversion.

\subsubsection{Polarised luminosity distributions with zero conversion distance}

Almost all of the physics analyses have been done with $b=0$. We will also 
conform to this practice. However, we would like to point out that taking 
$b=0$ may be too realistic and that taking 
$b \neq 0$ may have some advantages as 
it gives a peaked spectrum. In any case what is 
crucial is to make full use 
of the availability of polarisation. 
\begin{figure*}[hbt]
\caption{\label{spectre12}{\bf (a)} {\em The total luminosity spectra in the case 
of different combinations of the longitudinal polarisations of the 
linac electrons and the circular polarisations of the laser. The 
``classic" Weisz\"acker-Williams spectrum is shown for comparison. The spectra 
assume a distance of conversion, $b=0$.} 
{\bf (b)} {\em 
Projecting the contributions of the $J_Z=0$ and the $J_Z=2$ polarised 
spectrun in the peaked spectrum setting $2\lambda_e P_c=2 \lambda_e' P_c'=-1$.}}
\begin{center}
\mbox{\epsfxsize=14cm\epsfysize=9cm\epsffile{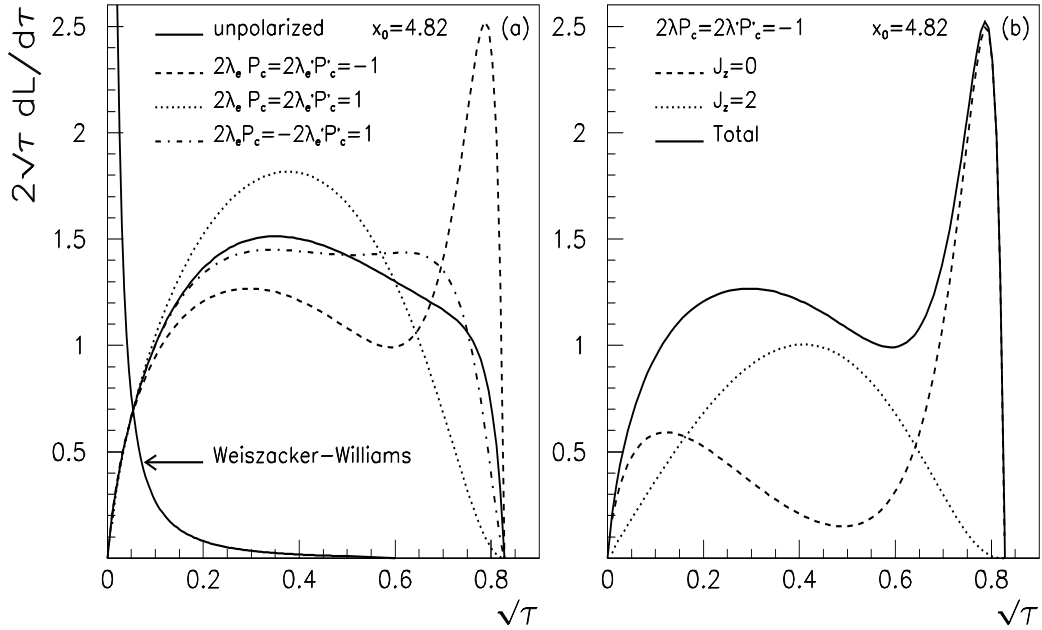}}
\vspace*{-1.cm}
\end{center}
\end{figure*}

\noi In Fig.~\ref{spectre12}a we compare the luminosity spectrum 
(as a function of the reduced \gag invariant mass) 
that one obtains by choosing different 
sets of polarisations for the two arms of the photon collider. First of all, 
in all cases and as advertised earlier one has a hard spectrum compared to 
the ``classic" Weisz\"acker-Williams spectrum. In case of no 
polarisation at all, one obtains a broad spectrum which is almost a step 
function that extends nearly all the way to the maximum energy (restricted by 
the value of $x_0$).  
The hardest spectrum is arrived at 
by choosing the circular polarisation 
of the laser ($P_c$) and the mean helicity of the electron ($\lambda_e$) 
to be opposite, i.e., $2 \lambda_e P_c=-1$, 
for {\it both} arms of the collider. In the case where both arms have 
$2 \lambdae P_c=+1$ the spectrum has a ``bell-like" shape
which favours the middle range values of $\sqrt\tau$. In the case where 
the two arms of the collider have an opposite value for the 
product $2\lambdae P_c$, the spectrum is almost identical to the one 
obtained in case of no polarisation. It is clear that for processes 
where we need the maximum energy, like those we have encountered with 
$W$ production processes or those with higher thresholds or again those 
where the indirect effects of some New Physics grow with energy, 
the ``peaked spectrum", $2\lambda_e P_c=2 \lambda_e' P_c'=-1$, is best. On 
the other hand, if one is looking for peaks in the \gag 
invariant mass like in the resonant-scalar production, 
it is best to take a scheme 
that explores uniformly the whole energy range.
In this case, the spectrum obtained without having any of the beams polarised could 
do the job. \\
\noi
However, in all situations one should always insist on 
having some polarisation. 
This is because polarised laser beams (which are easily obtained) 
and electron (which should not be too difficult) means that 
the colliding photons are in a preferred state of polarisation. This 
is crucially important in favouring some physics channels that occur 
in $J_Z=0$ for example rather than for $J_Z=2$. As one can read 
from Eq.~\ref{c20} the produced photon has a net mean helicity 
even if only either the electron or laser  beam is polarised. 
We show in Fig.~\ref{spectre12}b how 
the total luminosity is shared between the two states $J_Z=0$ and $J_Z=2$. 
In the ``peaked" set-up, i.e, 
$2 \lambda_e P_c=2 \lambda_e' P'_c=-1$ 
and in the case where both lasers 
are tuned to have a right-handed circular polarisation ($P_c=P_c'=+1$), 
one has the added advantage that the high-energy photons are produced 
mostly with the same helicity, therefore giving a $J_Z=0$ dominated 
environment, for short we will refer to this situation as the ``0-dom." case. 
The $J_Z=2$ tail almost disappears for $\sqrt{\tau}>0.7$. 
For some processes where the $J_Z=2$ is dominant, 
or if one wants to compare the $J_Z=2$ and the $J_Z=0$ on an 
``equal basis", one  
would also like to isolate the $J_Z=2$ at the expense of the $J_Z=0$ spectrum. 
We point out\cite{SanDiego} that this could be easily achieved by flipping {\em both} the 
electron and laser polarisations of {\em one} of the arms {\em only} 
while maintaining $2 \lambda_e P_c=-1$ (for a maximum of monochromaticity). 
In this case, the  $J_Z=0$ and  $J_Z=2$ spectra are simply 
interchanged. We will refer to this case as the 
``2-dom.", for short. 
For $W$ processes we have preferred, for reasons that should be clear by now, 
the peaked spectrum. 

\begin{figure*}[hbt]
\caption{\label{spectre34}{\em {\bf (a)} 
Projecting the contributions of the $J_Z=0$ and the $J_Z=2$ polarised 
spectrum in the ``broad" setting $2\lambda_e P_c=2\lambda_e' P_c'=1$ 
(with a conversion distance $b=0$). 
Thick lines are with a $100\%$ longitudinal polarisation for the electron 
while the light lines are for $50\%$ longitudinal polarisation. The lasers 
are taken to be fully right-handed. 
{\bf (b)} As in {\bf (a)} but for unpolarised electrons and where we have 
imposed a rapidity cut of $\eta<1$.}}
\begin{center}
 \mbox{\epsfxsize=14cm\epsfysize=9.cm\epsffile{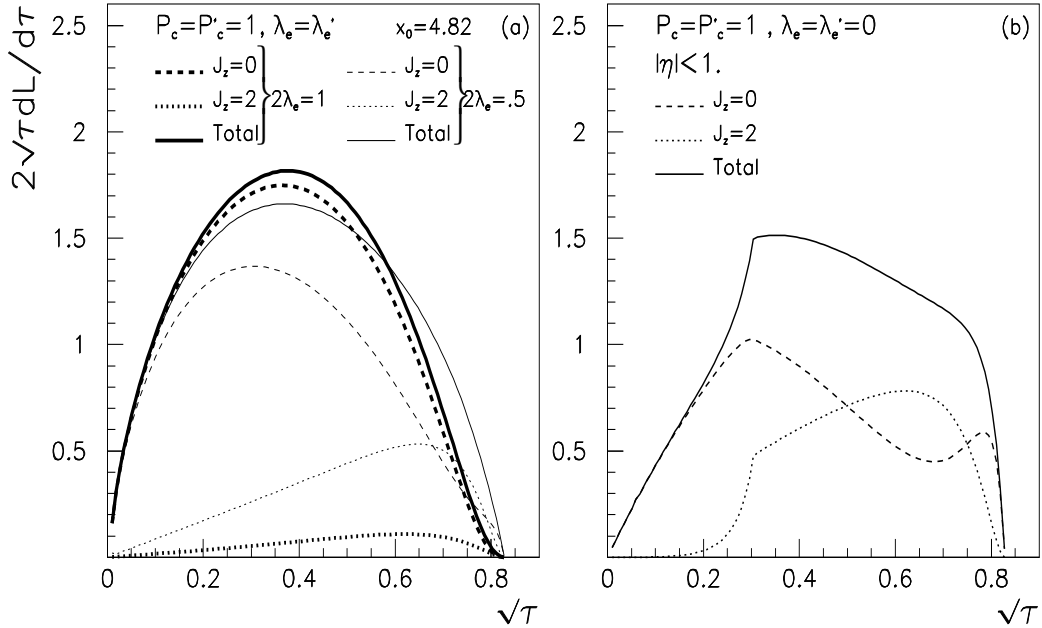}}
 \vspace*{-1.cm}
\end{center}
 
\end{figure*}

For the Higgs search, that is when we would like to keep an almost constant 
value for the differential luminosity, the ``broad" spectrum 
that favours the $J_Z =0$ is highly recommended. What is very gratifying is 
that with $P_c=P_c'=2\lambdae=2\lambdae'=1$ 
the whole spectrum is accounted for almost totally 
by the $J_Z=0$ spectrum(see Fig.~\ref{spectre34}a); the $J_Z=2$ contributes slightly only at the 
higher end. This near purity of the $J_Z=0$ is not much degraded if the 
maximum mean helicity of the electron is not achieved. We show on the same 
figure (Fig.~\ref{spectre34}a) 
what happens when we change  both $2\lambda_e$ and
$2\lambda_e'$ from $1$ to $.5$, keeping 
$P_c=P_c'=1$. There
 is still a clear dominance of the $J_Z=0$ especially for the lower values 
of the centre-of-mass energy. We would like to draw   attention to the fact 
that this effect, (increasing the 
$\frac{J_Z=0}{J_Z=2}$ ratio), 
can be further enhanced (when the maximal electron polarisation is not 
available) by imposing rapidity cuts. The 
point is that the mean helicity of the final photon, $\lambda_\gamma$, 
is non zero even in the case of 
no electron polarisation.
Now, if the energy 
factor 
multiplying $P_c$ in Eq.~\ref{c20} is the same for both photons, 
we would expect that 
the colliding photons have the same degree of polarisation hence producing 
a $J_Z=0$ state. This could be achieved by requiring small rapidities. In Fig.~\ref{spectre34}b 
we show the luminosity spectrum for the case 
where only the laser photons have the same maximal circular polarisation but 
where we have imposed a cut on the rapidity $\eta<1.$ We see that for 
small centre-of-mass energies $\tau<0.22$ 
we have a highly dominant $J_Z=0$ environment. The relevance of this 
observation will be fully exploited in the Higgs search section. 

\subsubsection{Polarised spectra with a finite conversion distance}

\begin{figure*}[hbt]
\caption{{\bf (a)} {\em The total luminosity spectrum with 
$2\lambda_e P_c=2\lambdae' P_c'=-1$ (``peaked spectrum") for different 
values of the conversion distance taking a spotsize $\sigma_e=200$~nm.}
{\bf (b)} {\em As in {\bf a} but 
with $2\lambda_e P_c=2\lambda'_e P_c'=1$ (``broad spectrum").}}
\begin{center}
\mbox{\epsfxsize=14cm\epsfysize=9.cm\epsffile{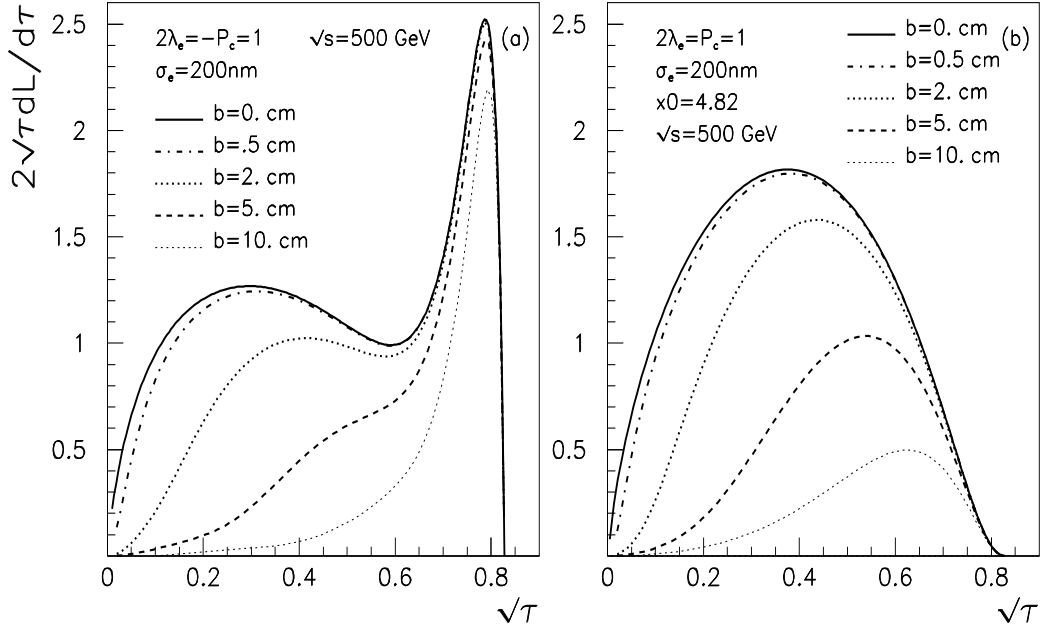}}
\vspace*{-1.cm}
\end{center}
\label{spectre56}
\end{figure*}

\noi For illustration, we take an electron beam 
with a Gaussian profile. As explained above, increasing the 
distance of conversion filters the high energy modes and therefore the 
spectrum becomes more monochromatic for large values of \gag centre-of-mass 
energy. For the discussion, we take the spotsize of the electron beam to 
be $\sigma_e=200$~nm, a \epm cm energy of $500$~GeV and consider 
a few values of the conversion distance. Note that exactly the same features 
are obtained for a smaller spotsize $\sigma_e=100$~nm and a proportionally 
smaller conversion distance. For the peaked 
spectrum, arrived at by having
 $2 \lambda_e P_c=2 \lambda'_eP_c'=-1$, the peaking 
is dramatically enhanced for a large conversion distance $b=10$cm 
($\rho_0\simeq 0.72$). This means for example that with a conversion 
distance of 
$5$cm or $10$cm, there is almost no 
luminosity below $\sqrt{\tau}<0.65$. This also means 
(see Fig.~\ref{spectre56}a) 
that the spectrum is a purely $J_Z=0$ peaked spectrum. This is 
the most ideal situation to study a $J_Z=0$ resonance if its mass falls in 
this energy range, {\it i.e}, $ 0.7 < \sqrt{\tau_{res.}} < 0.82 $. 
The $J_Z=2$ component that was present for the zero-distance 
of conversion is effectively eliminated for large distances 
$b>5$cm. Note that in this case if one could ``manage" 
with a conversion distance of $0.5$cm then we almost recover the $b=0$ 
spectrum. \\
The situation is not as bright for 
the broad spectrum case when the interest is on small $\sqrt{\tau}$, like 
the search of an intermediate-mass Higgs (IMH) at a $500$~GeV \epm. The nice features 
that were unravelled in the last paragraph (an almost pure $J_Z=0$ for 
small to moderate $\sqrt{\tau}$) are lost because the 
luminosity in the energy range of the IMH peak formation 
is totally negligible for conversion distances of order $\sim 5$cm or higher
(see Fig.~\ref{spectre56}b). 
If one could manage with a conversion distance below $2$cm then we may hope 
to keep the nice features of the ``broad" $J_Z=0$ scheme. 

\subsection{Measuring the luminosity}
As should have transpired from the previous paragraphs, 
the interaction region is quite complicated. Whether 
one chooses to convert the electron at some distance from the I.P, in 
which case one needs a 
strong electromagnetic field to sweep away unwanted soft electrons and 
introduce disruptions, 
or whether one chooses to hit very close to the I.P, in which case one 
has additional backgrounds (rendering the machine a mixture of \epm, $e\gamma$ 
and \gag) one expects some uncertainties in the true \gag luminosity. 
All this to say that it will be extremely important to {\it measure} the 
\gag luminosity. This measurement includes in fact various measurements. 
It will not be sufficient to measure the total luminosity but one should 
at least measure the differential \gag luminosity (as a function of the 
invariant \gag centre-of-mass energy). It will also be very desirable to 
have the double differential (in rapidity also) luminosity. Moreover, it is 
very important to ``reconstruct" the polarised $J_Z=0$ and $J_Z=2$ spectra 
and make sure that these measurements do not deviate much from the theoretical 
luminosity calculations. Otherwise, particular 
choices of the settings should be reconsidered if the true 
spectrum does not reproduce the ``theoretical" spectrum. 
For instance, in the Higgs search it should be 
important to know how much $J_Z=0$ one actually has in a realistic 
experimental set-up.\\
Which reaction could we choose as a luminosity monitor? 
Few ideas\cite{PhotonCol,TelnovHawai,Yasui} have been 
suggested but none has received the detailed investigation it deserves. First, 
the choice of the reaction will sensitively depend on the energy range one 
wants to cover. For a high energy machine that is going to be devoted to 
$W$ physics, as Fig.~\ref{allgamegam} shows, it is not \epm production that is a winner. 
As a general observation one should, in fact, not choose final states with 
electrons, 
especially at the lower energies, since there could be a contamination from 
non ``swept-away" electrons and/or non-converted electrons of the linac. In 
this respect, muons are best even if the corresponding 
total cross-sections are smaller. Note that it was suggested to use
the \epm final state in order to reconstruct the  
polarised components\cite{TelnovHawai}. Indeed, once a cut on the scattering
 angle $\theta_e^*$ in the \gag cm frame 
is imposed practically only the $J_Z=2$ is kept(see eq.~\ref{ggbbqed}).
 It is worth pointing that with a cut such that 
$|\cos \theta_e^*|<0.8$, it is just as well to use $\mu^+ \mu^-$. It should be 
remarked however, 
that the \epm or $\mu^+ \mu^-$ cross-section for cm energies above $300$~GeV 
is at least two orders 
of magnitudes less than $W$ pair production. The Novosibirsk group
\cite{PhotonCol} has 
proposed the 4-fermion processes $\gamma \gamma \ra \mu^+ \mu^- e^+ e^-$ 
and $\gamma \gamma \ra e^+ e^- e^+ e^-$ for the calibration. The disadvantage, 
apart from the presence of the electrons, 
is that these processes will not be able to measure 
the polarisation. From the experimental point of view, taking the muon 
($\mu \mu e e$) case 
as an example, for sub-100~GeV energies it rests to see if one could track 
down the muons whose average production angle will be below the $mrd$. 
The large total $4\mu$ cross-section has also the same shortcomings, although 
if we could track the muons this could be used to measure the total differential 
cross-section. The advantages of this total cross-section rest on the fact that
it is energy independent at high-energy and that it has a large value, 
$\sim 153pb$. \\
\noi 
We\cite{nousggvv} have pointed that the $W$ pair production should also qualify as a good 
luminosity monitor as one could convince oneself by glancing at 
Fig.~\ref{allgamegam} and~\ref{gammaee}. 
This has been taken up by \cite{Yasui}. 
However, a detailed investigation is in order. 
The needed analysis should be based on a full Monte-Carlo with four fermions 
in the final states that keeps the correlations between the decay products 
and that could allow to apply cuts in the laboratory frame 
directly on the observed fermions 
(acceptance factors etc...). 
It has already been shown how one can reduce the effect of non-resonant
diagrams to a negligible level\cite{Frankggww}.
A program within this spirit is being carried out
\cite{nousgglumi}. 
A drawback, unless one finds good variables, is that at large $WW$ invariant 
masses the cross-section does not favour a particular initial polarisation. 
Another hesitation in using this reaction is that we want to exploit it 
to uncover new physics that affects the tri-linear $WW\gamma$ as well as the 
quadrilinear $WW\gamma \gamma$ vertices and hence it might not be as unambiguous 
as the $4$ fermion QED-dominated processes. Nonetheless, we would like 
to argue that if one is to 
feel these effects one would probably need the full $W$ sample and therefore 
when measuring the luminosity this effect 
could be counted in the systematic error. Moreover, one expects this machine 
to be running after, or alongside, the high-energy \epm mode where these 
couplings (at least the tri-linear) would be well measured. Moreover, 
theoretical arguments (see later) indicate that anomalies would affect 
the longitudinal and central $W$'s rather than the preferentially transverse 
and forward $W$'s from the \sm.

 \section{A theoretical technical aside: The importance of a non-linear 
gauge fixing term for \gag electroweak processes} 

As Fig.~\ref{allgamegam} and~\ref{gammaee} 
make explicit, production of electroweak bosons 
is overwhelming especially at high-energies. Since the calculation 
of these processes involves a large number of diagrams it is 
highly recommended to simplify the computational task as much as possible. 
The complication not only arises from the large number of diagrams but also 
because  the non-Abelian gauge nature of the couplings and the ever 
increasing number of $W$ propagators renders the computation 
of even a single 
diagram arduous. Helicity technique methods do help and one can do better by 
combining these methods with a judicious choice of gauge-fixing. 
Calculating in the usual unitary gauge is rather awkward because of the 
cumbersome presence of the ``longitudinal" mode 
(``$k_\mu k_\nu$" term) of the various W propagators. 
A way out is to use a Feynman gauge. However, with the widespread choice 
of a linear gauge fixing term, 
this is done at the 
expense of having to deal with even more diagrams containing the unphysical 
Higgs scalars. An indisputable choice of gauge for photonic reactions or 
for processes involving a mixture of W's and photons is to quantize with a 
non-linear gauge fixing term \cite{nonlinear} 
and work with a parameter corresponding 
to the 
't Hooft-Feynman gauge. This type of gauges 
is also known as the background gauges. Recent developments in 
the calculation of QCD processes, especially the so-called string inspired 
organisation\cite{BernKosower}, 
 can be understood 
as being (partly) based on the exploitation of similar 
gauges. This type of gauges can also be efficiently used for 
tree-level electroweak processes\cite{nousgg3v}
 and not just for loop diagrams as has been customarily done till now. 
For a few of the vector boson processes in \gag 
($\gam \ra WW,Z\gamma,\gam; WW\gamma, WW\gamma\gamma,WWZ,WWH,..$)
one has, with this choice of gauge-fixing, 
the same number of diagrams as in the usual unitarity gauge save for the fact 
that we have no ``longitudinal" mode to worry about and that there are no 
diagrams 
with unphysical scalars since the virtue of this choice is that 
the vertex with the photon, the $W$ and the unphysical Higgs field 
does not exist. In any case the reduction in the number of diagrams is 
considerable. 
Of course, one has to allow for small changes in the vertices which turn out 
to have an even more compact form than in the usual gauges. For instance, 
all diagrams where the quartic $WW\gamma \gamma$ vertex appears are identically 
zero when the two incident photons have opposite helicities ($J_Z=\pm 2$). 

\noi With  $S^{\pm}$ being the unphysical Higgs bosons, 
the $W^\pm$-part of the linear gauge fixing condition

\beq
{{\cal L}}^{Gauge-Fixing}_{linear}\;=\;-\xi^{-1}
|\partial_\mu W^{\mu +}\,+\,i\xi M_W S^{+}|^{2}
\eeq
 

\noi is replaced by the ``constraint"

\beq
{{\cal L}}^{Gauge-Fixing}_{non-linear}\;=\;-\xi^{-1}
|(\partial_\mu\;+\;i e A_\mu\;+\;ig \cos \theta_W Z_\mu) W^{\mu +}\,+\,
i\xi M_W S^{+}|^{2}
\eeq

\noi where $\theta_W$ is the usual weak mixing angle. The 
best choice for $\xi$ is no doubt $\xi=1$. Note that in the recent 
calculations of the one-loop process $\gam \ra ZZ$, Jikia\cite{Jikiazz} 
and Dicus and Kao\cite{Dicus} have 
used two different variants of this gauge-fixing. 
We foresee this kind of choices to stand out for applications 
to W dynamics at a future $\gamma \gamma$ collider as it has proved to be 
for one-loop weak bosons induced amplitudes for photonic processes 
\cite{Z3gamus}.

 \section{Physics with $W$  pairs in \gag}

$W$ pair production has been a recurring theme and its importance 
can not 
be over-emphasized. The event sample is just too large! The reaction 
can be considered as the backbone on which one grafts even more bosons
at tree-level or through loops. This reaction was first studied by
Pesic\cite{Pesic73} within the context of 
the \sm and by considering $W$'s with no magnetic moment. 
Kim and Tsai\cite{KimTsai,Sushkov} 
extended the study to the case of a $W$ with an arbitrary 
value of the magnetic moment. A more detailed study by 
Tupper and Samuel\cite{TupperSamuel} followed. 
All these studies assumed a Weizs\"acker-Williams spectrum. 
Ginzburg {\it et al.} \cite{Ginzburg83} were first to study 
the effect of polarised photon beams. 

\subsection{Tree-level helicity amplitudes for $\gam \ra 
W^+W^-$ in the \sm} 
To understand the characteristics of the $WW$ cross-section it is best to 
give all the helicity amplitudes that contain a maximum of information on the 
reaction.
With $\la_1, \la_2 =\pm 1$ being the photon helicities, $\la_3,\la_4=
\pm 1$ the transverse helicities of the $W$'s and  $0$ their 
longitudinal mode, we have\cite{nousggvv}\footnote{
Yehudai\cite{Yehudai} had previously given the complete helicity
 amplitudes. 
We find sign differences in some of the standard model helicity 
amplitudes. This is due to an inconsistent labelling
of the polarisation vectors in\cite{Yehudai}. 
Our results have been confirmed by a recent calculation \cite{Zappala}.}
 \beq
{\cal M}_{\la_1 \la_2; \la_3 \la_4} = \frac{4 \pi \alpha}
{1-\beta^2 \cos^2 \theta} \;\;{\cal N}_{\la_1 \la_2; \la_3 \la_4} \;\;\; ; \;
\;\; \beta=\sqrt{1-4/\gamma} \; ; \; \gamma=s/\mww
\label{eq:mn}
\eeq
\noindent
where
\beqn
{\cal N}_{\la_1 \la_2; 0 0} &= &\frac{1}{\gamma}
\left\{ 
-4 (1+\la_1 \la_2) + (1-\la_1 \la_2) (4+\gamma) \sin^2 \theta  \right\}
\nonumber \\
{\cal N}_{\la_1 \la_2; \la_3 0} &=& \sqrt{\frac{8}{ \gamma}} \;\; (\la_1-\la_2) 
         (1+\la_1 \la_3 \cos \theta) \sin \theta  \;
\nonumber \\
{\cal N}_{\la_1 \la_2; \la_3 \la_4}&=& \beta (\la_1+\la_2) (\la_3+\la_4) + 
\frac{1}{2 \gamma}\left\{
-8 \la_1 \la_2 (1+\la_3 \la_4) +\gamma (1+\la_1 \la_2 \la_3 \la_4) 
(3+\la_1 \la_2) \right.  \nonumber \\
&+& \left.  2 \gamma (\la_1-\la_2)(\la_3-\la_4)\cos\theta
- 4 (1-\la_1 \la_2)(1+\la_3 \la_4) \cos^2\theta \right. \nonumber \\
&+& \left. \gamma (1-\la_1 \la_2)(1-\la_3 \la_4) \cos^2\theta \right\} 
\eeqn

\noi 
The unpolarised differential and total cross-sections follow
\beq
\frac{{\rm d}\sigma_{tot}}{{\rm d}\cos\theta} = 
\frac{3 \pi \alpha^2}{\mww}
\frac{\beta}{\gamma} 
\left\{ 1-\frac{4}{1-\beta^2 \cos^2\theta} \left(\frac{2}{3}+\frac{1}{\gamma}
\right) + 2 \left(\frac{4}{1-\beta^2 \cos^2\theta} \right)^2 \left(\frac{1}{3}+
\frac{1}{\gamma^2}\right) \right\}
\eeq

\beq
\sigma_{tot}=\frac{8\pi \alpha^2}{\mww} \beta 
\left\{ \left(1+\frac{3}{4 \gamma}+\frac{3}{\gamma^2}\right) 
- \frac{6}{\beta \gamma^2} \left(1-\frac{2}{\gamma} \right) 
{\rm Arcosh}\frac{\sqrt\gamma}{2} \right\}
\eeq
Very soon after threshold, the total $WW$ cross-section with either polarised
or 
unpolarised beams increases rapidly  with energy until it reaches a plateau, 
around $400$~GeV.
In all cases  the W pairs are
produced near the beam pipe and they are mostly transversely
polarised.
With a cut on the scattering angle so that
W's along the beam are rejected, the cross-section decreases with energy 
(see Fig.~\ref{eeggwwfig2}). 
Note that  in the $J_Z=0$ ( $\la_1=\la_2$) mode, 
the \sm does not produce W's with different helicities, so that either both W's 
are longitudinal or if transverse they have the same handedness. 
This property could be very useful   when looking
for signals of New Physics, in particular the ones associated with
the symmetry-breaking mechanism.

\subsection{Tests on the tri-linear anomalous couplings, the ``chiral 
lagrangian" approach and comparison with other machines} 
Although the bulk of the reaction as we 
alluded to in our introduction is due to the gauge transverse sector, 
the fact that there are so many $W$'s around (one is talking here about 
a sample of a million $WW$ pairs with a luminosity of $10fb^{-1}$) 
makes this reaction the ideal place 
to probe the mechanism of symmetry breaking by trying 
to reach the Goldstone Bosons, in a sense the longitudinal 
$W$'s. With so many transverse $W$'s one could 
also attempt to conduct precision tests on the electromagnetic 
couplings of the $W$. This aspect of the $W$ physics is more commonly referred 
to as the anomalous couplings of the weak bosons. \\
There is an extensive literature (see for instance
\cite{HPZH,BMT,Hawai} and references therein) on the effect of 
anomalous couplings 
defined through a set of operators that describe deviations from the \sm 
values of the $WWV$ ($V=\gamma,Z$) vertex. 
 Until very recently, 
the studies have been devoted to the tri-linear couplings as these could be 
probed very efficiently in $W$ pair production at the \epm machines or 
in $WZ/\gamma$ at the hadron machines. At higher energies, 
as at the NLC, novel quartic 
couplings could contribute non-negligibly to triple vector boson production 
or $WW$ scattering processes. As we have argued \cite{nousee3v,nousggvv} elsewhere, 
it is important to check for these quartic couplings 
which are more directly related to the scalar sector. 
Indeed, some of the quartic couplings could be 
the residual effect of a heavy scalar exchange at tree-level
while tri-linear 
couplings could be thought of as integrating-out heavy fields (at one-loop) 
or are the effect of mixing with other heavy vectors.
In fact, we should draw the analogy between these anomalous couplings 
and their study in $WW$ and $ZZ$ reactions with the approach that we have 
heard at the $DA \Phi NE$ sessions where people study, albeit in  
a totally different energy regime, various predictions of 
``models" of chiral perturbation theory. Our pions are the Goldstone bosons, 
of course. This analogy can be put out on more formal ground, but before 
doing so 
let us give the by-now ``classic" parametrisation of the anomalous tri-linear 
couplings. This is intended to those who are not familiar with the 
re-interpretation of these tri-linear couplings within 
a set of gauge invariant operators. Here, we will only address the issue of 
the \cviol and \pviol conserving couplings (for a discussion of
\cpviol violation in \ggww see\cite{belcou,MaMckellar}).
 The oft-used parameterisation of Hagiwara et al. \cite{HPZH}, 
(the {\em HPZH} parameterisation) for the $WW\gamma$ couplings and
their $WWZ$ counterparts, is
\beqn
{\cal L}_1&=& -ie \left\{ \left[ A_\mu \left( W^{-\mu \nu} W^{+}_{\nu} - 
W^{+\mu \nu} W^{-}_{\nu} \right) \;+\;
\overbrace{ (1+\mbox{\boldmath $\Delta \kappa_\gamma$} )}^{\kappa_\gamma} 
F_{\mu \nu} W^{+\mu} W^{-\nu} \right] \right.
\nonumber \\
&+& \left. cotg \theta_w \left[\overbrace{(1+ {\bf \Delta g_1^Z})}^{g_1^Z}
Z_\mu \left( W^{-\mu \nu} W^{+}_{\nu} - 
W^{+\mu \nu} W^{-}_{\nu} \right) 
\;+\;\overbrace{(1+\mbox{\boldmath $\Delta \kappa_Z$} )}^{\kappa_Z} 
Z_{\mu \nu} W^{+\mu} W^{-\nu} \right] \right.
\nonumber \\
&+& \left. \frac{1}{M_{W}^{2}} 
\left( \mbox{\boldmath $\lambda_\gamma$} \;F^{\nu \lambda}+
\mbox{\boldmath $\lambda_Z$} \; 
cotg \theta_w Z^{\nu \lambda}
\right) W^{+}_{\lambda \mu} W^{-\mu}_{\;\;\;\;\;\nu} \right\}
\label{hpzheqt}
\eeqn
\noi
The two couplings $\kappa_\gamma$ and $\lambda_\gamma$ are commonly 
associated with the 
magnetic and quadrupole moment of the $W$\cite{HPZH}.
Although here the interest lies with the $WW\gamma$ couplings, 
the corresponding
$WWZ$ couplings are   given to illustrate the
 fact that in \gag reactions one has a smaller
set of couplings to check than in $e^+e^- \ra W^+ W^-$. Hence,  we 
expect a better precision on the measurements (fewer parameters 
to fit). 
Furthermore, we will see that when   introducing  
well-motivated symmetry principles, one can relate
some $WWZ$ and $WW\gamma$ couplings.
Accepting this point of view of a constrained  set, the \gag mode 
can probe almost the same parameter space as the \epm. \\
\noi One of the symmetry principles that  we can, in no way, dare 
to do without is the local gauge symmetry. 
For instance, as it stands, the above Lagrangian~\ref{hpzheqt}   
  is not gauge invariant even in the QED sense. The $\lambda$ and 
$g_1^Z$ need appropriate accompanying quartic couplings. 
This Lagrangian has been written 
with the specific process $e^+e^- \ra W^+ W^-$ in mind.
Failing to render it 
explicitly gauge invariant, one can not apply it to  a process such as the one 
we are interested in: \ggww. 
There is a simple 
prescription on how to correct for this if we only want to maintain 
$U(1)_{QED}$. For instance, some time ago Aronson\cite{Aronson} has shown 
how we could write the appropriate operator accompanying the parameter 
$\lambda$. For \ggww, $\lambda_\gamma$ provides a $WW\gam$ vertex. We stress 
that this is {\em not a genuine quartic} coupling. \\
\noi
 In fact, in view of the beautiful LEP1 results that confirm the 
\sm to an unsurpassable degree of precision one should embed all the 
couplings within a set of $SU(2)\times U(1)$ gauge invariant operators. One 
ends up with a hierarchy of coupling based on the dimensionality of these 
extra (anomalous) operators. This view has, recently, rallied a large 
support. In writing these operators, we will also make an additional 
assumption motivated by the fact that the $\rho$ parameter is, to a very 
good approximation, equal to 1. We take this to be a consequence of the 
so-called custodial (global) symmetry, which reflects the fact that, 
in the absence 
of mixing with the hypercharge, the $W^\pm$ and $W^0$ have the same mass. 
Therefore, we will take the scalar sector to have this additional $SU(2)$ 
custodial symmetry. For details we refer to \cite{Hawai}. \\
There are two approaches to the construction. In the first approach one 
assumes the Higgs to be light, in which case one has a linear realisation of
symmetry breaking, while in the second approach there is no Higgs. 
With these few points spelled out, we arrive at the most {\em probable} 
set of yet-untested operators, within a linear \cite{BuchWy,Ruj} or a 
non-linear \cite{AppelquistLong,Holdom,Espriu,FLS,BDV,Feruglio,AppelquistWu} 
realisation of \ssb. \\ 

In Table~1 we have defined 
\beqn
\W_{\mu \nu}&=&\frac{1}{2} \left( 
\partial_\mu \W_{\nu}- \partial_\nu \W_{\mu} +\frac{i}{2} g [\W_\mu, \W_\nu] 
\right) 
=\frac{\tau^i}{2} \left(\partial_\mu W^{i}_\nu-\partial_\nu W^{i}_\mu
-g \epsilon^{ijk}W_{\mu}^{j}W_{\nu}^{k} \right) \nonumber \\
\B_{\mu \nu}&=&\frac{1}{2} \left( 
\partial_\mu B_{\nu}- \partial_\nu B_{\mu} \right)\tau_3 
\;\;\;\; \B_\mu=\tau_3 B_\mu
\eeqn
\noi 
(with $\W_\mu=W^{i}_{\mu}\tau^i$, the normalisation for the Pauli 
matrices is $\tr(\tau^i \tau^j)=2\delta^{i j}$).
$\Phi$ is the usual complex doublet 
with hypercharge $Y=1$ while in the 
non-linear realisation the (dim-0) matrix $\Sigma$ 
describes the Goldstone Bosons with the built-in custodial $SU(2)_c$ 
symmetry:
\beqn
\Sigma=exp(\frac{i \omega^i \tau^i}{v}) \;\; ; v=246~GeV\;\;
\;\mbox{ and} \;\;
{{\cal D}}_{\mu} \Sigma=\partial_\mu \Sigma + \frac{i}{2} 
\left( g \W_{\mu} \Sigma
- g'B_\mu \Sigma \tau_3 \right)
\eeqn

\begin{table*}[htb]
\caption{\label{linearvsnonlinear}
{\em The Next-to-leading Operators describing the $W$ Self-Interactions which 
do not contribute to the $2$-point function.}}
\vspace*{0.3cm}
\centering
\begin{tabular}{|l||l|}
\hline
{\bf Linear Realisation \hspace*{0.1cm}, \hspace*{0.1cm} Light Higgs}&
{\bf Non Linear-Realisation \hspace*{0.1cm}, \hspace*{0.1cm} No Higgs}\\
\hline
&\\
${{\cal L}}_{B}=i g' \frac{\epsilon_B}{\Lambda^2} (\cd_{\mu}
\Phi)^{\dagger} B^{\mu \nu} \cd_{\nu} \Phi$&
${{\cal L}}_{9R}=-i g' \frac{L_{9R}}{16 \pi^2} \tr ( {\bf B}^{\mu \nu}\cd_{\mu}
\Sigma^{\dagger} \cd_{\nu} \Sigma )$ \\
&\\
${{\cal L}}_{W}=i g \frac{\epsilon_w}{\Lambda^2} (\cd_{\mu}
\Phi)^{\dagger} (2 \times \W^{\mu \nu}) (\cd_{\nu} \Phi)$&
${{\cal L}}_{9L}=-i g \frac{L_{9L}}{16 \pi^2} \tr ( \W^{\mu \nu}\cd_{\mu}
\Sigma \cd_{\nu} \Sigma^{\dagger} ) $ \\
&\\
$\cl_{\lambda} = \frac{2 i}{3} \frac{L_\lambda}{ \Lambda^2} 
g^3 \tr ( \W_{\mu \nu} \W^{\nu \rho} \W^{\mu}_{\;\;\rho})$&
$\;\;\;\;\;\;\;\;---------\;\;\;\;$\\
&\\
$\;\;\;\;\;\;\;\;---------\;\;\;\;$&
$\cl_{1}=\frac{L_1}{16 \pi^2} \left( \tr (D^\mu \Sigma^\dagger D_\mu \Sigma) 
\right)^2\equiv \frac{L_1}{16 \pi^2} {{\cal O}}_1$ \\
$\;\;\;\;\;\;\;\;---------\;\;\;\;$&$
\cl_{2}=\frac{L_2}{16 \pi^2} \left( \tr (D^\mu \Sigma^\dagger D_\nu \Sigma)
\right)^2 \equiv \frac{L_2}{16 \pi^2} {{\cal O}}_2$ \\
& \\
\hline
\end{tabular}
\end{table*}

The phenomenological parameters are obtained by 
going to the unitary gauge which,  
in the non-linear case, corresponds to formally setting $\Sigma \ra${\bf 1}. 
Note that, in the 
non-linear realisation the counterpart of ${\cal L}_\lambda$ is 
relegated to a lower 
cast as it is counted as ${{\cal O}}(p^6)$:${\cal L}_\lambda \propto
\tr \left( \left[{{\cal D}}_{\mu},{{\cal D}}_{\nu} \right]
\left[{{\cal D}}^{\nu},{{\cal D}}^{\rho} \right]
\left[{{\cal D}}_{\rho},{{\cal D}}^{\mu} \right] \right)$. 
On the other hand, the operators ${\cal L}_{1,2}$ which represent genuine 
quartic couplings (they do not contribute to the tri-linear couplings) 
and involve a maximum number of longitudinal modes are 
sub-sub-dominant in the light Higgs scenario. Unfortunately, they 
do not contribute to $\gam \ra W^+W^-$ at tree-level. 

\noi By going to the 
physical gauge, one recovers the phenomenological parameters with the 
{\em constraints}:
\beqn \label{constraints}
\Delta\kappa_\gamma&=&
\frac{e^2}{s_w^2} \frac{v^2}{4 \Lambda^2}
(\epsilon_W+\epsilon_B )=
\frac{e^2}{s_w^2} \frac{1}{32 \pi^2} \left( L_{9L}+L_{9R} \right)
\nonumber \\
\Delta\kappa_Z&=&\frac{e^2}{s_w^2} \frac{v^2}{4 \Lambda^2}
(\epsilon_W -\frac{s_w^2}{c_w^2} \epsilon_B) 
=
\frac{e^2}{s_w^2} \frac{1}{32 \pi^2} \left( L_{9L} 
-\frac{s_w^2}{c_w^2} L_{9R} \right) 
\nonumber \\
\Delta g_1^Z&=&\frac{e^2}{s_w^2} \frac{v^2}{4 \Lambda^2}
(\frac{\epsilon_W}{c_w^2})=
\frac{e^2}{s_w^2} \frac{1}{32 \pi^2} \left(\frac{ L_{9L}}{c_w^2} \right)
\nonumber \\
\lambda_\gamma&=&\lambda_Z=\left(\frac{e^2}{s_w^2}\right)
L_\lambda \frac{M_W^2}{\Lambda^2}
\eeqn
In the numerical applications we will 
take $\alpha$ and ``$s_w$" at $M_Z^2$, i.e, in Eq.~\ref{constraints}
$e \ra e(M_Z^2)$ and 
$s_w^2 \ra s_Z^2=0.228$. Note that there is a one-to-one correspondence 
$L_{9L,9R}\leftrightarrow \eps_{W,B}$ for the $WWV$ parts. So, for two bosons
production or neglecting Higgs exchanges in $3V$ production, the two sets
 are 
equivalent (same constraints). We note that with this set of parameters 
$\gamma \gamma \ra W^+ W^-$ has one little drawback in the sense that 
$L_{9L,9R}$ are only probed through their sum: the vectorial 
combination $L_{9L}+L_{9R}$. 

\subsection{Tri-linear couplings in $W$ pair production}

We now turn to the analysis of the signal coming from the 
tri-linear couplings. We have derived the full set 
of the helicity amplitudes 
including both the $\Delta \kappa_\gamma$ and the 
$\lambda_\gamma$ terms. The full expressions are  given in the Appendix.
  There have, in the past, 
been a few studies including these couplings
but only Yehudai\cite{Yehudai} has given the 
helicity amplitudes. 
For the unpolarised differential cross-section, we recover the results of
Choi and Schrempp\cite{Choi} with either the anomalous $\Delta\kappa_\gamma$
or $\lambda_\gamma$ couplings.

\noi We will concentrate mainly on the $\kappa_\gamma$ 
terms as they  have a stronger link with the Higgsless scenario
\footnote{Our idiosyncratic choice is also motivated by the 
fact that a scenario 
of New Physics that admits a light Higgs (scalar) is,  in the \gag mode,
  better studied on top of the scalar resonance. }. 
The results on 
$\lambda_\gamma$ are shown in a pictural form and only succinctly commented 
upon. We should point out that the chiral Lagrangian route 
based on the set $L_{9}$ has been taken by \cite{Herrero}. In their 
calculation only the leading $W_L W_L$ contribution, in $\sqrt{s_{\gam}}$, 
has been retained 
by invoking the equivalence theorem in the chiral limit. We confirm their 
result in the limit of non-forward processes where the $M_W \ra 0$ limit
can be taken and where the leading terms can be unambiguously isolated at the 
amplitude level
\footnote{There has been a reappraisal \cite{Knetteret,Dobadoet} 
of the equivalence theorem 
as applied to effective Lagrangians with the conclusion that care should be 
taken when applying the equivalence theorem. In all our studies we avoided 
the use of the equivalence theorem.}
. Their results are in fact a direct adaptation of the chiral 
Lagrangian calculation of $\gam \ra \pi^+ \pi^-$.

The first comment 
we would like to make is that, for small values 
of the anomalous couplings, the sensitivity on the measurement of the 
coupling does not grow much with the increase of energy especially 
for the $\Delta \kappa_\gamma$ coupling. This is 
due to the fact that,  for small couplings, the main contribution to the 
cross-section comes 
from terms that are linear in the anomalous coupling. Unfortunately, 
the  interference between the \sm amplitude and the new coupling 
is not ``effective" in the sense that the energy enhancement brought by 
the anomalous coupling is 
tamed by the energy drop of the corresponding \sm amplitude. 
For instance, with $L_9$ ($\Delta \kappa_\gamma$) 
the leading, in $s_{\gam}$, contribution is from 
the $J_Z=0$ amplitude in $W_L W_L$. The latter, however, 
decreases as $1/s_{\gam}$ for 
the \sm part. This explains why the net effect is sensibly the same 
as in the $J_Z=2$, as far as the interference is concerned. 
To wit, for central $W$'s the leading terms in $W_L W_L$ are 
\beqn
{\cal M}^{\smx}_{++LL} \sim 
4 \pi \alpha \times \frac{-8}{\gamma \sin^2 \theta}\;\;\;\;
{\cal M}^{\smx}_{+-LL} \sim 4\pi \alpha \times 2 \;\;\; 
\mbox{\rm while} \nonumber \\
{\cal M}^{\Delta \kappa_\gamma}_{++LL} \sim 
4 \pi \alpha \times \gamma\;\;\;\;
{\cal M}^{\Delta \kappa_\gamma}_{+-LL} \sim 
-4 \pi \alpha \times 4\;\;\;\;
\eeqn
Therefore, for this particular coupling, 
if we do not reconstruct special correlations that would give information
on all the elements of the density matrix, going higher in energy
does not pay much. 
Unless, of course, the couplings are large enough that terms 
quadratic in $\Delta \kappa_\gamma$
play an important role. However, within the chiral Lagrangian 
approach, quadratic terms correspond to  higher order terms in 
the energy expansion as those mimicked by the interference terms 
between the $\cal{O}(p^6)$ and the lowest (universal) terms. 
\begin{figure*}[hbt]
\begin{center}
\caption{\label{kalamlog400}{\em Deviation in the various polarised cross-sections due to 
the anomalous couplings before folding with the spectrum. 
The effect of an angular 
cut on the total cross-section is also shown.}}
\mbox{\epsfxsize=14cm\epsfysize=9.5cm\epsffile{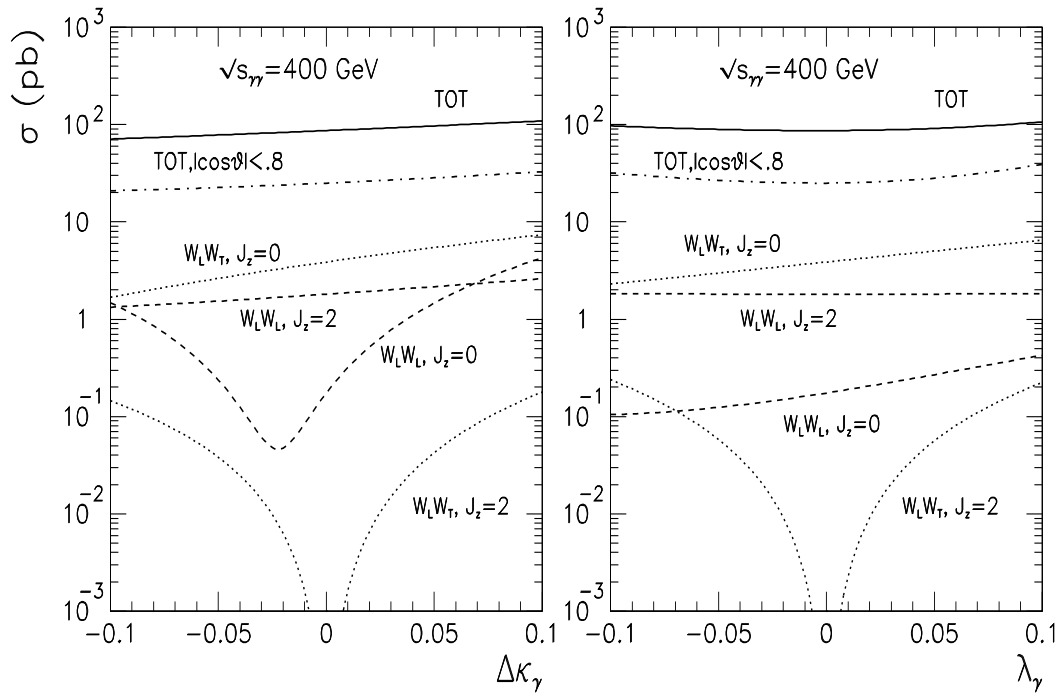}}
\vspace*{-1.cm}
\end{center}
\end{figure*}
The other prominent effect is in the 
production of $W_L W_T$ states in the $J_Z=2$ which is {\it absent} 
in the \sm at 
tree-level. Unfortunately, while this is a clear signature 
of the tri-linear term, the yield in this channel 
is even smaller than the $W_L W_L$. 
At the end of the day, the most troublesome feature is the large 
$W_T W_T$ production which is essentially of a gauge nature. 
Fig.~\ref{kalamlog400} shows that the largest deviation does occurs for $W_L W_L$ in the 
$J_Z=0$ channel. However, this channel is completely buried under 
the transverse modes even when angular cuts are imposed. \\

\begin{figure*}[t]
\begin{center}
\caption{\label{wwtotspec}{\em Variation of the cross-section as a function of 
$\Delta \kappa_\gamma$ including the unpolarised spectrum and the 
``0-dom". The effect of cuts is also shown.}}
\vspace*{-1cm}
\mbox{\epsfxsize=11.5cm\epsfysize=11.cm\epsffile{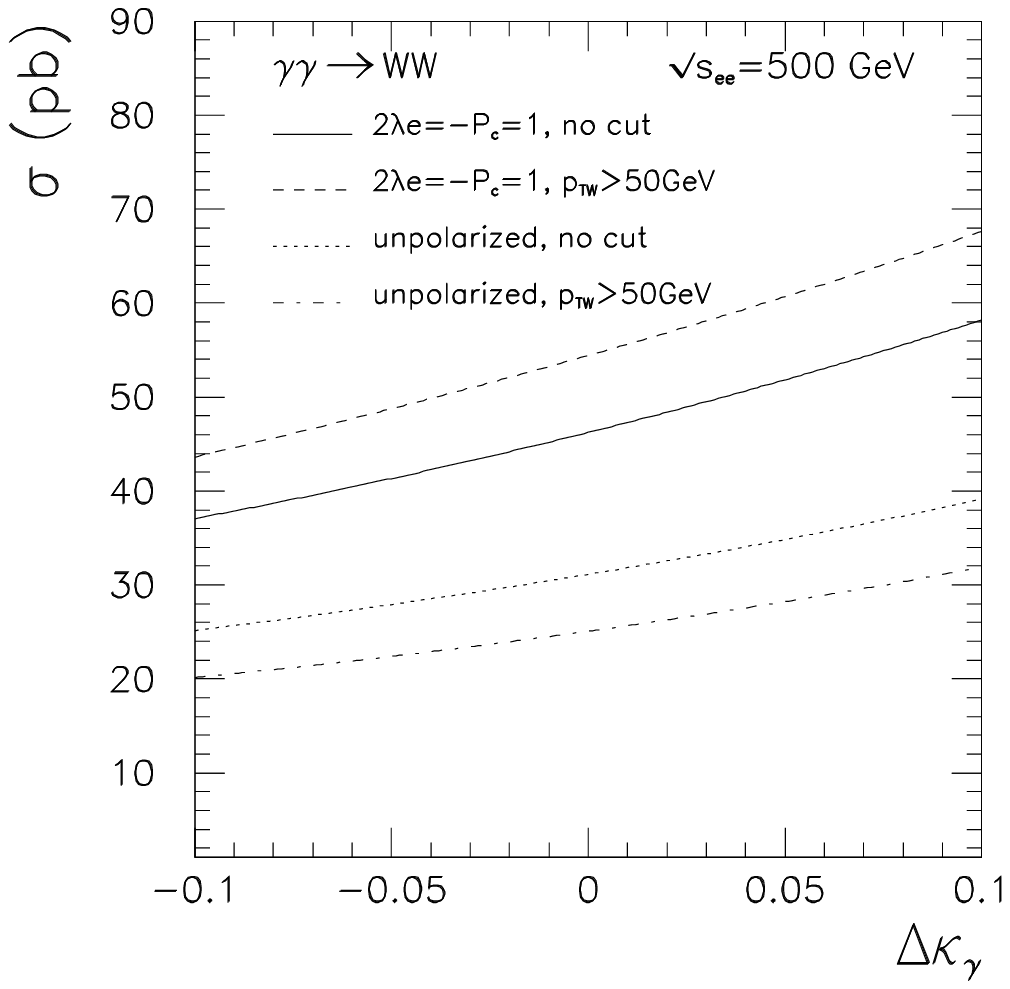}}
\vspace*{-1.cm}
\end{center}
\end{figure*}

Our analysis is preliminary, in the sense that ultimately 
we would like to generate the four-fermion final state from the decay of the 
$W$'s keeping the full spin-correlation. We would then apply cuts on the 
observed fermions. Nonetheless, we applied cuts in the laboratory frame by 
requiring $p_{T}^{W^{\pm}}>50$~GeV for 
$\sqrt{s_{ee}}=500$~GeV. 
We consider 
an unpolarised as well as a ``0-dom" spectrum Fig.~\ref{spectre34}. With the latter, due to the fact 
that the effect of $L_9$ does not sensibly grow with the energy, we do not 
gain much with polarisation whose effect is also to give a peaked 
spectrum at high-energy. This is seen in Fig.~\ref{wwtotspec}.
To extract the limit on $L_9$ ($\Delta\kappa_\gamma$) we only considered 
the sample of $WW$ events that does not consist of double leptonic channels 
nor of any $\tau$ events. We have taken an efficiency of $0.8$ and have not 
aimed at reconstructing the final $W$ polarisation. 
We take a luminosity of $\int \cl_{ee} =10fb^{-1}$. With these values 
(even with  $p_{T}^{W}>50$~GeV) the statistical error is too small compared 
to the systematic error that we have assumed to be $1\%$ with these cuts 
and the above clean sample. 
Taking a 3$\sigma$ deviation our results for both the polarised (``0-dom") 
spectrum Fig.~\ref{spectre12} and the unpolarised case are 
\beqn
  -9.6  &<&L_{9L}+L_{9R} \;<\;9.6 \;\;\mbox{for the unpolarised spectrum} 
\nonumber \\
 -10.4 &<&L_{9L}+L_{9R}\;<\;9.6 \;\;\mbox{\rm for the ``0-dom" spectrum}
\eeqn

These values are in qualitative agreement with the ones given by 
both Yehudai\cite{Yehudai} and by Choi and Schrempp \cite{Choi} 
(when interpreted as $\Delta \kappa_\gamma$). 
We have also compared these limits with the ones obtained in the \epm 
mode\cite{BMT} and from the LHC\cite{FLS} within the same 
constrained set of the dominant 
operators of the chiral Lagrangian\cite{Hawai}. 

\begin{figure*}[p]
\caption{\label{gg19}{\em Comparison between the expected bounds on the 
two-parameter
space $(L_{9L},L_{9R}) \equiv (L_W,L_B) \equiv (\Delta g_1^Z, 
\Delta \kappa_\gamma$ (see text for the conversions) at the NLC500, LHC 
and LEP2. The NLC bounds are from 
$e^+e^- \ra W^+W^-$\protect\cite{BMT}, $W^+W^-\gamma, W^+W^-Z$\protect\cite{nousggvv} 
(for the latter these are one-parameter fits) and 
$\gamma \gamma \ra W^+W^-$.
The LHC bounds are from $pp \ra WZ, W\gamma$\protect\cite{FLS}. 
We also show (``bars") 
the limits from a single parameter fit.}}
\begin{center}
\vspace*{-2.cm}
\mbox{\epsfxsize=15.cm\epsfysize=20.cm\epsffile{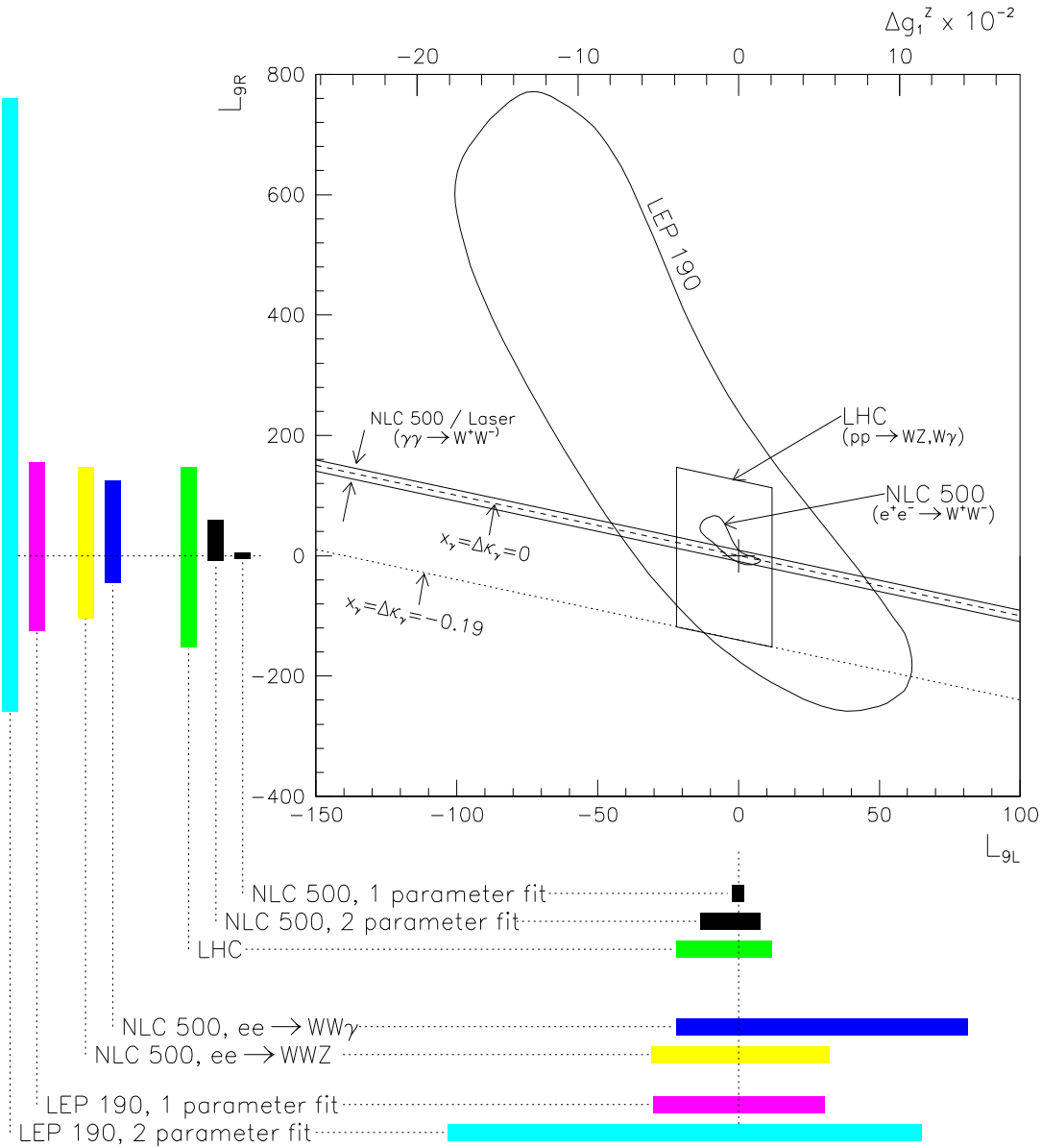}}
\vspace*{-1.cm}
\end{center}
\end{figure*}

The conclusion of the comparison, beside the fact that the 
next \epm at 500~GeV 
does much better than the LHC and improves considerably on the expected 
LEP200 bounds, is that the allowed parameter space 
spanned by the two $L_9$ would be further reduced (by about $50\%$) given 
the availability of the \gag mode as shown in Fig.~\ref{gg19}. 

\subsection{Effects of genuine quartic couplings and the scalar connection}
Within the spirit of the classification of the anomalous couplings based 
on gauge-invariance and scaling\cite{Hawai}, 
{\it genuine} quartic couplings, 
$WW\gamma \gamma$, only appear at the next-to-next-to-leading. 
We\cite{nousggvv,SanDiego}have looked 
at these operators maintaining the custodial symmetry. Of course, going to the 
next order one has many more operators contributing to the trilinear
couplings so that the above relations between 
the $WWZ$ and $WW\gamma$ couplings are no longer maintained. This means 
that one could not make a direct comparison with the two-parameter fits 
in \epm. Nonetheless, as far as tri-linear couplings are concerned, in the 
no-Higgs scenario, \gag measures only a collective $\Delta \kappa_\gamma$ 
 while 
$\lambda_\gamma$ is already  of higher order in the energy expansion and 
of a purely 
gauge origin. 
So, if the measurements at \gag are precise enough one may hope 
to also check the $WW\gamma\gamma$ couplings. There are in fact only two 
additional $WW\gamma\gamma$ Lorentz structures. 
In the non-linear realisation these can be embedded in the 
following sets: 
\beqn
{{\cal L}}_{0}^{2\gamma}&=-&\frac{L^{2\gamma}_{0}}{\Lambda^2} \left\{ 
k_{0}^{W} g^2 \tr(\W_{\mu \nu}\W^{\mu \nu}) + k_{0}^{B} g'^{2} 
\tr(\B_{\mu \nu}\B^{\mu \nu}) + 
k_{0}^{WB} g g'\tr(\W_{\mu \nu}\B^{\mu \nu})\right\} \times \nonumber \\
&&\tr(\cd^\alpha \Sigma^\dagger \cd_\alpha \Sigma)
\nonumber \\
{{\cal L}}^{c}_{2\gamma}&=-&\frac{L^{2\gamma}_{c}}{\Lambda^2} 
\left\{
k_{c}^{W} g^2 \tr(\W_{\mu \alpha}\W^{\mu \beta}) + 
k_{c}^{B} g'^{2} \tr(\B_{\mu \alpha}\B^{\mu \beta}) + 
k_{c}^{WB} gg' \tr(\W_{\mu \alpha}\B^{\mu \beta})\right\} \times \nonumber \\
&&\tr(\cd^\alpha \Sigma^\dagger \cd_\beta \Sigma)
\eeqn

\noi
For $\gamma \gamma$ reactions, by only making explicit the $U(1)_{QED}$  
symmetry, these operators map into the set that we have previously defined
through the Lagrangian
\beqn
 {{\cal L}}_{0}^{2\gamma}&=-&\frac{\pi\alpha }{4\Lambda^2} a_0
  g^2  F_{\mu \nu}F^{\mu \nu} 
 \tr(\cd^\alpha \Sigma^\dagger \cd_\alpha \Sigma)
\nonumber \\
{{\cal L}}_{c}^{2\gamma}&=-&\frac{\pi\alpha }{4\Lambda^2} a_c
  g^2  F_{\mu \alpha}F^{\mu \beta} 
 \tr(\cd^\alpha \Sigma^\dagger \cd_\beta \Sigma)
 \eeqn
with
\beqn
a_{0,c}=\frac{4e^2}{s_w^2} 
L^{2\gamma}_{0,c} \left(k_{a,c}^W + k_{a,c}^B + k_{a,c}^{WB} \right) 
\eeqn  
The custodial symmetry imposed on these couplings means that, in
leading order in $s$, they contribute 
in the same way to the $\gam \ra WW$ and to the $\gam \ra ZZ$, that is we 
have 
\beq
{\cal A}(\gamma \gamma \ra W^{+}_{L}W^{-}_{L})\;=\;
{\cal A}(\gamma \gamma \ra Z_{L} Z_{L})
\eeq
The ``neutral" operator, $L_0^{2\gamma}$, only contributes to the $J_Z=0\; 
(\la_1=\la_2)$ amplitude
\beqn
{\cal A}^{0}_{\la_1 \la_2;00}&=& a_{0}^{V} (1+\la_1 \la_2) \;\;
          (2 - \gamma ) \hspace*{0.5cm} {\rm and} \hspace*{0.5cm} 
{\cal A}^{0}_{\la_1 \la_2;\la_3 \la_4}=a_{0}^{V} \;\; 
 (1 + \la_1 \la_2)(1 + \la_3 \la_4) \;\; \nonumber \\
a_{0}^{W}&=&\pi \alpha \;\; \frac{s}{4 \Lambda^2}\; a_0 \;
\hspace*{0.5cm} {\rm ,} \hspace*{0.5cm} a_{0}^{Z}=\frac{a_{0}^{W}}{c_W^2}
\hspace*{0.5cm} {\rm and} \hspace*{0.5cm} \gamma=\frac{s}{M_{V}^{2}}
\eeqn

For the ``charged" operator both the $J_Z=2$ and $J_Z=0$ contribute. This
property already allows a possible separation of the effects of 
the two quartic couplings

\beqn
{\cal A}^{c}_{\la_1 \la_2;00}&=&a_c^V \; \left[
(1+\la_1\la_2) \;\;
          (2 - \gamma) + \frac{\gamma}{2}\;\; 
(1-\la_1\la_2) \sin^2\theta \right]
\nonumber \\
{\cal A}^{c}_{\la_1 \la_2;\la_3 0}&=&a_{c}^{V} \; 
\sqrt{\frac{\gamma}{2}}\;\; (\la_1-\la_2)(1+\la_3 \cos\theta) 
\sin\theta \\
{\cal A}^{c}_{\la_1 \la_2;\la_3\la_4}&=& a_{c}^{V} \; 
\left\{  (1+\la_1\la_2) (1+\la_3\la_4)\;+\; \right. \nonumber \\
& & \hspace*{1cm} (1-\la_1\la_2) \left. 
\left[ (1-\la_3\la_4 \cos^2\theta) + \la_1\la_3 (1-\la_3\la_4) 
\cos\theta \right]\right\} \nonumber \\
a_{c}^{W}&=&\pi \alpha \;\; \frac{s}{8 \Lambda^2}\; a_c\;
\hspace*{0.5cm} {\rm and} \hspace*{0.5cm} a_{c}^{Z}=\frac{a_{c}^{W}}{c_W^2}
\nonumber
\eeqn
Here ``effective" interference does take place in the $J_Z=2$ channel. 
As in the \sm,  neither of the quartic couplings 
produces $W$'s or $Z$'s with opposite-helicities in the $J_Z=0$ mode. This is 
therefore an obvious way, in $W^+W^-$,
to disentangle their effect from that of the anomalous trilinear couplings 
$\lambda_\gamma, \kappa_\gamma$, which do not share 
this property. Furthermore, the trilinear couplings 
 contribute only to $\gamma \gamma \ra W^+W^-$, 
whereas the $SU(2)$ symmetric quartic couplings contribute also to
 $\gamma \gamma \ra ZZ$. 
\noi
It is very important to have polarised photon beams although
disentangling the two couplings is possible even if no initial beam 
polarisation 
were available either by a careful scan of the angular distribution 
(the $a_0$ coupling contributes a flat distribution) or through measurements 
of the final $W$ and $Z$ polarisations. 

To extract limits on the two operators, we use the total cross-section
with polarised or unpolarised beams. 
 As for the analysis of the trilinear 
couplings, a luminosity ${\cal L}=10 fb^{-1}$ is assumed for
$\sqrt{s_{ee}}=500$~GeV and the efficiency for W pair
detection is taken to be 80\%. We ignore events with tau's or
two neutrinos in the final state and a cut on $|\cos\theta_W^*|<.7$ is imposed.
Assuming a 1\% systematic error and taking $\Lambda=1~TeV$, a 3$\sigma$
deviation  with unpolarised beams give the limits
\beqn
  -25 < a_0 < 9.4 \;\;\;\; -19< a_c < 10
\eeqn
These can be improved with the use of polarisation, even when taking a 
2\% systematic error to take into account the additional error introduced
by the measurement of the polarisation, we find at 3$\sigma$
\beqn
  -11 < a_0 < 5.5 \;\;\;\; -10 < a_c < 6.6
\eeqn
These bounds were obtained with the polarisation that
gave the optimum sensitivity to each operator. Keeping
 $2\lambda'_eP'_c=-1$ in both cases, we took $2\lambda_e P_c=-1$,
a  $J_Z=0$-dominated spectrum, for $a_0$ and $2\lambda_e P_c=1$,
 a  $J_Z=2$ spectrum for $a_c$. \\ 

\subsubsection{Reconstructing the final polarisations} 
A reconstruction of the W polarisation, allowing 
to pull out the longitudinal W's, could help in 
establishing as well as confirming signals  
from the symmetry-breaking sector
\cite{nousgglumi}. To be able to determine the specific 
sign of the transverse helicity of the $W$ one would use  
the leptonic ($e, \mu$) decays of the $W$ in order to 
reconstruct the angular distribution
of the charged lepton in the rest frame of the decaying W.
The problem is, of course, to boost back from the lab frame to the
rest frame. In $e^+e^- \ra W^+W^-$ (with little or no beamstrahlung) 
this is done simply from the knowledge of the beam energy that is  also the energy 
of the $W$. In \gag the energy must be reconstructed. In $WW$ events where 
only one of the $W$ decays leptonically this should be possible without 
much error since the missing $p_T$ of the neutrino could be inferred. 
On the other hand, for many analyses 
it may be sufficient to extract the differential 
cross sections for 
longitudinal and transverse vector bosons as well as the spin correlation that 
do not require charge identification. In the \gag mode one could, 
in this case, use the all-hadronic decays.

\subsubsection{Higgs resonance in \ggww?} 
If the Higgs is heavy it will decay preferentially into a $W$ pair 
(see Fig.~\ref{Branchingh}). One could then envisage to exploit its 
\gag coupling to produce it as a resonance\cite{Gunionzz,Richard}. The 
problem will be the incredibly huge $WW$ continuum. A recent\cite{Zappala} 
study has looked 
at this aspect by also taking into account the interference between the 
continuum and the $s-$channel Higgs exchange. A large destructive interference 
is found in the \ggww channel which implies that a Higgs Boson of mass 
$M_H\geq 200GeV$ is manifest as a resonant {\em dip} in the $M_{WW}$ invariant 
mass distribution. The conclusion is based on a setting that enhances the 
$J_Z=0$ component ($P_c=P'_c$ and $\lambda_e=\lambda'_e=0$) with  a cut 
on the scattering angle of the $W$ measured in the $WW$ cms and by using 
the all-hadronic decays. Even with the most optimistic value on the 
resolution of the $WW$ mass, $\Delta M_{WW}=5$GeV, and an integrated 
luminosity of 20$fb^{-1}$ the statistiscal significance of the resonant 
dip is not encouraging. Doubling the value of the resolution (which is still 
very hopeful) the signal disappears. Therefore, there seems   to be 
very little (if none) prospect to look for the Higgs in this channel.

 \section{\boldmath{ $\gam \ra ZZ$}} 
This is a very interesting process for a variety 
of reasons. First, it is an effect of totally quantum origin. 
In its ``pure gauge symmetric" manifestation 
with both  $Z$ transverse 
this reaction is to be likened to the scattering of light-by-light. 
In its ``symmetry breaking phase", {\it i.e.}, 
in the $Z_L Z_L$ case one could draw the analogy with 
$\gam \ra \pi^0 \pi^0$. 
In fact, the early interest in this 
reaction was in the production of the longitudinal modes. 
Within the \sm realisation of \ssb and for a Higgs 
with mass above $\sim 2M_Z$, this reaction would, at first, 
seem more advantageous than $WW$ for revealing such a scalar as  
a resonance\cite{Gunionzz}\footnote{The branching ratios as well as 
the total width of the Higgs as a function of its mass are shown in 
Fig.~\protect\ref{Branchingh}.}.
This is because the $WW$ continuum is overwhelming while \ggzz only 
takes place at one-loop. 
Thus, one expected this 
contribution to be too small so that $ZZ$ would qualify as an indicator 
of a clean Higgs signal or else 
as an unambiguous manifestation of New Physics. 
In this case, 
identifying the longitudinal modes with the Goldstone bosons, one is back 
to the same topic covered here at some length especially in the sessions 
dedicated to DA$\Phi$NE, that is the $\gam \ra \pi^0 \pi^0$. In the 
context 
of chiral perturbation theory, this reaction has received 
a particular attention\cite{Bijnens2pi0,Donoghue2pi0} 
because it has been regarded to be 
a good test for the application of the formalism. 
One has argued so, because as with \ggzz, 
the leading contribution to $\gam \ra \pi^0 \pi^0$ occurs at $\cal{O}(p^4)$ 
and at this order it is completely predictable solely from the universal 
$\cal{O}(p^2)$ Lagrangian and does not require any counterterm. 
As is known now, it turned out that a naive 
and straightforward application of the formalism (at least to lowest 
order) was misleading\cite{Morgan,Truong2pi0} 
and did not quite reproduce the data\cite{Crystal}. \\
While this is a first warning if one were tempted to copy {\it mutatis 
mutandis} the Goldstone Boson sector of $QCD$ to the longitudinal 
$ZZ$ in \gag, by all means there is ``more to it" in the $ZZ$ than 
the longitudinals: it is crucial to check the importance of the 
transverse. 
\subsection{\ggzz in the \sm and the Higgs resonance}
\begin{figure*}[p]
\begin{center}
\caption{\label{ggzzee}{\em Cross section for ZZ production 
after convoluting with different spectra. The cut on the Z scattering 
angle is shown. The three combinations of the final Z polarisations are 
shown. The LL curves are indicative of the Higgs whose mass can be 
read off from the position of the bump/peak: 
$M_H=300,500,800,1000$ and $\infty$. From \protect\cite{Jikiazz}.}}
\vspace*{-1.5cm}
\mbox{
\mbox{\epsfxsize=8cm\epsfysize=10.cm\epsffile{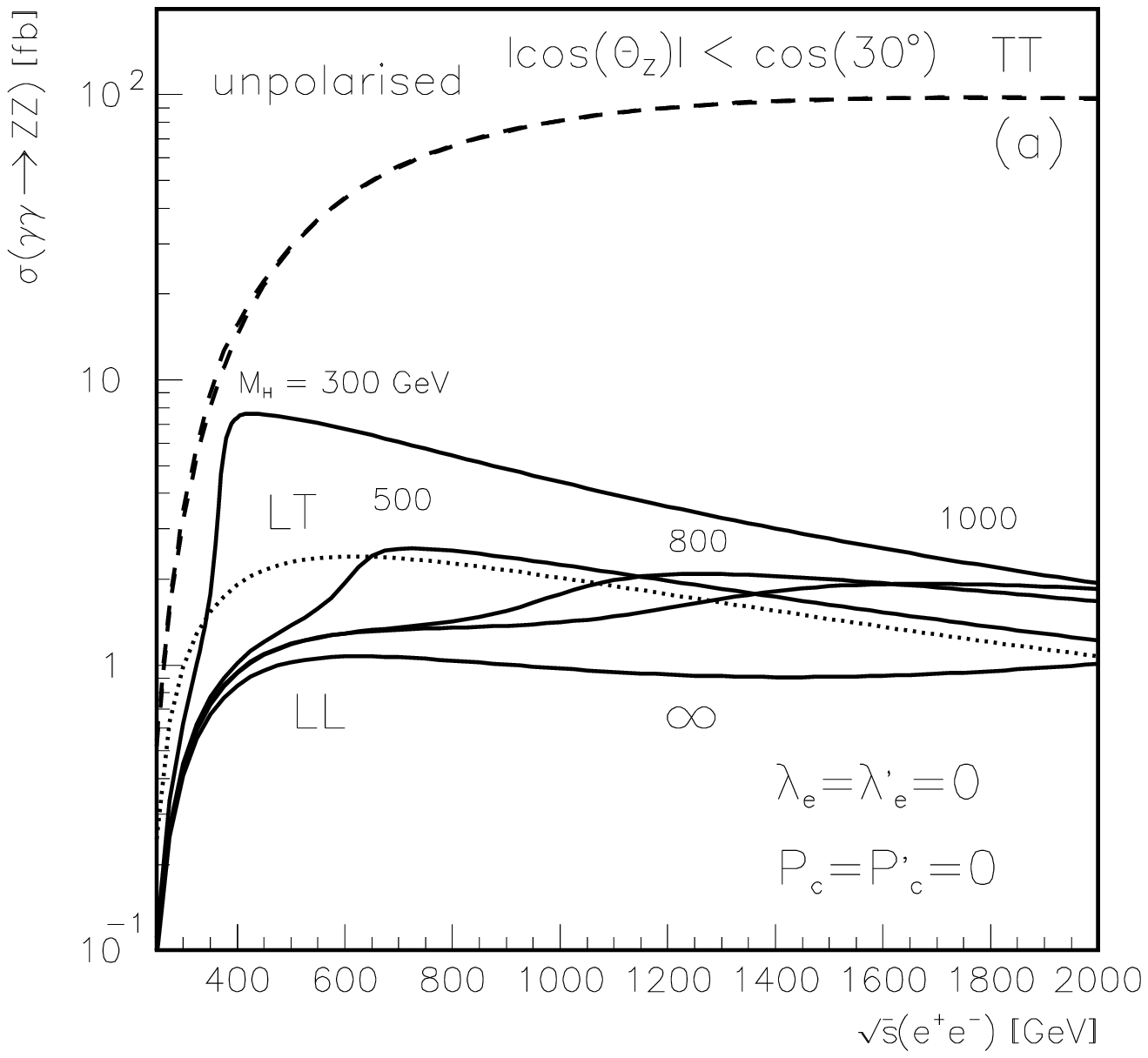}}
\mbox{\epsfxsize=8cm\epsfysize=10.cm\epsffile{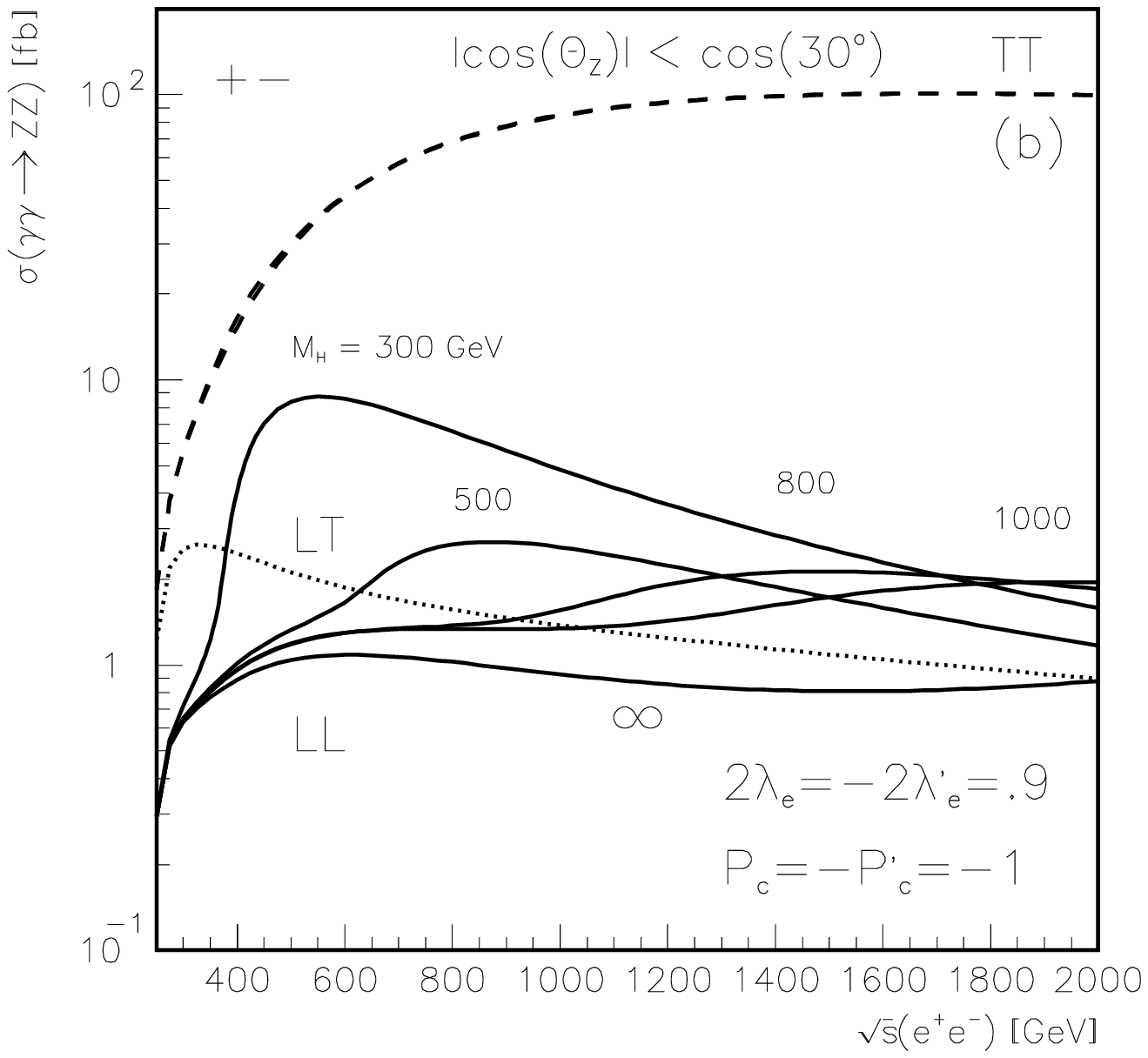}}}
\vspace*{-3cm}
\mbox{\epsfxsize=8.5cm\epsfysize=10.5cm\epsffile{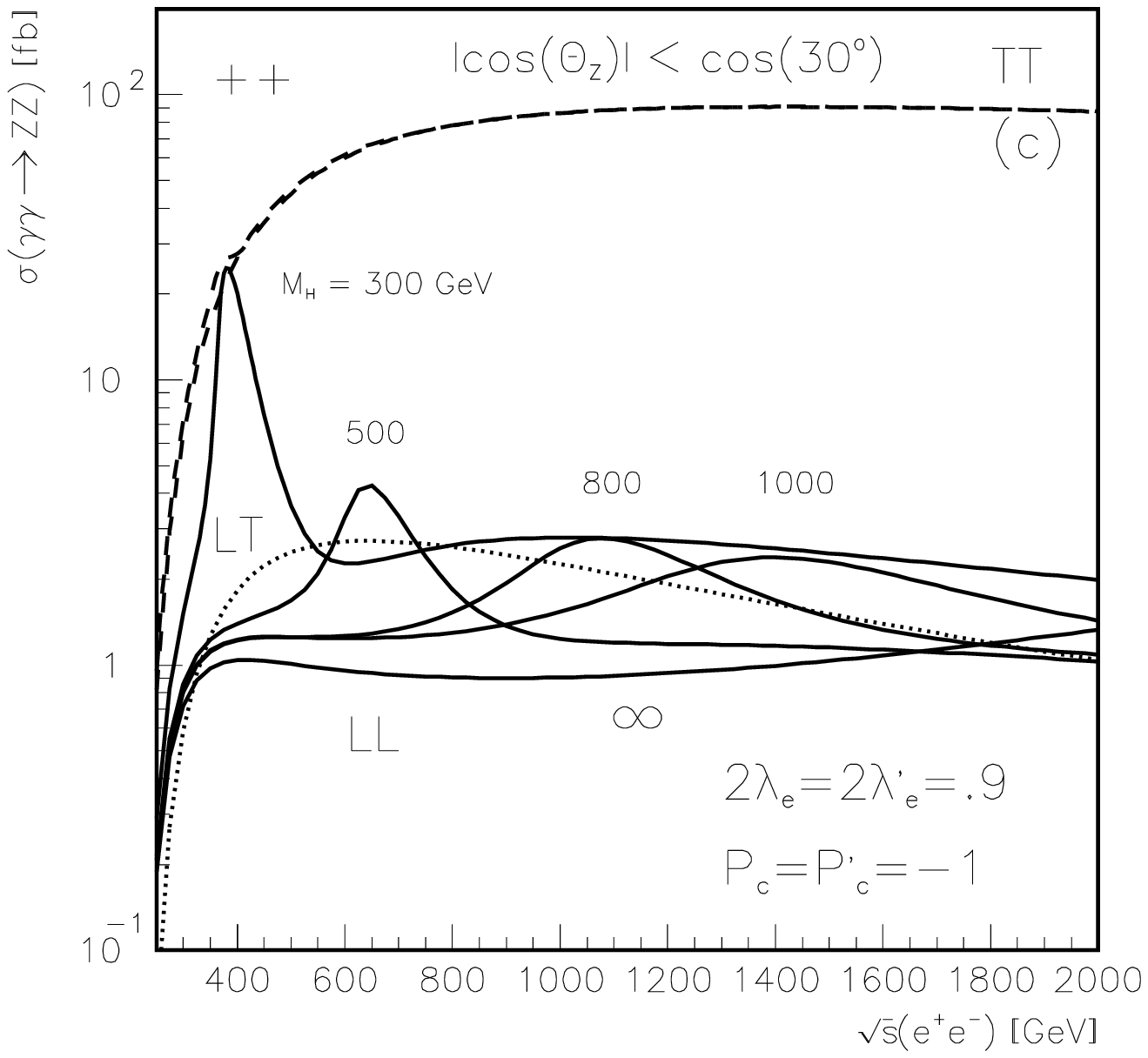}}
\vspace*{-1.cm}
\end{center}
\end{figure*}

\begin{figure*}[hbt]
\begin{center}
\caption{\label{ggzzmzz1} {\em Searching for the 
Higgs in the invariant $ZZ$ mass distribution at \epm cms 
of  400 and 500GeV.
The three combinations 
of the final polarisations as well as the sum over all final helicities 
are shown. The illustrative masses of the Higgs are shown. 
From \protect\cite{Jikiazz}.}}
\vspace*{-1.cm}
\mbox{
\mbox{\epsfxsize=8cm\epsfysize=10.5cm\epsffile{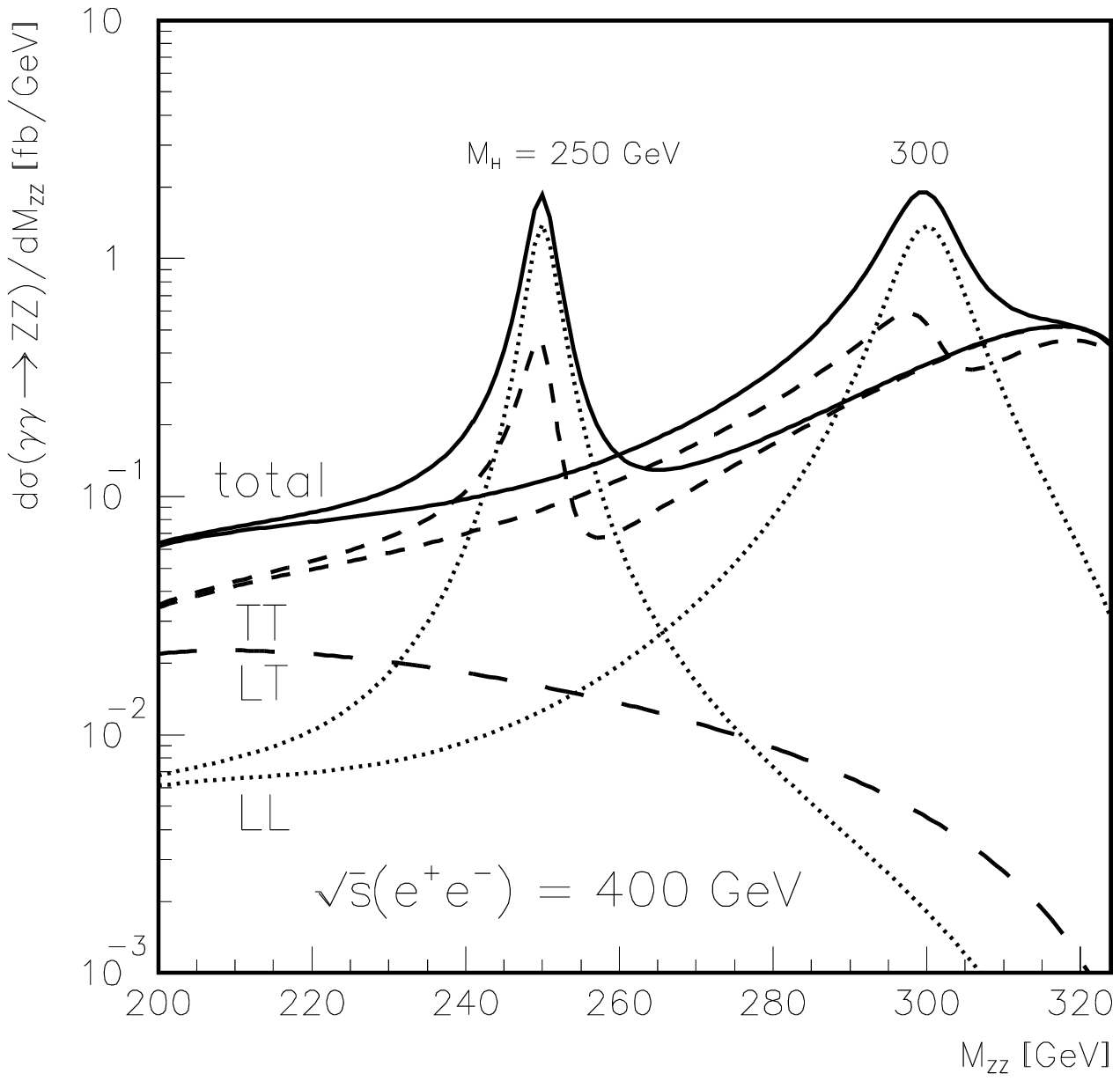}}
\mbox{\epsfxsize=8cm\epsfysize=10.5cm\epsffile{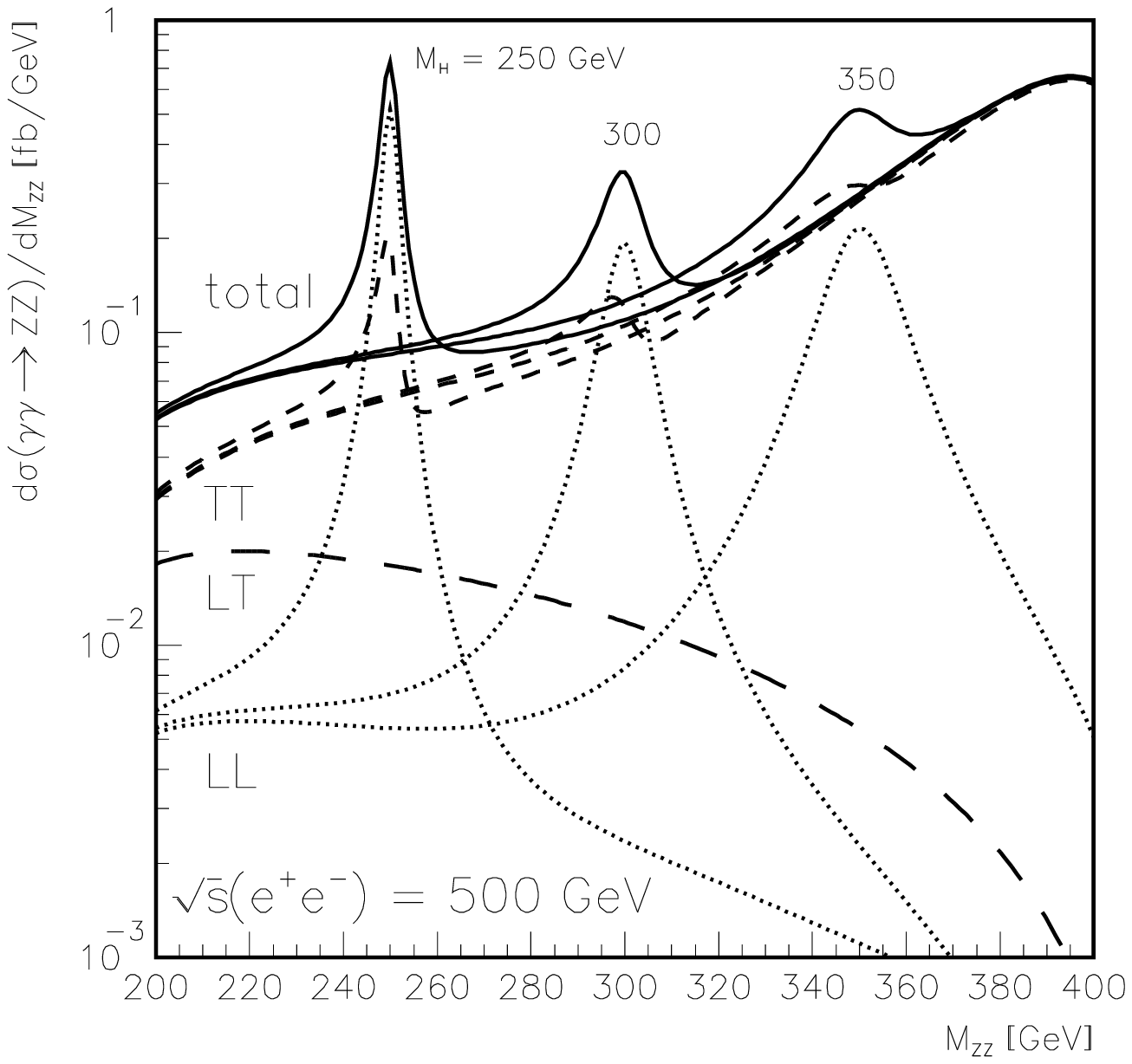}}}
\vspace*{-1.cm}
\end{center}
\end{figure*}

\begin{figure*}[hbt]
\begin{center}
\caption{\label{ggzzmzz2}{\em As in the previous but 
for 1.5TeV \epm. The infinite mass Higgs is also shown. 
From \protect\cite{Jikiazz}.}}
\vspace*{-1.cm}
\mbox{\epsfxsize=12.5cm\epsfysize=12.5cm\epsffile{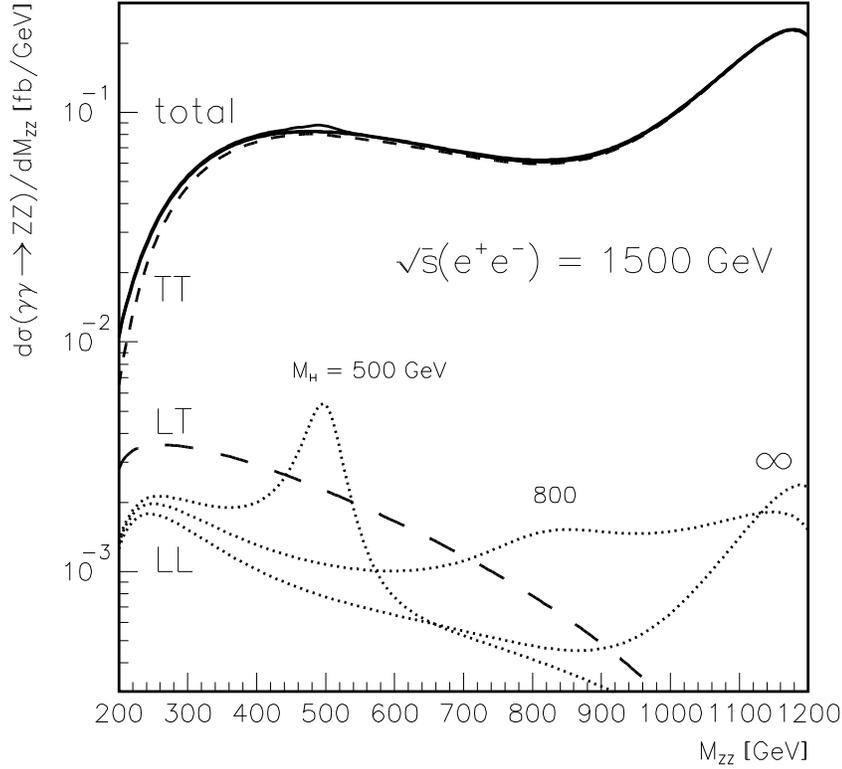}}
\vspace*{-1.cm}
\end{center}
\end{figure*}

The first full calculation of the one-loop \ggzz within the \sm 
is a beautiful computation 
by Jikia\cite{Jikiazz} whose results have been repeatedly 
confirmed\cite{Bajc,Berger,Dicus,Veltman}. As one might fear by having 
the ubiquitous \ggww in mind, 
even for this one-loop 
example it turned out that, once again, the 
transverse modes are overly dominant, especially at high energy. \\
\noi As we mentioned, 
when one examines the transverse modes it is almost as if one were 
studying the scattering of light-by-light. The most 
interesting novel aspect, compared to the pure case of QED, is that 
in the context of the electroweak interaction one has a pure non-Abelian 
contribution mediated by $W$ loops. 
This contribution has been looked at 
some time ago\cite{Sirlingg,Megg}. Already then, we stressed 
\cite{Megg} the importance of 
doing such calculations 
within the background gauge. 
Another similar process 
that has also been calculated within the same spirit is 
the decay of the $Z$ into three photons
\cite{Z3gamus,Z3gam2,Z3gamnigel,Jikiagz}. 
Jikia and Tkabladze have also recently calculated the $\gam \ra \gam$\cite{Jikiagg} 
and $\gam \ra \gamma Z$ \cite{Jikiagz}
cross section in the context of a high energy \gag collider.\\
\noi 
Although the $\gam \ra 
\gam$ and $Z\ra 3\gamma$ processes are
quite attractive and have a clear connection with \ggzz, the longitudinal 
modes for these processes 
are insignificant and decouple quite trivially. Physics-wise 
\ggzz is much ``richer" because the longitudinal modes play a crucial role. 
Our discussion on the characteristics of the \ggzz 
reaction within the \sm and the Higgs resonance in $ZZ$ is based 
almost entirely 
on the results of Jikia\cite{Jikiazz}. 
 The main conclusions of the study of \ggzz are summarised in 
Fig.~\ref{ggzzee},\ref{ggzzmzz1},\ref{ggzzmzz2} that we have 
borrowed from \cite{Jikiazz}. 
All the figures are with $m_t=120GeV$. To produce central events 
a cut on the $Z$ production angle measured in the 
\gag cms is imposed: $30^0 <\theta_Z<120^0$. 
Since, from the point of view of the \sm, 
the motivations for studying this reaction lie primarily 
on its ability 
to bring out the heavy Higgs peak formation\cite{Gunionzz,BordenHiggs},
 the issue of polarising 
the initial state is central. One must aim at creating a 
$J_Z=0$ environment. And, if possible, reconstruct the longitudinal 
polarisation of the $ZZ$ that signals the Higgs while trying to reduce 
the transverse. \\
 
\noi 
As is the hallmark with \ggww, one sees (Fig.~\ref{ggzzee}) 
that once again the transverse modes, 
either in the $J_Z=0$ or $J_Z=2$ state dominate, by far, the cross-section 
especially as the energy increases. This is due essentially to the $W$ 
loops. We effectively have that the $WW$ produced in \gag rescatter 
into $ZZ$. This shows, from a different perspective, 
how \ggww permeates so many aspects of $W$ physics in 
the \gag collider. \\
\noi It is clear 
that unless one recreates a $J_Z=0$ there would be little hope of extracting 
a Higgs signal apart, perhaps, if the Higgs mass is 
below $\sim 300GeV$ and the \epm energy below $500$GeV. 
Even if the ``0-dom" is realised (see Fig.~\ref{ggzzee}c), such that 
the $J_Z=0$ part of the \gag spectrum is peaked for $\sqrt{s_{\gam}}$ 
values around 
the Higgs mass, one observes that Higgs masses 
below $\sim 350$GeV could emerge above the $Z_TZ_T$ continuum. 
For instance, 
a Higgs mass of $500GeV$ which would require an optimum set-up in the 
``0-dom" environment that corresponds to $\sqrt{s_{ee}}=700GeV$ would 
already be completely buried under its intrinsic $Z_T Z_T$ background. 
Therefore, before addressing the issue of the identification of 
the $ZZ$ final state, one learns that a high energy \epm 
machine with $\sqrt{s_{ee}}$ in TeV range would not be of 
much help in revealing the resonant 
Higgs formation.

As to what happens in a Higgs search at a given energy, 
the results are displayed
in Fig.~\ref{ggzzmzz1}. 
The laser set-up is such that one filters out a predominant 
$J_Z=0$. In the calculation of Jikia this filtering is obtained through what 
we have called a ``0-dom" spectrum that is peaked towards the maximum 
$\hat{s}_{\gam}$ (see Fig.~\ref{spectre12}). 
This may not be the optimal choice of polarisation as we could arrive 
at a $J_Z=0$ environmemnt in a flat spectrum set-up that would be 
more appropriate if one were searching for the Higgs whose mass would not 
have been determined by another means
\footnote{As compared to the IMH this seems less justified as Higgs masses 
with $M_H>2 M_Z$ would be easily discovered at the LHC.}. 
Anyhow, we see that at $\sqrt{s_{ee}}=400GeV$, the situation is quite bright 
even with the ``0-dom" set up:
the Higgs resonance is clearly evident over the $TT$ continuum all the way 
up to the kinematic limit. With $\sqrt{s_{ee}}=500GeV$ in the ``0-dom" 
set-up it becomes already difficult to extract a Higgs with $M_H\sim350GeV$. 
It has been shown\cite{Dicus} that the situation improves for these values 
of the Higgs mass and the energy if we change the polarisation 
settings. In fact, if we take what we have called the polarised 
``broad spectrum" (see Fig.~\ref{spectre34}a) 
so that for small $M_{ZZ}$ one has 
a dominant $J_Z=0$, one could still see a peak in the $M_{ZZ}$ 
invariant mass for Higgs masses below $400$GeV at a $1TeV$ \epm machine. 
However, it is not clear whether the statistical significance of the signal 
in this case is satisfactory as the luminosity in the broad spectrum 
has been uniformally spread over a large energy range\cite{Dicus}. 
From the perspective of observing the Higgs resonance beyond TeV 
\epm energies, the situation as displayed in Fig.~\ref{ggzzmzz2} 
becomes totally hopeless as the 
transverse $ZZ$ are awesome. \\

So far, the pictural description of the $H$ resonance, if any, lacks 
a careful investigation of the backgrounds to $ZZ$. The problem is 
that the largest branching fraction of a visible Higgs through 
$ZZ$ is into $4$ jets. This 
signature will be utterly buried under the huge $WW$ that materialise into 
4 quarks. A quick look at Fig.~\ref{gammaee}, that also shows the $300$GeV 
Higgs resonance in $ZZ$, should be convincing. 
In the $4$ jets topology the only glimmer of hope that we foresee 
is the decay of (at least one) of the $Z$ into $b$'s. This 
requires a very good rejection of charm to disentangle the $Z$ from the 
$W$. A very nice signature of the $ZZ$ that has a relatively good branching 
fraction is when one of the $Z$ decays into neutrinos. The latter produces 
a large missing transverse energy, recoiling against an invariant 
mass $\sim M_Z$\cite{nousggvv,Gunionzz,SanDiego}. Basing the analysis of this signature 
on counting rate should help at small $\sqrt{s_{ee}}$. However, since we can 
not reconstruct the invariant $ZZ$ mass with these decays, it seems to be
difficult to carry this analysis to higher energies. Would a distribution
in the transverse mass improve the situation? \\
\noi The cleanest signal that 
allows an invariant $ZZ$ mass reconstruction 
is when at least one of the $Z$ decays leptonically (one should only consider 
$e$ and $\mu$) while the other decays visibly. Unfortunately, 
this corresponds 
to a small combined branching fraction of about $10\%$ of all $ZZ$ events. 
Nonetheless, it seems to be possible to extract a good signal for low enough 
\epm cms energy and Higgs masses\cite{Bowser}.  It 
 has been pointed out\cite{Bowser} that 
$f \bar{f} Z$ final state could fake the signal. Fortunately  
this background could be easily disposed off by requiring a central 
$f \bar f$ pair that reconstructs the $Z$ \cite{Bowser}. 

\subsection{Looking for \ssb scenarios in \ggzz}
As mentionned earlier, the reaction \ggzz was not considered to be 
interesting just from the point of view of the Higgs resonance 
but also from the perspective of checking the symmetry breaking. The nice 
proposals all predict an excess of $Z_L Z_L$ events at high $M_{ZZ}$ invariant 
masses, but the analyses were done 
with the tacit assumption that the excess was an excess over 
very negligible ($\sim${\em zero}!) $ZZ$ events. Looking at 
Fig.~\ref{ggzzmzz1} once more, it is clear that the conclusions should 
be reconsidered and the optimism somehow 
watered down. \\ 
Nonetheless, we think that with the proper 
inclusion of the \sm contribution adapted to the Higgsless limit, this reaction 
should still be useful. After all, we were able to 
extract competitive limits on the $L_9$ 
parameters of the chiral Lagrangian description of the $WW\gamma$ 
anomalous couplings through the reaction \ggww whose cross section is 
so much larger than the $Z_T Z_T$.  Furthermore, 
since the limits on the various parameters 
that were extracted fom the $ZZ$ process are quadratic in these 
parameters, we do not expect that the taking into account of the $Z_T Z_T$ 
mode changes the limits by more than an order of magnitude.\\
When adapting the \sm results to the search of unconventional 
scenarios of \ssb that do not admit a Higgs, 
the reference value of $Z_L Z_L$ should correspond to 
the large Higgs mass limit\cite{BoosJikia,Jikiazz,Veltman}, 
$M_H^2 \gg s_{\gam} \gg \mww,\mzz$,
that can also be more easily obtained by 
using the equivalence theorem or the chiral limit of the properly 
``translated" 
$\gam \ra \pi^0 \pi^0$\cite{Bijnens2pi0,Donoghue2pi0,Herrero}. 
\beqn
{\cal{M}}_{++LL} \ra \frac{\alpha^2}{s_w^2} \frac{s_{\gam}}{2M_W^2} 
\equiv 8\pi \alpha \frac{s_{\gam}}{(4\pi v)^2} 
\label{universalggzz}
\eeqn

\noi Once this universal part ($\cal{O}(p^4)$) 
has been included it turns out that all models of symmetry breaking 
can be described in terms of their contribution to 
the helicity amplitudes 
with only the two parameters $a_0$ 
and $a_c$ that we introduced in the previous section\cite{nousggvv}. 
These should be seen as corrections of order $\cal{O}(p^6)$ in the energy 
expansion. The corresponding operators that describe $\gam \ra \pi^0 \pi^0$ 
are the ones first discovered by Terent'ev\cite{Terentev}. 
Models that correspond to the residual 
effect of heavy charged scalars contribute to $a_0$
\cite{Donoghue2pi0,gg2pi06}, 
as well as 
the effect of a very heavy neutral scalar ($s_{\gam} \gg M_S^2$) 
coupled in a chiral invariant way \cite{Ramirez,Rosenfeld}. 
Note also that, in effect, the universal term in Eq.~\ref{universalggzz} 
contribute to the same leading helicity amplitude as $a_0$. 

\noi It should be realised that if one wished to include the effect 
of a not-so heavy scalar, which in a sense would be some unconventional 
Higgs\cite{nousggvv}, one would modify our non-resonant 
formulae by providing for the 
propagator of the scalar 
(in a sense $\Lambda^2 \ra s-\Lambda^2+i\Gamma\Lambda)$. 
One could even accommodate the model of 
Chanowitz\cite{Chanowitz} that would be interpreted 
in the limit that the above unconventional scalar is  a light 
Higgs having enhanced couplings, as a result of some 
ultra-heavy states. 
In any case, for all these models there would be 
no contribution to the $J_Z=2$ amplitude\footnote{Deviations in the \ggzz, 
$\gam \ra Z\gamma$ and $\gam \ra \gam$ due to an anomalous operator in the 
linear (light Higgs) realisation of symmetry breaking have been analysed 
recently 
\protect\cite{Fernandggzz}. We wish to point out that 
since the sole effect of this operator 
is a modification of the coupling $HVV$($V=\gamma,Z,W$),
in principle, we could also test this operator via Higgs production 
in \epm either in the Bjorken process or the fusion mechanism.}. 

This is not the case of the 
$a_c$ models that would show up both in the $J_Z=2$ and the $J_Z=0$. 
The case of heavy axial fermions that could be the analog of the nucleons 
in the chiral picture could also be made\cite{nousggvv,gg2pi06}. 
We would then identify this case with $a_0=-a_c$. Models 
with heavy vector resonances $V$ constructed on the hadron mould 
\cite{Bando,Ecker} like the so-called BESS model\cite{BESS} 
could not contribute in their minimal version to \ggzz. In this particular 
scenario
one can not just rely on the equivalence theorem to replace the longitudinal 
$Z$ by the ``pion" field and apply the result of $\gam \ra \pi^0\pi^0$ 
\cite{gg2pi06,gg2pi06Ko,Rosenfeld} that gives the same pattern $a_0=-a_c$. 
One needs to break \cviol for the effective $\gamma V Z$. This entails 
that the model\cite{Rosenfeld2} contains anomalies and loses much of its 
attraction. \\
\noi The case with $a_c=-2a_0$ would only contribute to the $J_Z=2$. 
We should also note that at high enough energy some unitarisation of the 
amplitudes should be provided\cite{Abbasabadi} and that for some values 
of the parameters of $\cal{O}(p^6)$ Lagrangian one has a softening 
of the growth of the $J_Z=0$ amplitude for the two longitudinal vector 
bosons.

\subsection{Limits on the $\cal{O}(p^6)$ operators in $ZZ$.} 
Before the results of the \sm cross section for \ggzz were available, 
the investigations of the symmetry breaking 
mechanisms 
\cite{nousggvv,Chanowitz,Peskinzz,Herrero} 
concluded that this channel was an ideal testing ground 
as compared to $WW$. Unfortunately the optimism has faded somehow, 
since the \sm provides a non negligible rate completely dominated by $Z_TZ_T$. 
As we argued above, even when taking into account the \sm 
contribution, the $ZZ$ still remains a good window for the New Physics. 
A naive estimate will suffice to get a good idea of 
what can be achieved in the ZZ channel. 
First of all, the \sm $Z_L Z_L$ contribution in the very heavy Higgs mass limit 
(the ``universal" term) does not pose much problem as it is quite small at 
all energies (around $1fb$, see Fig.~\ref{ggzzee}). 
To have a rough idea on how the limits 
are changed it will be sufficient to include the \sm $Z_T Z_T$ contribution 
that is not affected by the effects we are looking at. We have thus 
based our discussion 
on the total cross-section only. Another idea that deserves more study 
should exploit the fact that the $TT$ cross section does not depend much 
on the $J_Z$ of the initial two photons. Therefore, by constructing 
an asymmetry such as ($\Delta \sigma_{02}=\sigma_{J_Z=0} -\sigma_{J_Z=2}$), in a peaked spectrum 
set-up, Fig.~\ref{spectre12}, 
one should reduce the \sm background. This, of course, assumes that 
the New Physics is such that it does not contribute equally 
to the two $J_Z$ as is the case with $a_0$ or $a_c$ taken separately. 
This also assumes that one can ``afford" sharing the luminosity between the two 
modes while maintaining a good signal.

For our tentative estimate based on the total cross section we only 
considered the visible unambiguous $ZZ$ signature with one $Z$ decaying
hadronically and the other leptonically, with a cut $\cos\theta^*_Z<.866$. 
The ``0-dom" (Fig.~\ref{spectre12}) setting was assumed so that to reproduce 
the cut and settings of Jikia\cite{Jikiazz}. 
The criterion of observability was based on requiring 
$3\sigma$ statistical deviation from the standard cross-section. 
Taking $\Lambda=1TeV$, we obtain the limits
\beqn
|a_0| &<&2 \;\;\;\;\; |a_c| < 5 \;\;\;\; \mbox{with}\;\;\sqrt{s_{ee}}=500GeV\;;\; \int\cl=10fb^{-1}
\nonumber \\ 
|a_0| &<&0.3 \;\;\; |a_c| < 0.7\;\;\;\; \mbox{with}\;\; \sqrt{s_{ee}}=1TeV \;;\; 
\int\cl=60fb^{-1}
\eeqn
\noi
These limits are only a factor $2$ (at 500GeV) worse than what we have obtained 
by neglecting the \sm contribution and about a factor $4$ worse at $1$TeV. 
Nonetheless these limits are better than what we would obtain in the 
\epm mode through three vector production\cite{nousee3v} 
or in the $e\gamma$ mode \cite{Eboliquartic}.

 \section{$WWZ$ and $WW\gamma$ Production}
\renewcommand{\thequation}{\thesection-\arabic{equation}}
\setcounter{equation}{0}
As already pointed out in the introduction, triple vector production 
at \gag is very important. In this section we would like to point at some  
interesting aspects of these cross-sections. We note that with these 
reactions one is totally in the non-abelian gauge sector of the \sm. 
\subsection{$\gamma \gamma \rightarrow W^+ W^- \gamma$
\protect\cite{Ottawa,nousgg3v}} 
\begin{figure*}[hbt]
\begin{center}
\caption{\label{wwgwwz}{\em a) Polarised cross-sections for $WW\gamma$ production. 
The subscripts $0,2$ refer to the two-photon state. b) $WWZ$ with unpolarised 
photons. When the particle is not indicated this means that we have summed 
on all its polarisations. We also show the ratio $Z_L/Z_T$}}
\mbox{
\mbox{\epsfxsize=7.5cm\epsfysize=9.5cm\epsffile{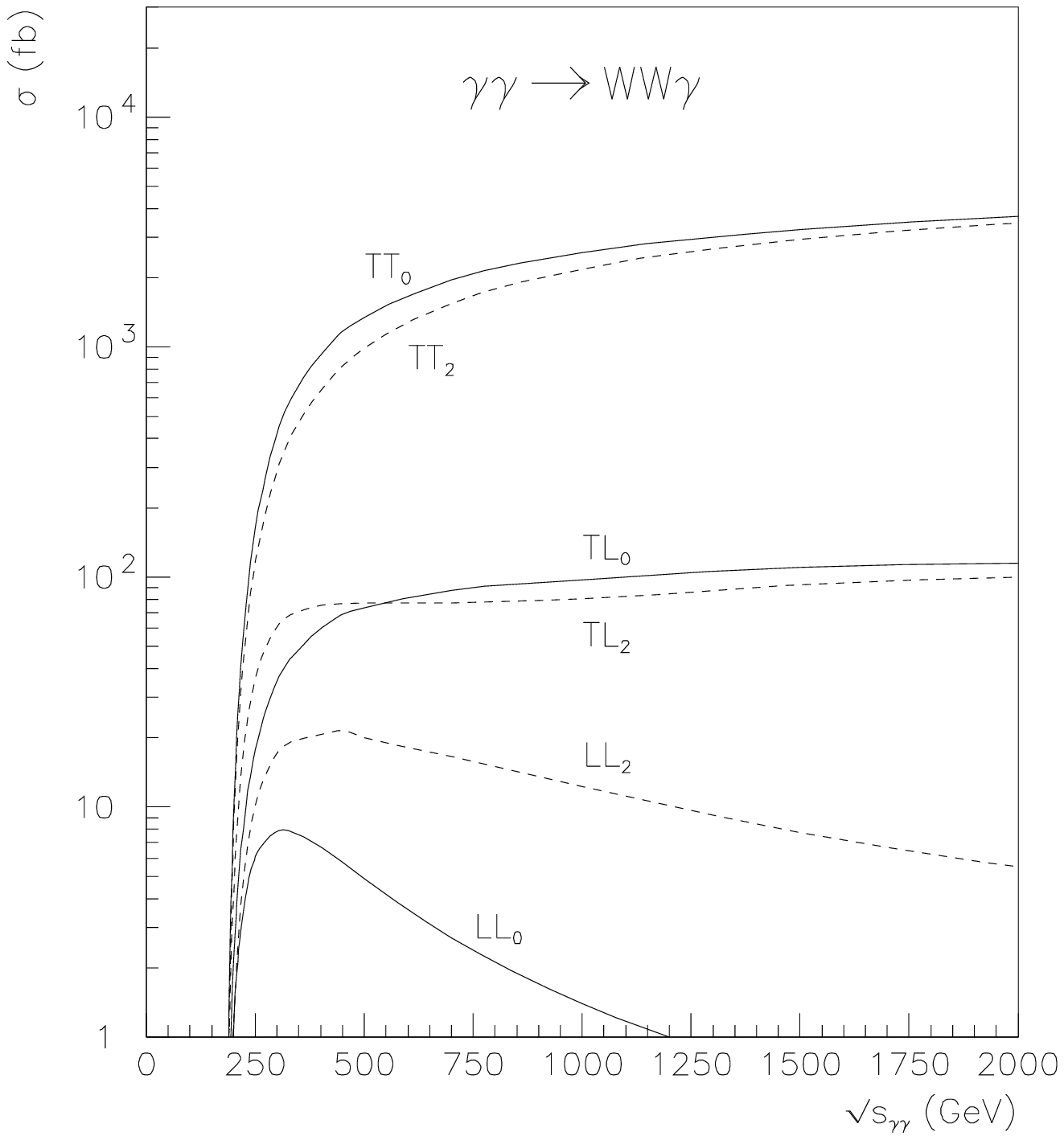}}
\mbox{\epsfxsize=7.5cm\epsfysize=9.5cm\epsffile{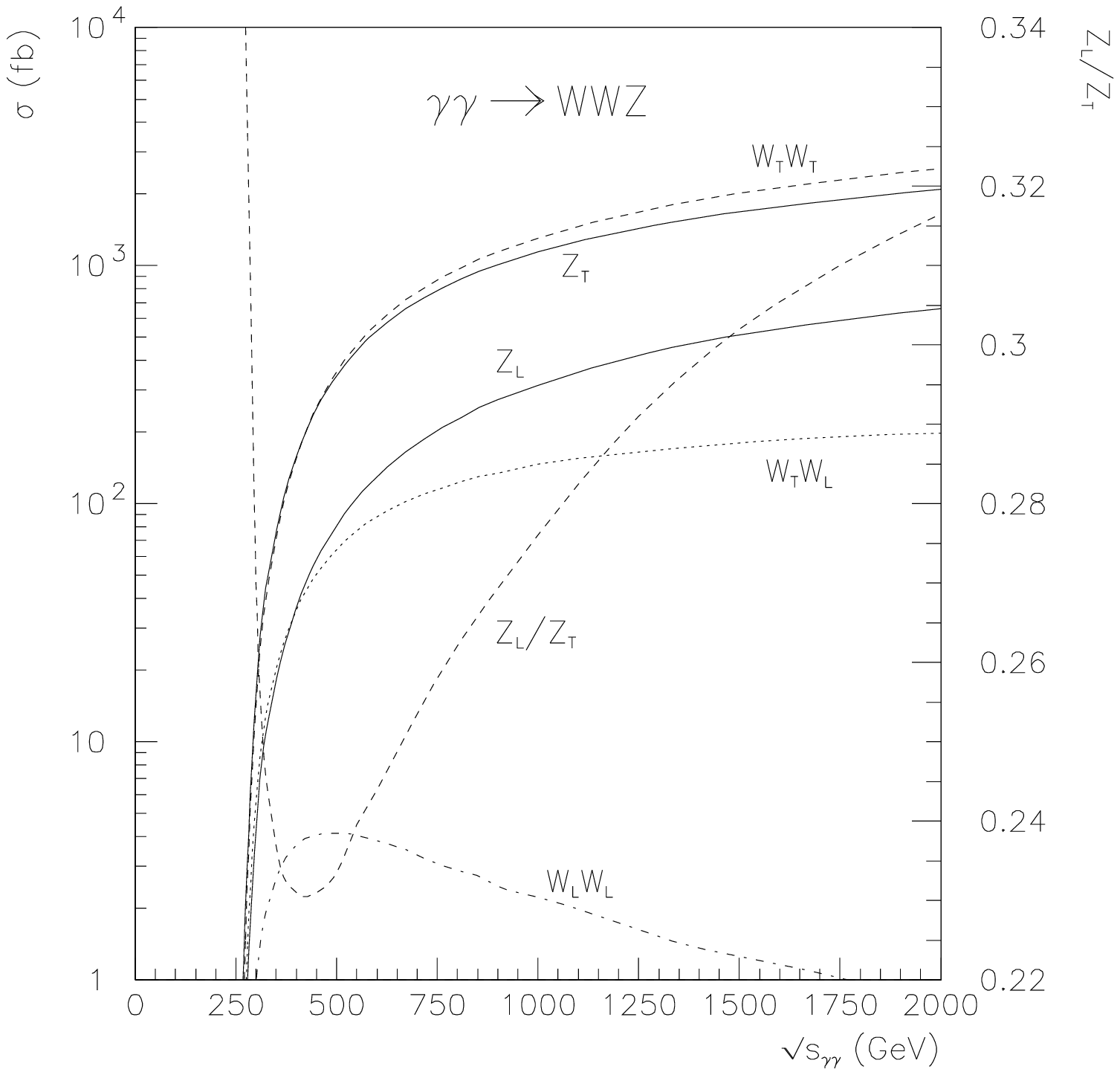}}}
\vspace*{-1.5cm}
\end{center}
\end{figure*}

For the $WW\gamma$ final state, a cut on either  the final photon energy 
or transverse momentum is 
required. With a fixed cut $p_T^\gamma > 20$~GeV for all centre-of-mass 
energies, the cross-section increases with energy (see Fig.~\ref{wwgwwz}a). 
At $500$~GeV one reaches a 
cross-section of about $1.3$pb. This is about $1.6\%$ of the 
$WW$ cross-section at the same energy.
The $J_Z=0$ obtained when both photons have the
same helicity slightly dominates over the $J_Z=2$ ($1.5$pb versus $1.1$pb). 
At $\sqrt{s}_{\gamma \gamma}=2$~TeV the cross-section with the same 
$p_T^\gamma >20$~GeV cut reaches $3.7$pb. The logarithmic ($\log^2 s$) growth 
can be understood
on the basis that this cross-section can be factorised in terms 
of $\gamma \gamma \ra WW$, which is constant at asymptotic $M_{WW}$ 
invariant masses, times the final state photon radiator which contains 
the logarithmic $s$ dependence. We note that this logarithmic 
increase only concerns the production of transverse $W$. When both 
$W$'s are longitudinal ($W_L W_L$) the cross-section decreases. This can also 
be traced back to the fact that $\gamma \gamma \ra W^+_L W^-_L$ decreases 
with energy. We find that the $W_L W_L$ fraction of all $W$'s is about  
only $1\%$ at $500$~GeV, $0.3\%$ at $1$~TeV and a mere $0.07\%$ at $2$~TeV. 
It must be noted that the bulk of the cross-section occurs when all 
final particles are produced at very small angles: this is a typical example 
of multiparticle production in the very forward region. For instance, increasing
the $p_T^\gamma$ cut and at the same time imposing a pseudorapidity cut 
on the photon, the $WW\gamma$ yield, as shown in Table~\ref{wwgres}, drops
considerably, especially at higher energies. The reduction is even more 
dramatic when 
we put an isolation cut between all the particles and force them away from 
the beam. With these strictures the cross-section decreases with energy 
(See Table~\ref{wwgres}).

\begin{table}[htb]
\caption{{\em Cross-section for 
$\gamma \gamma \rightarrow W^+ W^- \gamma$ (in fb) at different 
energies, including various cuts.}}
\begin{center}
\begin{tabular}{|l|c|c|c|c|} 
\hline
$\;\;\;\;\;\;\;\;\;\;\;\;\;\;\;\sqrt{s}_{\gamma \gamma}$(TeV) &0.5&1&1.5&2 \\
&&&&\\
type of cut&&&& \\ 
\hline
cut 1:  $p_T^\gamma >$ 20~GeV     &1254&2469&3195&3678\\
cut 2:  $p_T^\gamma >40$~GeV$\times\sqrt{s}_{\gamma \gamma}$(in TeV) &1254&1434&1258&1050\\
cut 3: cut 2 and  $|y^\gamma|<2$  &1235&1373&1159&930 \\
cut 4: cut 2 and $\cos$(between any two particles)$<0.8$ &201&86&47&32 \\
\hline
\end{tabular}
\end{center}
\label{wwgres}
\end{table}

\noindent While the $W_L W_L$ production is very much favoured 
in the $J_Z=2$ mode, $W_T W_T$ and $W_T W_L$ 
productions (which are by far the largest 
contributions) are slightly more favoured in the $J_Z=0$ channel 
(see Fig.~\ref{wwgwwz}a).
This is the same behaviour as in the two body process \ggww. 

\subsection{$\gamma \gamma \rightarrow W^+ W^- Z$\protect\cite{nousgg3v}}
\noi Contrary to the previous reaction one can calculate the 
total cross-section. It exhibits an interesting behaviour 
at TeV energies. 
One notes that already 
at $1$~TeV the triple vector boson production is larger that top pair 
($m_t \geq 130$~GeV) 
and charged heavy scalars production (with $m_{H^{\pm}}\simeq 150$~GeV) as 
shown in Fig.~\ref{gammaee}. 
At $2$~TeV the total $WWZ$ cross-section is about $2.8$pb 
and exceeds the 
{\em total} electron-positron pair production! This is a typical example 
of the increasing importance of multiparticle production in weak interactions 
at higher energies, purely within perturbation theory 
\footnote{As opposed to the surmise of large $W$ multiplicities 
due 
to topological effects at extremely high-energies. 
For a recent review see \cite{Ring}.}. The rising of the cross-section 
with the centre-of-mass energy is essentially from the very forward region 
due to the presence of the ``non-annihilation" diagrams with the (spin-1) 
$W$ exchanges. A similar behaviour in \epm reactions is single vector 
boson production. What is certainly more interesting in 
$\gamma \gamma \ra W^+W^-Z$ is the fact that it has a purely 
non-abelian origin. 
It may be likened to $gg \ra ggg$ in QCD except that we do not 
need any infrared cut-off (the $W$ and $Z$ mass provide a natural cut-off).

\noindent 
The bulk of the cross-section consists of both $W$ being 
transverse as is the case with the ``parent" process 
$\gamma \gamma \ra W^+W^-$. 
While the total cross-section is larger in the $J_Z=0$ than in the $J_Z=2$, 
the production of all three vector bosons being longitudinal occurs mainly 
in the $J_Z=2$ channel and accounts for a dismal contribution. For instance, 
the ratio of $LLL/TTT$ (three longitudinal over three transverse) in the case 
of unpolarised beams amounts to a mere $2$per-mil at $500$~GeV and drops to 
$0.1$per-mil at $2$~TeV. 
Nonetheless, the total \footnote{i.e., taking into account all polarisation 
states of the $W$.}  production of longitudinal $Z$'s 
as compared to that of transverse $Z$'s 
is not at all negligible. In fact, between $500$~GeV and $2$~TeV this ratio 
increases from about $23\%$ to $32\%$ (see Fig.~\ref{wwgwwz}b). 
This is somehow counterintuitive 
as one expects the longitudinal states to decouple at high energies. 
The importance of $Z_L$ production (in association with $W^+_T W^-_T$) is,  
however, an ``infrared" rather than an ``ultraviolet" phenomenon in this 
reaction: the $Z$ is not energetic. 
First, one has to realize that the 
$\gamma \gamma \ra  W^+W^-Z$ amplitude 
is transverse in the momentum of the $Z$, $q$, as is the case with the photons 
in $\gamma \gamma \ra W^+W^- \gamma $. 
With $k_1$ and $k_2$ being the momenta of the photons, the longitudinal 
polarisation vector of the $Z$, with energy $E_Z$, writes

\beq
\label{eq:polar}
\epsilon_{\mu}^L =\frac{1}{\sqrt{E_Z^2-M_Z^2}} \left(\frac{E_Z}{M_Z}q_\mu 
\;-\; \frac{M_Z}{\sqrt{s}} (k_1+k_2)_\mu \right) \; ; \;\ 
\epsilon^L.\epsilon^L =-1
\eeq

\noindent 
The transversality of the amplitude means 
that the leading (``ultraviolet" $\propto E_Z$) part does not contribute. 
Only the 
``infrared" part $\propto M_Z$ does. 
This contribution should vanish in 
the limit of vanishing $M_Z$. 
However, the amplitude, in analogy with what happens in $WW\gamma$, has the 
infrared factor 
$1/E_Z$ and the ``soft" term in Eq.~\ref{eq:polar} contributes. 
Furthermore, more importantly, the bulk of 
the cross-section is from configurations where both $W$ are transverse 
(see Fig.~\ref{wwgwwz}.b) and 
all three particles go down the beam. In the limit of 
vanishing masses this topology leads to collinear divergences
\footnote{In the limit $M_V \ra 0$, 
added to the divergence in $\gamma \gamma \ra W^+W^-$, there is 
the collinear divergence when the $Z$ and a $W$ are collinear.}. 
In this dominating configuration, in the exact forward direction, 
the longitudinal $Z$ 
contributes maximally. At the same time, angular momentum 
conservation does not allow the $Z$ to be transverse when all final 
particles are down the beam (with $p_T=0$) and both $W$ are transverse. 
So the ``maximal collinear enhancement" is not as operative for the 
transverse $Z$ 
as it is for longitudinal $Z$ when both $W_T$ are at zero $p_T$. 
However, as soon as one moves away from these singular configurations, the 
longitudinal $Z$ does decouple and the ``smooth" mass limit may be taken. 
This is well rendered in Table~\ref{wwzres} 
which displays the ratio of $Z_L/Z_T$ without any cut and with the inclusion 
of cuts. The most drastic of these cuts is when we impose angular separation 
cuts between the final particles and 
force them to be away from the beam, with the effect that 
the $Z_L/Z_T$ decreases with energy and gets dramatically smaller. 
\begin{table}[tbh]
\caption{{\em Cross-section for 
$\gamma \gamma \rightarrow W^+ W^- Z$ and ratio of longitudinal over 
transverse $Z$ (L/T) including various cuts. ``all" means that we require 
the final particles to be separated \underline{and} to be away from the beam 
by an angle $\theta$ corresponding to $\cos \theta <0.8$.}}
\vspace*{0.3cm}
\begin{tabular}{|l||c|c||c|c||c|c||c|c||} 
\cline{2-9}
\multicolumn{1}{c||}{} &\multicolumn{2}{c||}{$\sqrt{s}_{\gamma \gamma}=500$~GeV}
&\multicolumn{2}{c||}{$\sqrt{s}_{\gamma \gamma}=1$~TeV}
&\multicolumn{2}{c||}{$\sqrt{s}_{\gamma \gamma}=1.5$}
&\multicolumn{2}{c||}{$\sqrt{s}_{\gamma \gamma}=2$~TeV} \\
\multicolumn{1}{c||}{} 
&$\sigma$(fb)&L/T &$\sigma$(fb)&L/T &$\sigma$(fb)&L/T &
$\sigma$(fb)&L/T \\
\hline
no cut&428&$24\%$&1443&$27\%$&2195&$30\%$&2734&$32\%$ \\ \hline
$\cos(WZ)<0.8$&368&$19\%$&1025&$18\%$&1321&$19\%$&1465&$19\%$ \\ \hline
$\cos(W\gamma)<0.8$&164&$18\%$&232&$17\%$&186&$17\%$&145&$16\%$ \\ \hline
$\cos(\;{\rm ``all"})<0.8$&115&$11\%$&140&$8\%$&105&$6\%$&77&$5\%$ \\ \hline
$E_Z>150$~GeV&184&$11\%$&1032&$21\%$&1700&$25\%$&2220&$28\%$ \\ \hline
\end{tabular}

\vspace*{0.3cm}
\label{wwzres}
\end{table}
\noindent
On the phenomenological side, the study of this reaction is important 
as, especially for $M_H \sim M_Z$, it is a background to Higgs detection
through $WWH$ production\footnote{
Two calculations of the $WWZ$ cross section have 
appeared\cite{Booseg,Eboligg3v}. They deal with the total 
 unpolarised cross section. In \cite{Booseg} we find complete agreement 
at all energies. In \cite{Eboligg3v} we agree after folding with the 
unpolarised luminosity spectrum at $s_{ee}=500$GeV and 
$1$TeV and taking 
into account the same value of the parameters $\alpha$ and 
$sin^2\theta_W$. 
With the same changes our result for the unpolarised $WW\gamma$ are also 
confirmed by \cite{Eboligg3v}.}.

 \section{Search for the \sm Intermediate-mass Higgs}
\renewcommand{\thequation}{\thesection-\arabic{equation}}
\setcounter{equation}{0}
One of the most attractive motivations for doing physics with very energetic
photon beams is the unique capability of this mode for producing 
a scalar particle as a resonance. In our context this is the Higgs. 
This resonant Higgs structure 
is out of reach in the usual \epm mode. However, as we 
stressed in our general introduction, the coupling of the Higgs to two 
photons only occurs at the loop-level. Therefore, on the one hand, the 
rate of production is not as large as with the resonant $Z$, say.    
On the other hand, a precision measurement of the 
$H\gamma \gamma$ coupling is an indirect way of revealing all the massive 
charged particles that would be present in an extension of the \sm. These 
heavy quanta 
would not decouple and would therefore contribute substantially 
to the production rate in \gag. \\

To conduct the precision measurements 
of the $H\gamma \gamma$ coupling one needs to sit at the Higgs resonance 
and collect enough luminosity.
For the IMH, this would mean operating within a narrow energy range
that could be much below the highest accessible energy.
 Therefore, the question arises whether 
one should tune the \gag machine such that one obtains a spectrum 
that is peaked at the Higgs mass and that has very little spread.
We willl refer to this scheme as the narrow-band low-energy  \gag
collider.
This is certainly 
possible, however this choice will preclude the study of a plethora 
of interesting weak processes and thus it rests to see whether 
such a decision would be made. For instance, 
the most 
important mass range at the NLC is the intermediate mass Higgs, IMH, 
$M_Z < M_H <140$~GeV. This mass range is, in a sense the preserves of the NLC, 
since it is going to be extremely difficult (if at all) to cover this 
range at the LHC. \\
\noi The unease with the choice of 
a peaked \gag spectrum is that, within this narrow domain of energy, 
obtained from an \epm with $\sqrt{s_{ee}}=500$~GeV or higher 
\footnote{The choice of this 
scheme tacitly assumes that the mass of the Higgs has been determined in the 
\epm mode and that $\sqrt{s_{ee}}=500$~GeV would, in any way, be available.}
it will not be possible to reach the 
$WW$ threshold (that seems to be a good luminosity monitor) 
and other $W$ reactions that offer a rich physics program. 
This could include 
the direct production of some of those particles that would only 
be probed indirectly 
in $H\gamma \gamma$. Of course, one may argue that these would be necessarily 
produced in the \epm but in view of the 
known universal character of the production mechanism in \gag,
they may be better studied. Not to mention that it is not excluded 
that the \gag mode when operated in  
the full range of energy can access scalar particles that would kinematically 
be out of 
reach in the \epm mode. This could happen if in \epm 
they can only be produced in association with another 
heavy  particle. The \cpviol-odd Higgs of the minimal supersymmetric 
model is such an example\cite{Gunionzz}. \\
It is certain that a narrow-band low-energy  \gag collider has its merits 
especially if it is achieved with high luminosity, since precision tests 
on the nature of the light Higgs may be performed. To some 
degree, this would be a spin-0 version of a polarised LEP1! Moreover, as we will see, 
with a low-energy scheme many backgrounds are drastically suppressed. 
Investing enough running time in such a mode to  be able to switch 
between different polarisation settings (circular/linear polarisation,...) one 
could, for instance, {\em directly} test the parity of the Higgs
\cite{Peterparity,Gunionparity} or perform \cpviol tests from probing 
the $H\gam$ coupling\cite{Gunion}.  
These are undoubtedly quite interesting studies to do, but we should 
stress that they do call for very high luminosities and would be done 
at the expense of a rich program. In addition, keeping in mind that 
this ``narrow-band" scheme presupposes that it is in the \epm mode 
that the mass of the Higgs has been determined and used to tune the laser, 
the \epm would also give a good 
clue on some of the above issues that one wants scrutinized in the peaked 
\gag mode. 
For instance, the parity 
of the scalar will, in a large degree, be inferred from its rate of 
production in the \epm. A spin-$0$ either standard or supersymmetric 
produced through the $VVH$ vertex is \cpviol even. As pointed out 
in 
\cite{Gunionzz} it may also happen that  a measurement of the $h^0\gam$ 
if not very precise would not provide much more insight. This could occur 
if in the \epm 
mode of a $500$GeV only the lightest Higgs of the minimal supersymmetric 
model is discovered while the other susy particles are above threshold. 
Eventually, the choice will depend 
on the priority of the time and how the \epm collider is operating. 
For instance if we will have two interaction regions, one devoted to 
\gag physics as suggested in\cite{Wiik}, then one should search in both 
modes with a later (long) run dedicated to precision measurements. \\

\begin{figure*}[htb]
\begin{center}
\caption{\label{Branchingh}{\em 
Some of the \sm branching ratios of the Higgs as a function 
of the Higgs mass calculated 
with a top mass of $150GeV$. Also shown (thick lines) 
the total width and the \gag width.}} 
\vspace*{-1.2cm}
\mbox{\epsfxsize=12.5cm\epsfysize=12cm\epsffile{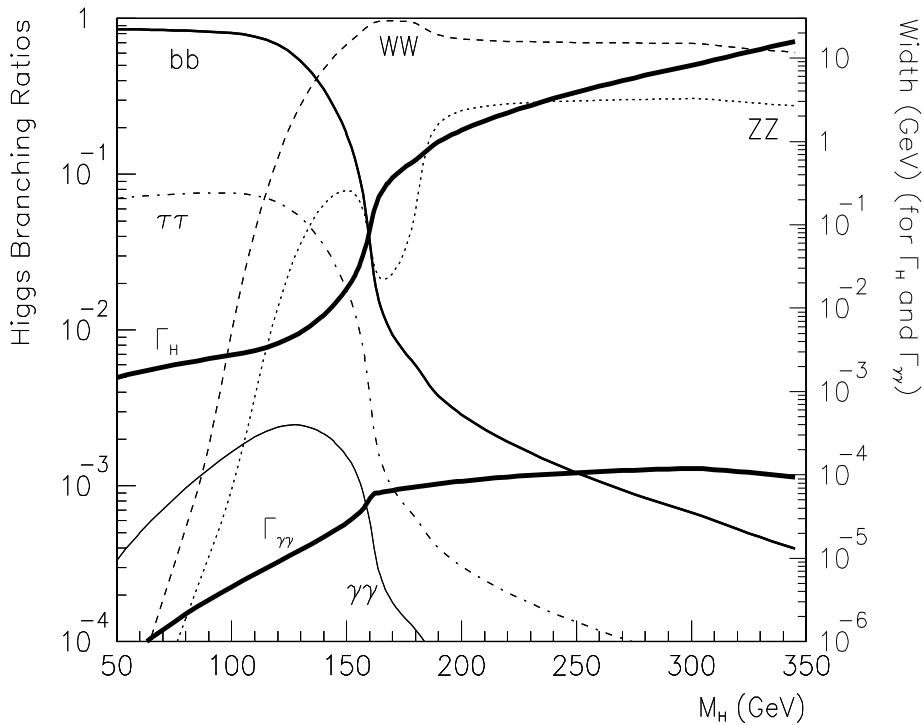}}
\vspace*{-.5cm}
\end{center}
\end{figure*}

Anyway, we feel it is essential to address the issue of 
whether 
the intermediate mass Higgs  could be observed as a resonance 
in a setting with a spectrum that 
allows a whole and self-contained physics program  to be conducted. \\
\noi The issue of seeing the heavy Higgs in the $WW$ and 
$ZZ$ channels has been addressed in  sections 6 and 7 and discussed in the 
talk of D.~Zappala and H.~Veltman\cite{Zappala,HeleneTalk}. 
In this section we only reassess the \gag discovery potential of the 
\sm Higgs, readdress the question of different backgrounds and how one could 
reduce their effect to a minimum. A few investigations of this 
aspect have been done with different emphasis and approaches
\cite{Gunionzz,Richard,Borden,BordenHiggs,Halzen}. \\
\noi The IMH, as is known, will decay predominantly into a 
$b \bar b$ pair and has an extremely narrow width 
(see Fig.~\ref{Branchingh}). This width, 
$\Gamma_H=\Gamma_{total}$, is 
of order of a few MeV. 

In the rest frame of the Higgs, the fermions are produced isotropically in 
the $J_Z=0$ state. The corresponding cross-section is described by 
the Breit-Wigner formula\cite{Gunionzz,BordenHiggs}
\beqn
\sigma{(\gamma \gamma \stackrel{H}{\longrightarrow} b \bar b})=
\frac{8 \pi \Gamma(H\ra \gam) \Gamma(H\ra b\bar{b})}{(\she -\Mhe)^2+
\Gamma_H^2 \Mhe^2} (1+\lambda_1 \lambda_2)
\eeqn
The exact expressions for the branching ratios of the Higgs and the 
$\Gamma(H\ra \gam)$ width that we will use in our analysis 
are taken from\cite{Abdel}\footnote{We thank Abdel 
Djouadi for providing us with the Fortran code.}. A top mass of $150$~GeV 
has been assumed. 
\begin{figure*}[htb]
\vspace*{-1.cm}
\begin{center}
\caption{\label{ggh1}{\em The Higgs signal into $b \bar b$ and its QED background}} 
\vspace*{.2cm}
\mbox{\epsfxsize=11.cm\epsfysize=3cm\epsffile{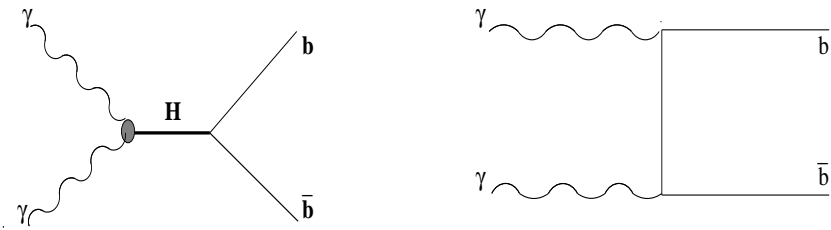}}
\vspace*{-.5cm}
\end{center}
\end{figure*}
\noi 
One obvious background is the direct QED $\gam \ra q \bar q$ production 
where $q=b$. One should also worry about other light quark flavours if no 
$b$-tagging is possible, charm causing much problem. 
A glance at the expression of the differential cross-section gives the clue 
as to how one could efficiently suppress this background. For the quark 
of charge 
$e_f$ and with $N_c=3$ we have, in the \gag \underline{cms} with 
$\theta^*$ being 
the $q$ scattering angle and $x=\cos\theta^*$:
\begin{eqnarray} 
\frac{{\rm d}\sigma_{QED}}{{\rm d}x}=\frac{2 \pi \alpha^2 e_f^4 N_c \beta}
{\she (1-\beta^2 x^2)^2} \left\{(1+\la_1 \la_2)(1-\beta^4)+
(1-\la_1 \la_2)\beta^2 (1-x^2)(2-\beta^2 (1-x^2))\right\}\nonumber\\
\label{ggbbqed}
\end{eqnarray} 
It is clear that the bulk of the cross-section is from the 
extreme forward-backward region. A modest cut on $\cos\theta^*$ 
will reduce the continuum substantially and will almost 
totally eliminate its 
$J_Z=0$ contribution (note its $(1-\beta^4)$ chiral factor). 
Therefore, 
choosing a spectrum with a predominantly $J_Z=0$ component\cite{Gunionzz} and
applying a cut on  $\cos\theta^*$ should do the trick. It is instructive 
to note that the $J_Z=2$, because of angular momentum conservation, 
vanishes in the exact forward region. \\

\noi 
It has recently been pointed out\cite{Halzen} 
that, unfortunately, this is not the whole 
story. Owing to the fact that the photon has a hadronic 
structure\cite{Witten} 
 it can 
``resolve"\footnote{This terminology has been introduced 
by Drees and Godbole
\cite{Manuel}.} into a gluon with some spectator jets   
left over. One then has to worry 
about $q \bar q$ production through $\gamma g$. 
\begin{figure*}[htb]
\begin{center}
\vspace*{-.5cm}
\caption{\label{ggh2}{\em Bottom pair production through the 
``once-resolved" photon.}}
\vspace*{.2cm}
\mbox{\epsfxsize=13.cm\epsfysize=2.5cm\epsffile{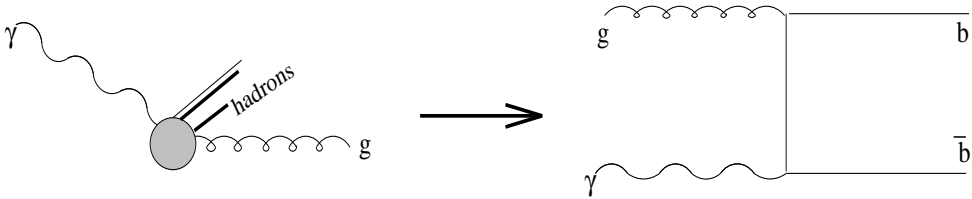}}
\vspace*{-.5cm}
\end{center}
\end{figure*}
Although the photon transfers only a small fraction, 
$x_g$, of its energy to the gluon, at $\sqrt{s_{ee}}\sim 500$~GeV 
the overall energy can be 
large enough for this gluon to combine with a photon so that the subsystem 
energy is enough for the IMH production. At $500$~GeV with a spectrum with 
$x_0=4.82$ so that $\sqrt{s_{\gam}^{max}}\sim 400$~GeV, this is an annoying 
background. Of course, if the photon transferred 
  all of its polarisation to the gluon then there will be not much 
problem as we will be in the same situation as with the polarised 
\gag initiated 
process. Unfortunately, we expect the polarisation to be diluted in 
the transfer. 
In principle, one could discriminate the gluon initiated processes from 
the direct production through the presence of the spectator jets. However, 
it seems to be very difficult to tag these spectator jets in the \gag 
environment\footnote{For a more optimistic view, see \cite{BordenHiggs}.}. 
As rightly 
remarked by  \cite{Halzen}, in the $\gamma g$ initiated process, the gluon 
has in general much less energy than the photon since the gluon 
distribution inside the photon comes essentially from the low $x_g$ 
region. This will lead to a larger boost of the $q \bar q$ 
system along the photon direction leading to a system with 
a much larger rapidity than in the direct processes. Thus the 
authors of \cite{Halzen} have suggested to apply a cut to reject $b$'s with 
large rapidities. It is to be noted that the twice 
resolved (both photons resolving into gluons), 
contribution, is negligible at $500$~GeV and {\it a fortiori} at lower 
energies. We have not included the ``2-resolved" contribution. \\

When the mass of the Higgs is around that of the $Z$ 
we have another non-negligible background\cite{Ilyazbb}: $Z$ radiation 
off a fermion pair while the external fermions go down the beam undetected.
 The $Z$ subsequently decays in $b \bar b$. 
 We will not cover this particular case here. We concentrate on two cases:
$M_H=120,140$~GeV. 
 
\begin{figure*}[htb]
\vspace*{-1.cm}
\begin{center}
\caption{\label{ggh3}{\em 
The fake $Z \ra b\bar b$ in \gag 
through the untagged fermions. Note that the first diagram, $\gam \ra Z$,  
is forbidden by Yang's theorem.}}
\vspace*{.2cm}
\mbox{\epsfxsize=11.cm\epsfysize=3cm\epsffile{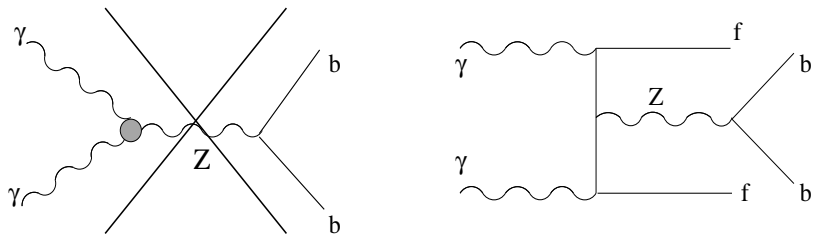}}
\vspace*{-.5cm}
\end{center}
\end{figure*}

We have reanalysed the issue of the background suppression 
and considered different \gag spectra.
 To illustrate the fact that the resolved photon problem is much less severe 
at lower \epm cms we also take the case of $\sqrt{s_{ee}}=300$~GeV. 
We also differ from\cite{Halzen} 
in the set of cuts. First, instead of a cut on $\cos\theta^*$ we required 
that both quarks have a transverse momentum such that  
 \beq
p_T^b > max(30~GeV,0.375 M_{b\bar b}) 
 \eeq
with $M_{b\bar b}$   the  invariant mass of the $b \bar b$
 system. 
To reduce the resolved photon contribution we have applied 
cuts on the reconstructed 
energy fractions of the initial $\gamma$ or $g$, $x_g^{min}$ and 
$x_g^{max.}$. 
\beq
x_g=\frac{1}{\sqrt{s_{ee}}} \left( (E_{b}+E_{\bar b}) \pm 
\sqrt{(E_{b}+E_{\bar b})^2-M_{b\bar b}^2} \right)
\eeq
The resolved contribution has on average a much smaller $x_g^{min}$ and a larger 
$x_g^{max}$. We have therefore cut on $x_g^{min}$ 
and $x_g^{max}$ and have found that the optimised values for the cuts 
were 
\beqn
x_g^{min}>\sqrt{\widetilde{M_{b\bar b}}^3/y_{max}} \;;\;
x_g^{max}<\sqrt{y_{max}\widetilde{ M_{b\bar b}} } \;\;;
\widetilde{M_{b\bar b}}= M_{b\bar b}/\sqrt{s_{ee}} \;;\; 
y_{max}=\frac{x_0}{1+x_0}=0.83 \nonumber \\
\eeqn
For the gluon distribution inside the 
photon we have taken the GRV\cite{GRV} parameterisation 
with $Q^2=(60~GeV)^2$. This $Q^2$ is such that $Q^2 \sim (\mh/2)^2$.
We would like to point out that the effect of this cut 
is another lucky strike: besides allowing to 
get rid of large rapidities it improves 
on the rejection of the $J_Z=2$ from the {\em direct} $b \bar b$ 
process. This is so because it forces the 
two photons to have equal energies. Therefore, even if no electron polarisation 
were available and if $P_c=P'_c$, each of the
colliding photons will have the same degree of polarisation (Eq.~\ref{c20})
hence providing a dominant $J_Z=0$ setting. This is in fact the message that 
Fig.~\ref{spectre34}b conveys. 

\noi Moreover, we emphasize the importance of the 
resolution on the invariant $b\bar b$ mass. To simulate the resolution 
we have introduced a Gaussian smearing onto the Higgs signal. We consider two 
values for the resolution: $\Delta M=5$~GeV and $\Delta M=10$~GeV. The 
conclusions depend critically on the would-be achieved resolution. 

\noi The issue of $b$-tagging was also investigated.  
We have generated both $c\bar c$ 
and $b \bar b$ final states, considering two different possibilities:
no $b$-tagging and a realistic efficiency of $\epsilon_b$=.47
and $\epsilon_c$=.11 as would be the case with  
a micro-vertex detector with a modest performance.  
Our meaning of no $b$-tagging is that there is a total confusion between 
$b$ and $c$ only but that the other light flavours cannot be confused. 
Needless  to say that if
\begin{figure*}[p]
\begin{center}
\vspace*{-1.5cm}
\caption{\label{h500}{\em The fate of the Higgs resonance at $500$~GeV 
assuming different polarisation settings, resolutions and $b$-tagging 
efficiencies. The contribution of the resolved photon is shown as
a dashed line.}}
\vspace*{-2.cm}
\mbox{\epsfxsize=17.cm\epsfysize=22cm\epsffile{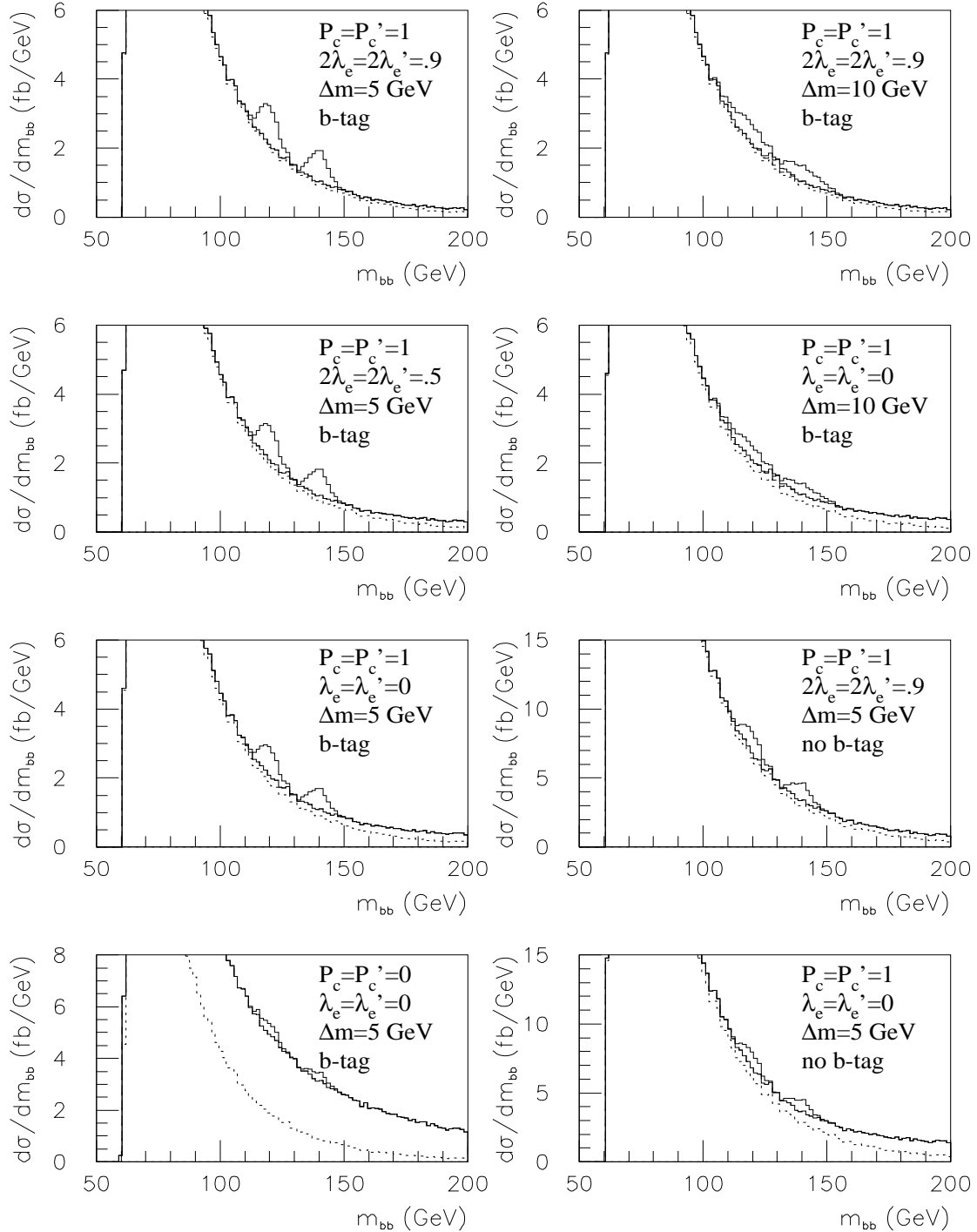}}
\end{center}
\end{figure*}
\noi
one has no distinction at all between the flavours 
the situation is much worse 
than in the case of no $b$-tagging (in fact it is hopeless). 
We applied the single-jet tag strategy\cite{eevvh} which gives 
a global efficiency of 
$\epsilon_{b\bar{b}}=$.72 and $\epsilon_{c\bar{c}}=$.21. Probably 
by the time (or even before) this machine is built one could achieve 
better efficiencies, but it is not clear how these detectors will 
perform in the \gag mode. 

\noi Since we need a broad spectrum for the search, with a dominance of the 
$J_Z=0$ we will take $P_c=P_c'=1$ with $2\la_e P_c=2\la_e' P_c'\geq 0$ and 
compare  with the case of no polarisation. 

\subsection{The case of a 500~GeV \epm}
The first remark is that, as shown on Fig.~\ref{h500}, even after the cuts 
have been applied the bulk of the background is due to the resolved 
contribution (dotted line). With a good resolution (5~GeV) and a realistic 
$b$-tagging efficiency together with $90\%$ longitudinal polarisation 
for the electrons we obtain a good signal with a significance 
$\sigma=S/\sqrt{B}=4.7$ ($\sigma=5.3$) for $M_H=120$~GeV ($M_H=140$~GeV). These 
numbers and the following values of the significance assume 
an integrated luminosity of 
$10fb^{-1}$ only. This is slightly 
reduced when only $50\%$ $e^-$ polarisation is achieved. What we find is that 
there is not much further reduction in the significance 
if no electron polarisation were available.
This result, as explained above, is due to the fact that the cuts filter 
photons with sensibly the same energies and hence we are in a setting that 
corresponds to the spectrum in Fig.~\ref{spectre34}b. In the case of no polarisation at 
all, which is very unlikely for the lasers, the situation seems hopeless. 
The change in the resolution from $5$~GeV to $10$~GeV 
has  a more dramatic effect. Even when 
the electron polarisations are at $90\%$ the nice peak structure 
has almost disappeared. Assuming $b$-tagging, the above significances change 
to $\sigma=3.5$ ($4.1$) for $120$~GeV ($140$~GeV). 
Further reduction 
occurs with no linac polarisation. \\
In the case of no $b$-tagging at all 
the event sample increases but the significance of the signal 
diminishes. A signal is clearly seen with a good resolution and $90\%$ 
$e^-$ longitudinal polarisation. For $M_H=120$~GeV we get $\sigma=3.9$ 
and even better for $M_H=140$~GeV ($\sigma$=4.7
(5~GeV) ). This significance 
does not degrade dramatically if no linac polarisation 
were available. However, with a resolution of $10$~GeV even with optimum polarisation 
the situation is desperate. \\
\noi To conclude, we stress the primary 
importance of 
aiming at having a good resolution. The polarisation settings seem to be 
achievable without much problem since an excellent degree of $e^-$ 
longitudinal 
polarisation is not absolutely essential, though helpful. Good (present-day 
LEP performances) $b$-tagging 
must be provided. In fact, a good $\mu$vx could even help in a better 
reconstruction of the invariant $b\bar b$ mass. It rests to see how 
a $\mu$vx would perform in the unusual environment of the laser scheme. 

\begin{figure*}[p]
\begin{center}
\vspace*{-1.cm}
\caption{\label{h300}{\em The fate of the Higgs resonance at $300$~GeV 
assuming   $b$-tagging and different  resolutions. 
The dashed line shows the
resolved contribution. }}
\vspace*{-2.cm}
\mbox{\epsfxsize=17.cm\epsfysize=22cm\epsffile{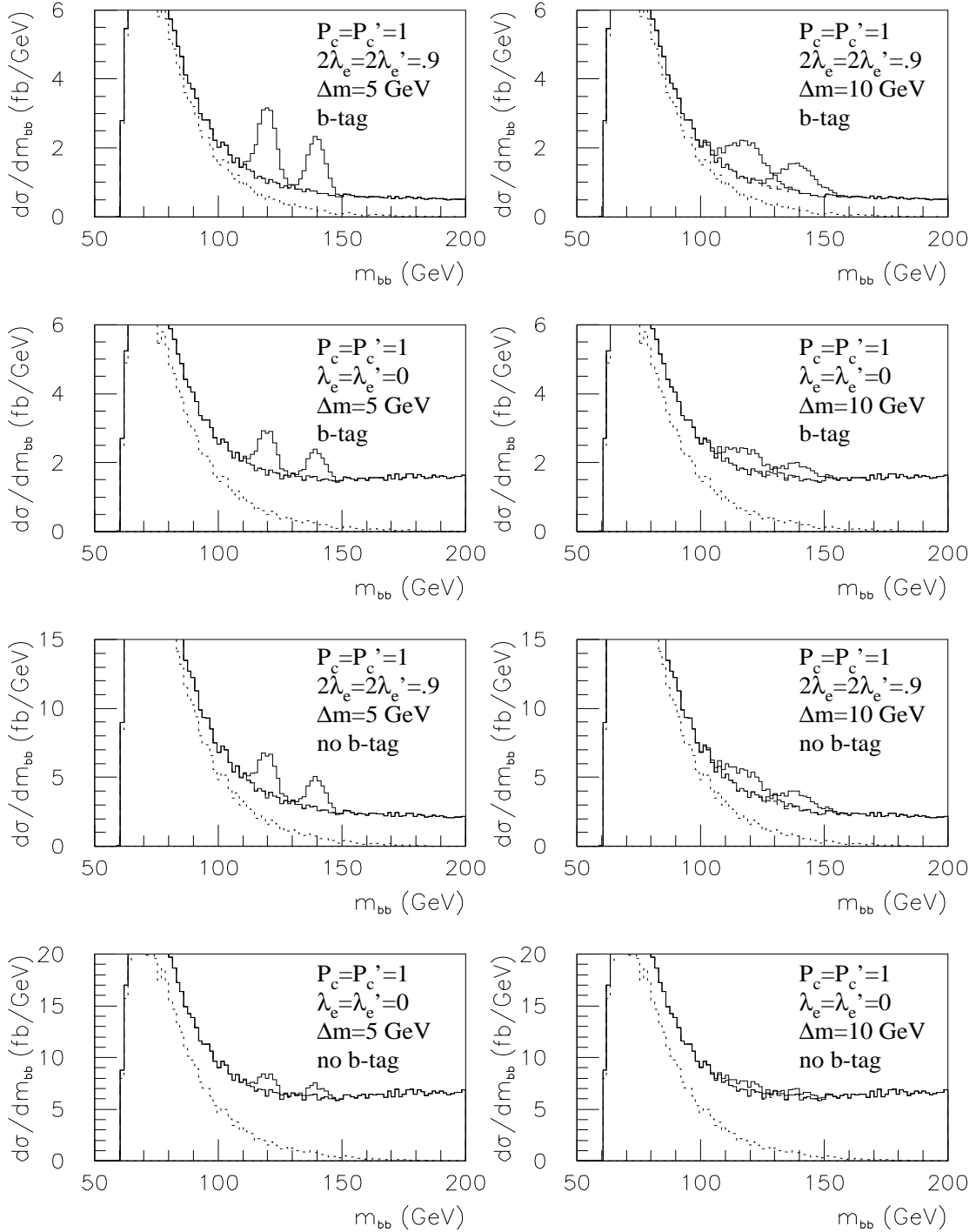}}
\vspace*{-2.cm}
\end{center}
\end{figure*}

\subsection{The case of a 300~GeV \epm}
Here the situation is far better: see Fig.~\ref{h300}. This is 
mainly because the resolved 
contribution has dropped. With $90\%$ longitudinal electron 
polarisation and 5~GeV resolution we obtain beautiful peaks. These peaks 
are still very clear with no electron 
 polarisation. Even when $c$ and $b$
are totally confused, provided one has a good resolution, the resonant 
structure is evident even when the electrons are unpolarised. In this non 
optimal case we obtain $\sigma=4.1$ for $M_H=120$~GeV 
(with $\int \cal{L}=10fb^{-1}$ only). 
Taking a larger resolution (10~GeV) the signal has an excellent 
statistical 
significance if a very good degree 
of longitudinal polarisation for the 
$e^-$ ($2\la_e=2\la_e'=90\%$) can be achieved together with $b$-tagging. 
In the case of no 
$b$-tag the same polarisation setting and $\Delta M=10$~GeV 
leads to a good significance $\sigma=7.2$ (6.5) for 
$M_H=120$~GeV (140~GeV). Unfortunately, even 
at $300$~GeV, the IMH is lost if no $b$-tag is provided and if the resolution 
is large without much electron polarisation. But this is the most pessimistic 
scenario. \\

In view of these results it is important to realise that at 
$300$~GeV $e^-$ polarisation is a top priority and that we could, in this case, 
make do with a not so good resolution, $\Delta M=10$~GeV. At 500~GeV, we could 
not, in all cases,  afford having such values of the resolution. 
On the other 
hand, although polarisation is very helpful it is not as important as at 
lower energies. This is easily explained by the fact that as we go higher 
in energy the contribution of the resolved photon becomes more and more 
important. This contribution is not killed by the choice of 
polarisation as is the direct contribution 
whose yield is more important at lower energies.

 \section{Non resonant Higgs production: the importance 
of the $WWH$ process}
\subsection{Comparison with other Higgs production mechanisms} 

\begin{figure*}[hbt]
\begin{center}
\vspace*{-.5cm}
\caption{\label{ggwwh1}{\em Comparison between different mechanisms of 
Higgs production in \epm, $e\gamma$ and \gag at $500$~GeV and $1$~TeV. 
No folding with the luminosity spectrum has been performed. \epm 
processes are, in the figure, only characterised by their final state.}}
\vspace*{-1.cm}
\mbox{\epsfxsize=14.5cm\epsfysize=11.5cm\epsffile{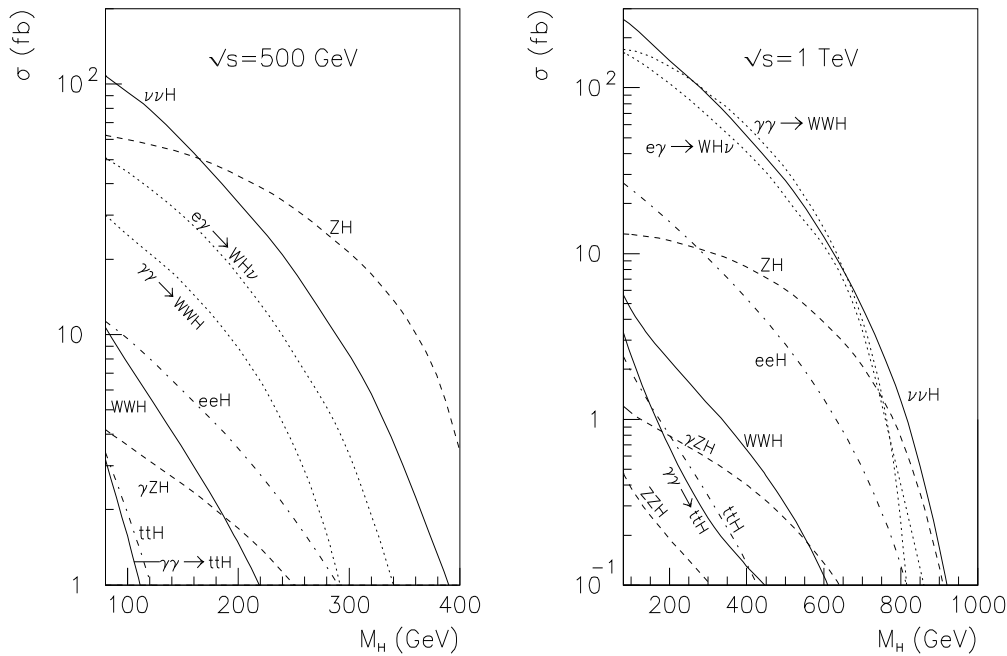}}
\vspace*{-2.cm}
\end{center}
\end{figure*}

Unless one chooses a very monochromatic spectrum designed to cover 
a narrow band around the mass of the IMH, the nice resonance structure 
of the Higgs  is not  so conspicuous. As we have seen, this 
production suffers from large backgrounds at  $500$~GeV and for 
higher energies it seems extremely laborious if not impossible to extract a 
signal. For a Higgs 
heavier than about $400$~GeV at $\sqrt{s_{ee}}\geq 500~GeV$, 
it seems to be impossible to unravel 
a peak formation in the most favourable channel: $ZZ$. 
As in the case of the IMH, the situation gets worse as the \epm energy 
increases. 
It is, therefore important to see whether other mechanisms for Higgs 
production are possible and survive over their corresponding backgrounds. 
Once again, the backbone $WW$ reaction should 
provide for an efficient Higgs production. 
The fact that the WW  cross-section is
so large and that the Higgs couples preferentially to the weak 
bosons, one  expects $WW$ to trigger a good Higgs yield. We have found
\cite{Ottawa,nousgg3v}
that at high enough energy this is one of the most 
important reactions for Higgs production even when 
compared to the usual \epm mode. We concentrate essentially 
on the case of the IMH.  \\
\begin{figure*}[hbt]
\begin{center}
\vspace*{-.5cm}
\caption{\label{ggwwhfeyn}{\em 
Typical diagrams for $WWH$ production in
\gag. Diagram a) is of the fusion type. }}
\vspace*{-2.2cm}
\mbox{\epsfxsize=15cm\epsfysize=6.5cm\epsffile{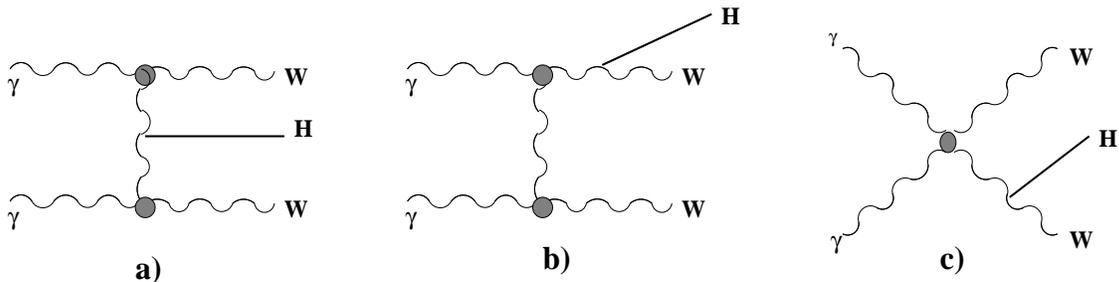}}
\vspace*{-1.cm}
\end{center}
\end{figure*}
\noi
We find that for a Higgs mass 
of $100$~GeV we 
obtain a cross-section, before folding with the
photon luminosity spectrum, of about $20$fb at $\sqrt{s_{\gam}}=500$~GeV.
The $WWH$ cross-section quickly rises to yield $\simeq 400fb$ at 
$2$~TeV (for $m_H=100$~GeV). The importance  of this mechanism 
at TeV energies is best illustrated by contrasting it with top pair 
production (see Fig.~\ref{gammaee}). 
For $m_H=100$~GeV and $m_t=130$~GeV 
the two processes have the same threshold energy 
and lead to the same final state ({\em IMH} 
decays predominantly into $b \bar b$). 
While at $\sqrt{s}_{\gamma \gamma} \simeq 500$~GeV top pair production is 
almost two-orders of magnitude larger than $WWH$, the latter which is a third 
order process is twice as large at $2$~TeV. Nonetheless, the $WWZ$ 
cross-section is about an order-of-magnitude larger than the ``{\em IMH}-$WWH$" 
for all centre-of-mass energies. 
Another non resonant Higgs mechanism that 
has recently been suggested 
\cite{ggtth} 
is Higgs production in association with a top pair in analogy 
to $t \bar t H$ production in hadron machines. Unfortunately, the {\em IMH} 
yield 
does not exceed $1-3fb$ (for $m_t\leq 150$~GeV). The $t\bar t H$ cross-section 
decreases very slowly with $\sqrt{s}_{\gamma \gamma}$. \\
To see the importance of this process we first 
compare it, before folding with  a  luminosity spectrum, 
with other Higgs 
processes that take part in the \epm environment (including $e\gamma$ 
and of course \gag modes). 
This is shown in Fig.~\ref{ggwwh1}. For \epm the standard dominant reactions 
are the Bjorken process and the $WW$ fusion processes. Other \epm processes 
(Higgs in association with two vector bosons) have been studied but 
they yield smaller cross-sections\cite{eevvh,BargerH2v}. \\
\noi
An efficient mechanism for Higgs production in an $e \gamma$ 
environment is through $e \gamma \ra \nu W H$ \cite{egnuwh,Boosegnuh}. 
 Comparing at the \underline{same}
$\sqrt{s}_{\gamma \gamma}$ and $\sqrt{s}_{e \gamma}$ centre-of-mass, 
in the {\em IMH} case, 
the cross-sections for $WWH$ start becoming larger than those
of $e \gamma \ra \nu W H$ for energies around $700$~GeV. At lower energies
the $e\gamma$ mode benefits from a larger phase space 
(see Fig.~\ref{allgamegam} and Fig.~\ref{ggwwh1}). \\
\noi
In the {\em IMH} case, taking for illustration $M_H=80$~GeV, 
at $500$~GeV, $\sigma(\gamma \gamma \ra WWH)\simeq 30$fb which is 
by only a factor $2$ smaller than $\sigma(e \gamma \ra \nu WH)$ and 
a factor 
$3.3$ compared to the dominant $WW$ fusion process in $e^+e^-$. 
On the other hand, $\sigma(\gamma \gamma \ra WWH)$
is larger than all the $VVH$ ($WWH, ZZH, ZH\gamma)$ processes 
in \epm 
by at least a factor $3$\cite{eevvh}. Higgs production from top bremstrahlung ($t\bar{t} H$ 
final state), 
either in \epm \cite{eetth} or \gag \cite{ggtth} 
is abysmally small. At $1$~TeV 
our $\gam \ra WWH$ 
process becomes very comparable to $e\gamma \ra \nu WH$ and is only 
about a factor 2 smaller than the dominant
$WW$ fusion process in \epm. 
Nonetheless, the fact that in $\sigma(\gamma \gamma \ra WWH)$,
unlike the $WW$ fusion in $e^+ e^-$ or the corresponding one at 
$e \gamma$, all final particles 
can be observed or reconstructed 
(hence alleviating the lack in 
energy constraints) makes this reaction worth considering especially at a 
TeV $\gamma \gamma$ collider. But of course, this statement tacitly assumes 
an ideal monochromatic \gag collider. We will now turn 
to more realistic photon luminosity spectra. Before so doing, it is 
worth pointing out that an almost equal number of $H$ is produced in the 
$J_Z=0$ or the $J_Z=2$ with both $W$ being essentially transverse.

\subsection{Folding with the luminosity spectra}
\begin{figure*}[hbt]
\begin{center}
\vspace*{-.5cm}
\caption{\label{ggwwh2}{\em As in the previous figure but where 
we have included the luminosity spectra. For all \gag and 
$e \gamma$ processes we have considered 
a broad spectrum with unpolarised laser and $e^-$. In the case of $WWH$ at 
\gag we also show the effect of a ``$J_Z=0$-dominated" spectrum (see text).}}
\vspace*{-1.cm}
\mbox{\epsfxsize=15.5cm\epsfysize=11.5cm\epsffile{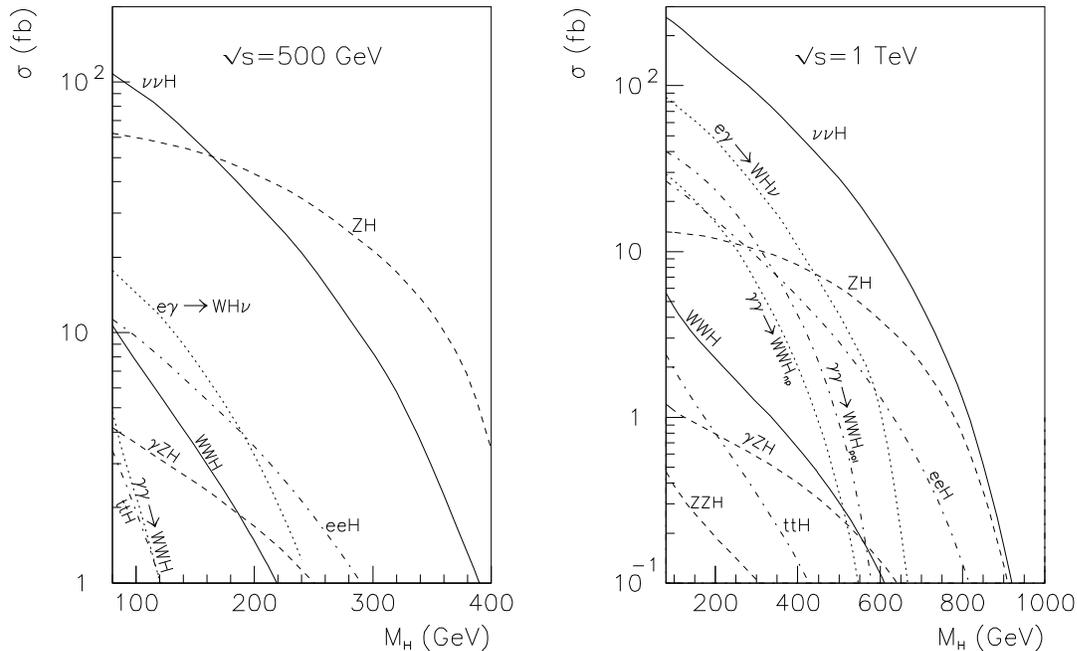}}
\vspace*{-2.5cm}
\end{center}
\end{figure*}

With $\sqrt{s}_{ee}=500$~GeV, the inclusion of the spectra changes 
the $WWH$ yield 
significantly due to the fact that the maximum 
$\sqrt{s}_{\gamma \gamma} \simeq 400$~GeV  leaves a small phase space for the 
{\em IMH}. 
Even when we choose the polarisation of the primary beams to give the 
peaked $J_Z=2$ dominating spectrum, 
the cross-section does not exceed $4$fb 
and
is therefore almost two orders of magnitude below the $WW$ fusion process in 
the \epm mode and an order of magnitude smaller than $\nu W H$ production in the
$e \gamma$ mode (See Fig.~\ref{ggwwh2}). \\
\noi The situation is much more favourable at $1$~TeV. Up to 
$M_H \simeq 300$~GeV this mode produces almost twice as many Higgses as the 
conventional Bjorken process. For $M_H=100$~GeV and choosing a setting 
which gives a ``0-dom", we obtain $\sim 37.5$fb (compared to $37.2$ in 
the ``2-dom") and 
$26.3$fb with no polarisation for the primary beams. 
The advantage of a polarised spectrum is undeniable. 
The two cross-sections for $\nu W H$ (in $e \gamma$)
and $H \nu \nu$ (in \epm) are respectively about 2 and 5 times larger in the 
{\em IMH} 
case. A comparison between the variety of Higgs production modes in the 
NLC(1TeV) environment is shown in Fig.~\ref{ggwwh2} which clearly brings out the 
importance of $WWH$ in \gag.

  \noindent
Considering the large $WWZ$ yield, $b$ tagging is almost necessary for the 
{\em IMH} 
search. Another dangerous background, even with $b$-tagging is
due to top pair production: $\gamma \gamma \ra t \bar{t} \ra W^+W^- b \bar{b}$. 
For instance, at $\sqrt{s}_{ee}=500~GeV$, this is about two-orders of magnitude 
larger 
than $WWH_{\hookrightarrow b \bar b}$. Fortunately, one can eliminate this huge 
contamination by rejecting all those $WWH$ events where 
the \underline{simultaneous} cuts on the invariant mass of the two $Wb$ is 
such that the $Wb$ does not 
reconstruct the top mass (within $15$~GeV)

\beqn
m_t - 15~GeV < M_{W^+b} < m_t + 15~GeV \;&\mbox{{\rm and}}&\;
m_t - 15~GeV < M_{W^{-}b'} < m_t + 15~GeV \nonumber \\
&\mbox{{\rm or}}& \nonumber \\
m_t - 15~GeV < M_{W^+b'} < m_t + 15~GeV \;&\mbox{{\rm and}}&\;
m_t - 15~GeV < M_{W^-b} < m_t + 15~GeV \nonumber \\
\hspace*{2cm}
\eeqn

\noindent 
The reason we try both combinations $W^+b$ or $W^+b'$ is that we do not want 
to rely on charge identification, for the $b$ especially, which necessarily 
entails a reduction in the $b$ sample (and hence our signal). A good vertex 
detector should be sufficient 
\footnote{We have not tried to cut the $t \bar t$ by demanding that 
$m_{b\bar b}=M_H\pm10$~GeV, as the cut above is very efficient. Moreover, 
based on our previous analysis of $WWH$ in \epm \cite{eevvh}, the $Wb$ cut 
was by far more efficacious.}. In carrying the vetoing in our Monte-Carlo 
sample we made the Higgs decay isotropically in its rest frame. The 
effective loss at $500$~GeV is about a mere $0.3$fb while at $1$~TeV, where 
we have a ``healthy" cross-section, the percentage loss is only about $4\%$ 
for all choices of the polarisation. Table \ref{tablewwh} shows the cross-sections taking 
a Higgs mass of $M_H=100$~GeV with $Br(H \ra b \bar b) \sim 80\%$,
$|\cos\theta|<0.8$ for all angles between two particles  and
assuming $m_t=150$~GeV. 
 
\begin{table}
\caption{\label{tablewwh} 
{\em Cross-section in fb for three-boson (and $t \bar t$) 
productions with $M_H=100GeV$ and $m_t=150$~GeV. The $WW\gamma$ includes 
a $p_T^\gamma$ cut of $20GeV$ at
$500$~GeV and $40GeV$ at $1TeV$. ``non-top" means all Higgs events with Higgs 
decaying into $b \bar b$ and where the simultaneous $Wb$ invariant mass has 
been applied as explained in the text. ``direct" means that we have not 
taken into account top pairs produced through the gluons inside the photon, 
i.e., the ``resolved" photons contribution has not been considered.}}
\vspace*{.3cm}
\begin{tabular}[htb]{|l||c|c|c||c|c|c||}
\cline{2-7} 
\multicolumn{1}{l||}{}&\multicolumn{3}{c||}{500~GeV}&\multicolumn{3}{c||}{1~TeV}
\\ \cline{2-7}
\multicolumn{1}{l||}{}&non pol.&0-dom.&2-dom&non pol.
&0-dom.&2-dom\\ \hline
$WWH$&1.0&1.7&2.24&26.3&37.5&37.2 \\
$WWH_{\hookrightarrow b \bar b}$&0.8&1.4&1.8&21&30&29.8\\
$WWH_{\hookrightarrow b \bar b}$ ``non-top"&0.7&1.1&1.5&20.2&28.9&28.7 \\
$WWZ$&24.2&55.3&36.9&342&473&408 \\
$WWZ_{\hookrightarrow b \bar b}$&3.6&8.3&5.53&51.3&70.9&61.2 \\
$WW\gamma$&205&321&272&483&592&560\\
$t \bar t$ (``direct")&207&458&250&620&525&687 \\
\hline
\end{tabular}
\end{table}
Once the ``faked" top events have been dealt with, 
the $WWZ_{\hookrightarrow b \bar b}$ do not bury the signal 
(for $M_H \sim M_Z \pm 10$~GeV).
These $WWZ$ can be further reduced by judiciously switching the 
``2-dom." setting, both at $500$~GeV and at $1$~TeV. Although at the former 
energy 
the event rate is probably too small to be useful, at $1$~TeV, in the ``2-dom.", 
we have, after including the cuts and the branching fractions into $b$, $30$fb 
of signal compared to $60$fb from $WWZ$. With one $W$, at least, decaying into 
jets and not taking into account decays into $\tau$'s, the number of $WWH$ 
with the contemplated integrated luminosity of ${\cal L}=60fb^{-1}$ will be 
about 1400 events. Even if one allows for an overall efficiency of 
$50\%$ this is a very important channel to look for the Higgs. There is one 
background which we have not considered, the 
$W^+W^- b \bar b$ final state with  $b \bar{b} \ra W^+ W^-$ as a 
sub-process. We expect this to be 
negligible once one puts a high $p_T$ cut on both $b$'s and require 
$m_{bb} \sim M_H$. \\

To conclude,  this new mechanism of Higgs production 
in a $\gamma \gamma$ mode of $\sim 1~TeV$ \epm collider 
is a very promising prospect. 
The oft discussed intermediate mass Higgs production, as a narrow 
resonance in \gag collisions, relies on a spectrum which is peaked 
around the Higgs mass in a $J_Z=0$ dominated setting. The extensive study 
in \cite{Borden} finds that with $\int {\cal L}_{ee}=10fb^{-1}$, one expects 
between about $500$ Higgs events for $M_H \sim M_Z$ to about $600$ events for 
$M_H \sim 140$~GeV. 
This is  $2-3$ times more than what we get 
with $WWH$ at $\sqrt{s}_{ee}=1$~TeV if only $\int {\cal L}_{ee}=10fb^{-1}$ 
is assumed.  However, as we have 
stressed repeatedly, the resonance scheme means that 
the available $\gamma \gamma$ invariant mass covers a very narrow, and 
in the case of the {\em IMH}, low range of energies. Hence, while allowing 
a precise study of the $H\gamma \gamma$ coupling it forbids the 
study of a wealth of interesting processes in the $\gamma \gamma$ mode of the 
NLC. Higgs detection through $WWH$ at $1$~TeV will be one aspect among a 
variety of studies of weak processes ($WW, ZZ, WWZ$,...{\it etc}).  
At $500$~GeV this mechanism does not offer much prospect with a 
nominal integrated luminosity of $10fb^{-1}$. However, in this case we have shown that 
with a broad spectrum -provided optimal values for the mass resolution, the 
polarisation and the tagging efficiencies are achieved- 
the Higgs can be observed through the resonant 
mechanism. Therefore, as far as the IMH is concerned, it will always be 
possible to discover the Higgs in a \gag machine without having recourse 
to the peaked-narrow spectrum. The latter, of course, assumes a knowledge 
of the Higgs mass. 

 \section{The $W$ content of the photon and the prospect of 
$WW$ scattering in the \gag mode} 

\renewcommand{\thequation}{\thesection-\arabic{equation}}
\setcounter{equation}{0}

\begin{figure*}[hbt]
\begin{center}
\caption{\label{ggwwsplit}{\em A photon of helicity $\lambda$
``splits" into a longitudinal $W$ that carries a fraction
$y$ of its momentum and a spectator $W$ with helicity $\lambda'$.
The longitudinal $W$ then takes part in the hard process.}}
 \vspace*{0cm}
 \mbox{\epsfxsize=15.5cm\epsfysize=6cm\epsffile{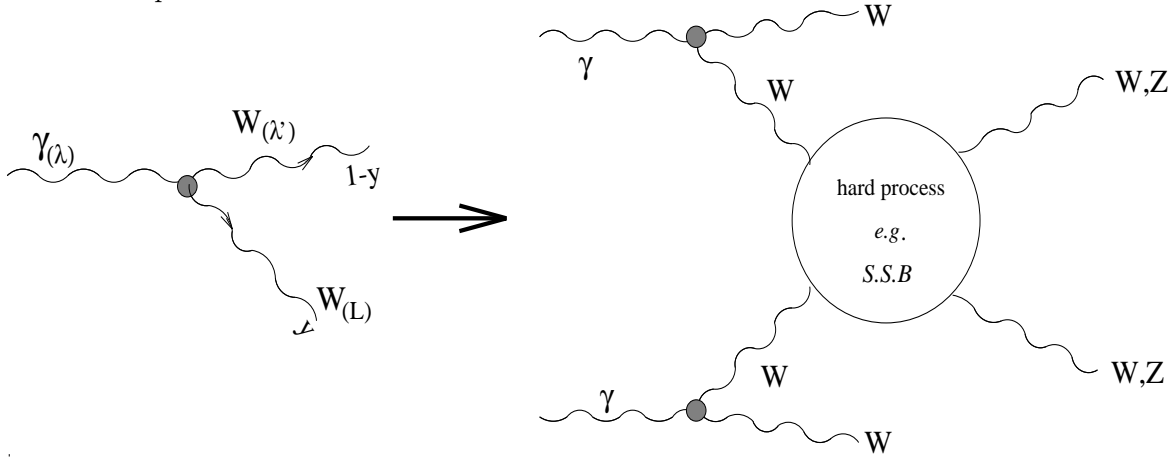}}
 \vspace*{-.5cm}
\end{center}
\end{figure*}

We have seen that with $W$ pair production we could investigate the 
symmetry breaking sector of the \sm through, for instance, 
the chiral Lagrangian 
parameterisation. This reaction, together with the $ZZ$ process, 
could be regarded as the testing ground of 
possible rescattering effects in $WW\ra VV$ ($VV=WW,ZZ$) that originate 
from the symmetry breaking sector. Unfortunately, 
at high energies, {\it i.e.}, at high $VV$ invariant masses, where 
the effect of the New Physics would be most evident, one has to fight 
extremely hard against the pure gauge sector that produces large 
cross sections for transverse $W$ and $Z$. \\
\noi At the $pp$ collider(s) and in the 
usual \epm mode one often relies, at TeV energies, 
on $WW$ scattering processes. Ideally, the symmetry breaking sector would be 
most efficiently probed through the reactions $V_L V_L \ra V_L V_L$ but then 
one needs a source of longitudinal 
vector bosons. Lacking such a source, the $V_L V_L$ 
subprocesses are embedded 
in a large class of diagrams which, not only have not much to do with 
the physics one wants to get at, but 
also complicate the computational task. 
It has become customary to rely on some approximation
\cite{CahnEWA,DawsonEWA,ChanowitzEWA,KaneEWA,LindforsEWA,RolnickEWA}, 
with various degree 
of accuracy, whose aim is to isolate 
the interesting subprocess and convolute the cross section with an effective 
luminosity of longitudinal vector bosons. The latter are regarded 
as partons or constituents of the light fermion. The approximation 
(effective $W$ approximation or EWA
\cite{CahnEWA,DawsonEWA,ChanowitzEWA,KaneEWA,LindforsEWA,RolnickEWA}) is really an 
adaptation of the Weisz\"acker-Williams approximation\cite{EPAWW}, 
the effective photon approximation (EPA). 
Still, in the 
context of non-Abelian QCD, this has to be paralleled with the 
splitting or structure function language. The latter analogy is more to the 
point for the 
application to the \gag collider where we would like to find out the splitting 
of the photon into $W^\pm$ that emanates solely from the non Abelian 
part (as in $g\ra gg$). 
A novelty, here, being of course the longitudinal $W$ 
content (see Fig.~\ref{ggwwsplit}).\\
An expression for the longitudinal $W$ content inside the photon has been 
derived \cite{egnuwh}. It was found that the photon 
has a relatively larger $W_L$ component than the electron. 
In view of this encouraging result, it has been pointed out that
one could hope\cite{Munich92,BrodskyHawai}, 
at TeV energies, 
to study $WW$ scattering more efficiently than in the \epm mode.
Here, we will argue that great care should be exercised in 
the use of the expression derived in\cite{egnuwh} 
to applications to other processes. We will then present a new 
set of polarised structure functions of the $W_L$ content of the 
photon. In a nut-shell, 
we have identified a $Q^2$ dependent part in the splitting function 
that is not interpreted as such in \cite{egnuwh}. 

Having in view the extraction of the longitudinal $W$ component of the 
photon, it is worth recalling the extraction of the analog component 
of the light fermion. The problem in the fermion case 
is simplified because 
it is possible to find processes that are described by a single diagram. 
This diagram is then amenable to an interpretation in terms of structure 
functions. In the all-bosonic sector and especially when 
one is dealing with the longitudinal vector bosons it is difficult 
to isolate 
a single diagram since it is not gauge 
invariant by itself. As a consequence 
one has to tackle the problem of unitarity first and extract the part 
of the single diagram which is well behaved and reflects the dominant 
contribution of the whole set of diagrams.\\
\noi Another aspect common to both the fermion and the 
photon splittings is that the validity of the effective boson 
approximation exploits the fact that 
the parton virtuality is small (so that its propagator is large) and that 
the subprocess (the hard scattering part) has a smooth limit in the 
small virtuality. The smooth mass limit in Abelian QED exists and explains 
the success of the EPA. 
Unfortunately, the off-shell hard scattering 
diagrams in the electroweak theory do not have a ``smooth 
mass limit" and are not element of the
$S$-matrix. This is especially acute for the ``longitudinal" states 
where the very delicate gauge theory cancellations are no longer operative. 
This problem of the virtuality versus unitarity that is present in the 
fermion case of the EWA 
is also unavoidable in the photon to $W$ splitting. 
This first aspect of the unitarity problem has to do with 
the bad high energy behaviour of the off-shell hard scattering process
\cite{KleissEWA}. 
In the photon case, there is another potential 
complication with unitarity which concerns the splitting part and has to do 
with the spectator $W$. 
Take a polarised photon that splits into two $W$'s. The case 
where both $W$ are transverse, $\gamma \ra W_T W_T$, 
should be quite similar to the gluon splitting 
into two gluons, $P_{GG}$\cite{Altarelli-Parisi}. This component has not much 
incidence on what 
we want to study, the \ssb mechanism. As for the $W_L$ component,
it consists of two parts: one when the spectator $W$ is transverse,
$\gamma \ra W_L (W_T)$ and one 
when it is longitudinal ($\gamma \ra W_L (W_L)$). 
The extraction of the latter poses more problems. 
The reason is the following. 
In the diagrammatic identification of the spectator $W_L$, this is an on-shell 
external $W_L$. As with most processes that involve vector bosons, 
the problem of unitarity is more severe with $W_L$ than with $W_T$. 
Subtle cancellations between different 
diagrams have to take part before arriving at a unitary result. 
Therefore, in general, even when one isolates 
the single diagram that should capture the essence of the 
EWA ($t$-channel diagram), there is some ``kneading" to do. 
One should first subtract 
any potential ultraviolet part that it may contain. 
When this is done, the ``universality" of the remaining 
``collinear-sensitive" term is not so obvious. \\

In\cite{egnuwh} the $W_L$ content of the photon is extracted, 
in a heuristic 
way, by studying the process $\gamma W_L \ra H W$. From the high-energy 
asymptotic cross section of this process
\beqn
\sigma(\gamma W_L \ra W H)\stackrel{\mwmw \ll s,\mhmh}{\rightarrow}
\frac{\pi \alpha^2}{s_w^2 M_W^2} 
\left(1-\frac{\mhmh}{s}\right) \left\{1+\frac{M_H^4}{2s^2} \left(-2 + 
\log\frac{(s-\mhmh)^2}{s\mwmw} \right)\right\} \nonumber \\
\label{gwhwapprox}
\eeqn
the following $W_L$ distribution inside the photon is identified 
\beq
D_{W_L/\gamma}(y)=\frac{\alpha}{\pi} \left\{\frac{1-y}{y} +\frac{y (1-y)}{2} 
\left(-2 + \log\frac{s(1-y)^2}{\mwmw} \right)\right\}
\label{harddis}
\eeq
where $y$ is the momentum fraction of the photon transferred to the 
$W$ (see Fig.~\ref{ggwwsplit}).
Note that in this result, one has summed over all helicity states of the 
spectator $W$. \\
We have rederived the asymptotic form $\sigma(\gamma W_L \ra W H)$, 
our aim being 
to isolate the different helicity states of the spectator $W$ and to 
check that, indeed, there is a subtle cancellation in the $W_L$ spectator 
(the outgoing $W$). 
In order not to unduly complicate the cancellation mechanism, we have chosen 
(once again) to work in the non-linear gauge since we keep the same number of 
diagrams as in the unitarity case while the $W$ propagator is as in the 
t'Hooft-Feynman gauge. Power counting (in the ultraviolet sense) 
is trivial and shows that when the would-be spectator $W$ is transverse, 
taking only the $t-$channel 
diagram suffices to reproduce the leading term in the 
cross section. When the spectator $W$ is longitudinal 
it is crucial to subtract a part that corresponds to the annihilation diagram 
in order not to violate unitarity. 
For the transverse spectator $W$ 
one could then identify 
\beqn
D_{W_L/\gamma_\lambda}^{(W_\lambda)}(y)=\frac{\alpha}{\pi} \frac{1-y}{y} 
\label{fstructl}
\eeqn
For this part we arrive at the same identification as \cite{egnuwh}. 
We note that in this case the photon transfers its helicity, $\lambda$, to 
the spectator $W$. The contribution from the $W$ with the opposite helicity 
, $-\lambda$, 
is non leading. The important observation is that 
$D_{W_L/\gamma_\lambda}^{(W_\lambda)}(y)$ is of exactly 
the same form and strength 
(allowing for the overall strength of the $We\nu_e$ coupling) 
as the $W_L$ distribution inside the electron:
\beq
D_{W_L/e}^{(W_\lambda)}(y)=\frac{\alpha}{4\pi s_W^2} \frac{1-y}{y} 
\eeq
This is already a hint that 
$D_{W_L/\gamma_\lambda}^{(W_\lambda)}(y)$ qualifies as a universal 
distribution. 
The remaining $(W_L)$ contribution can be read-off from 
the $M_H^4$ term in Eq.~\ref{gwhwapprox}. If one makes the same 
identification as the 
authors of ref.~\cite{egnuwh} one would arrive at 
\beq
\widetilde{D}_{W_L/\gamma_\lambda}^{(W_L)}(y)=
\frac{\alpha}{\pi} \frac{y (1-y)}{2} 
\left(-2 + \log\frac{s(1-y)^2}{\mwmw} \right)
\eeq
We would like to argue that this expression could be 
misleading when  applied to other processes. First of all, it does not have 
the $y\leftrightarrow (1-y)$ symmetry as it should for $\gamma \ra W_L W_L$. 
In fact, the argument of the logarithm, which is precisely the non-symmetric 
part, depends on the kinematics and the nature of 
the hard sub-process. The argument of the logarithm comes solely 
from integrating over the virtuality $Q^2= M_W^2 -(k_\gamma -k_W^{spec.})^2$; 
where $k_W^{spec.}$ is the momentum of the spectator $W$ and 
$k_\gamma$ that of the incoming photon. It is best to 
write the logarithm in Eq.~\ref{gwhwapprox} as 
\beq
\log \left( \frac{(s-M_H^2)^2}{s M_W^2} \right) =\log( 
\frac{Q^2_{max.}}{Q^2_{min.}}) 
\eeq
where $Q^2_{max.,min.}$ are the minimum and maximum value of $Q^2$. 
For the validity of the EWA in the photon splitting we have warned that 
the virtuality must be very small, so that in the identification 
$Q^2$ should be kept small. We note that our observation has some similarities 
with the so-called modified 
EPA \cite{EPAmodified}. We thus suggest 
to replace 
\beq
\log( 
\frac{Q^2_{max.}}{Q^2_{min.}}) \longrightarrow \log( 
\frac{Q^2}{Q^2_{min.}}) \longrightarrow \log( 
\frac{Q^2_p}{M_W^2}) 
\eeq
where as a further approximation we have taken $Q^2_p$ to  be a typical 
$Q^2$ of the hard subprocess under study. 
Therefore, we arrive at the approximate parameterisation with a fixed $Q^2_p$ 
\beq
D_{W_L/\gamma_\lambda}^{(W_L)}(y,Q^2_p)=
\frac{\alpha}{\pi} \frac{y (1-y)}{2} 
\left(-2 + \log\frac{Q^2_p}{\mwmw} \right)
\label{strucfq2}
\eeq

\vspace*{1cm}

Because of the various subtleties involved in the extraction of 
the $W$ component inside the photon it is worth reconsidering the issue 
by studying another process. We have looked at \ggwwht. We leave the full 
details for another paper and only present the main points. \\
\begin{figure*}[htb]
   \begin{center}
   \vspace*{-.5cm}
  \caption{\label{wwhapprox1} a) {\em The ratio $\sigma_{TT}/\sigma_{total}$ in the case 
$M_H=400$~GeV (full line) and $M_H=100$~GeV dashed line. The total refers to 
the exact numerical result including all the diagrams and summing over all 
polarisations. The lines ``all diag." are for the exact $TT$ result. 
The fusion refers to taking only the fusion diagrams with 
the $W$ being transverse while the ``approx." curves are the approximate 
analytical expression for $TT$ (see text)}.  
{\em b) Comparing the exact 
$TT$ and $TL$ cross sections for $WWH$ with $M_H=400$~GeV with the result 
of using the structure function. For $TL$ we show the result of taking 
the structure function with $Q^2_p=M_H^2$ ($EWA(Q^2)$) as well as the harder 
distribution
``EWA'' \protect\cite{egnuwh}.}}
 \vspace*{-1.cm}
\mbox{   
\mbox{\epsfig{file=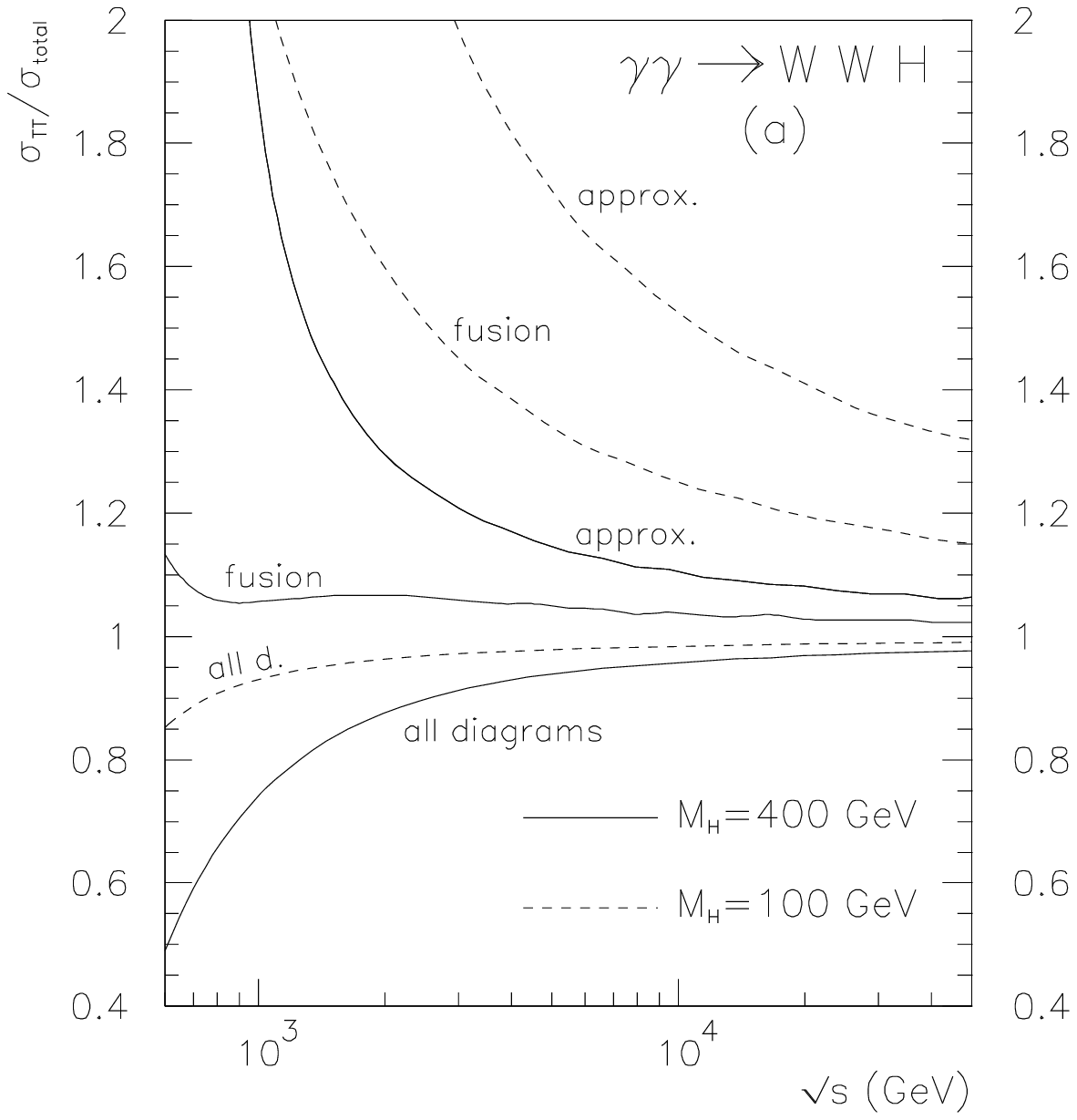,width=8.cm,height=10cm}}
\hspace*{-.6cm}
\mbox{\epsfig{file=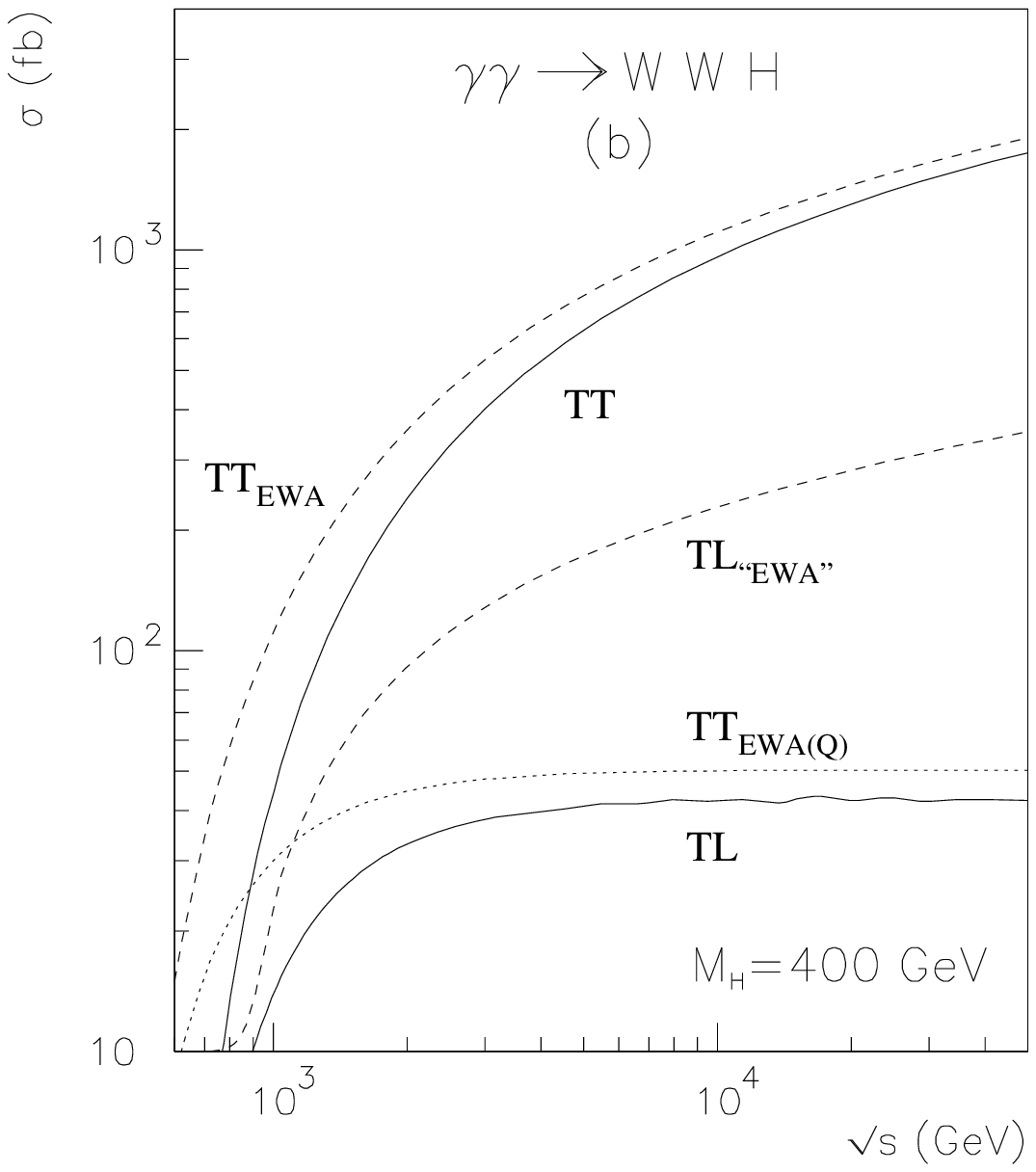,width=8.cm,height=10cm}}}
   \end{center}
   \vspace*{-1.cm}
 
\end{figure*}
\noi To be able 
to formulate the problem in terms of a distribution function, we will consider 
the particular case of a heavy Higgs of mass $400$~GeV that couples 
almost exclusively to the longitudinal vector bosons. We recall that 
\ggwwht cross section is dominated, at high energy, by the production of both 
$W$ being transverse. $W_T W_L H$ represents a small fraction, about 
$10\%$ for a Higgs mass of $400$~GeV and for $\sqrt{s_{\gam}}>2$~TeV. The $LL$ 
contribution is dismal. This is a good sign since the contribution
that requires the most delicate cancellation between the different 
diagrams, $W_L W_L H$,  is the smallest\cite{nousgg3v}. 
Therefore, at high energy and in the case of a heavy Higgs (that we expect 
to couple almost exclusively to the internal quasi-real longitudinal $W$'s) 
we can afford to keep only the fusion diagrams 
and restrict ourselves to the transverse modes of the final $W$. 
We find that this is indeed the case (see Fig.~\ref{wwhapprox1}a). At 
energies around $2$~TeV, Fig.~\ref{wwhapprox1}a shows that our expectation 
is borne out by the result of the exact calculation. Indeed, we see that 
the 
$TT$ are dominating and that taking only the $TT$ fusion diagrams reproduce 
the total cross section extremely well. Note, however, that there is a slight 
overestimate of the cross section by only keeping the fusion diagrams. In this
figure we have plotted the ratio $\sigma_{TT}/\sigma_{total}$ where 
$\sigma_{total}$ is the exact result (all diagrams included) with all 
final polarisation summed over. 
Fig.~\ref{wwhapprox1}a
 shows also the case of a light (IMH) Higgs where we learn 
that, as expected, the approximation of keeping only the fusion diagrams 
is not as good especially at lower energies. Nonetheless, the $TT$ cross 
section dominates (with all diagrams included) and almost reproduces 
the total cross section. \\
 \noi Having isolated the dominant topology, the fusion diagrams, 
it becomes easier to find an analytic approximation for $W_T W_T H$. 
The analysis is 
quite simple in the non linear gauge. By neglecting non-leading $M_W$ 
and $P_T^W$ terms we find, independently of the Higgs mass,
\beq
\sigma_{\ggwwh}\sim\sigma_{\gam \ra W_T W_T H}\sim
\frac{\alpha}{4 \pi \sin{\theta_W}^2} \left\{
(1+\frac{\mhmh}{s})\ln{\frac{s}{\mhmh}}-2(1-\frac{\mhmh}{s})\right\} 
\sigma_{\ggwwe}
\label{ggwtwthapprox}
\eeq
We show in Fig.~\ref{wwhapprox1}a how well this analytic 
approximation of the fusion
diagrams performs ($\sigma_{total}$ once again is the exact result). 
We see that for the $400$~GeV Higgs the approximation is quite good, 
already at $2$~TeV it is within $10\%$ of the exact result and gets 
better with increasing energy. For the IMH, only the order of magnitude is 
reproduced by using the approximation. \\

In fact, once we have brought the problem down to the fusion diagrams 
and the transverse external $W$'s, the situation is quite similar to 
the study of  heavy Higgs production in $e^+ e^- \ra \nu \bar{\nu} H$ 
which is a pure fusion process\cite{Cahnnunuh}. \\
\noi 
 Of course, the above formula~\ref{ggwtwthapprox} being for 
transverse external $W$ we can only extract the part of the structure 
function that corresponds to a 
transverse spectator $W$: $D_{W/\gamma_\lambda}^{(W_\lambda)}$. 
It is easy to obtain that the leading term is 
indeed given by Eq.~\ref{fstructl}.
Therefore, this gives more credence to the fact that this part can be 
considered universal. 
We show in Fig.~\ref{wwhapprox1}b how good the use of the structure function 
with the transverse spectator $W$ is by comparing it with the {\em exact} 
$W_T W_T H$. 
We have not attempted to extract the 
$W_L W_T$ part from this process, however we have tried the
$Q^2$ parameterisation 
suggested above (Eq.~\ref{strucfq2}). We have taken the typical $Q^2_p=\mhmh$ for the hard process 
since this is the only scale at hand, if we take $M_W$ to be 
an ``infrared" scale. 
The ``$Q^2_P$-EWA" 
is excellent and reproduces the energy dependence of the cross
section very well, this is not the case of the harder distribution suggested 
in \cite{egnuwh} as  Fig.~\ref{wwhapprox1}b makes clear. 

As a conclusion we would like to emphasize that care must be taken 
in applying the structure function formalism to $WW$ scattering in a \gag 
option. The encouraging aspect is that there is indeed more 
$W_L$'s inside the photon than in the electron due to the additional 
degree of freedom provided by the longitudinal spectator $W$. Unfortunately, 
the bulk of the contribution is from the transverse spectator part 
that has 
exactly the same distribution as that of the $W_L$ inside the electron. The 
strength of the latter ($W_L$ inside the $e^-$) is in fact a little larger than 
the former ($W_L$ inside the $\gamma$) because of the slightly larger $We\nu$ 
coupling.\\
\noi An application 
of the $W$ content inside the photon to models of a strongly interacting Higgs 
has been done in \cite{Cheungstrong} by taking 
the very hard 
distribution, Eq.~\ref{harddis}. 
Apart from the choice of the distribution, the most worrying 
aspect that we foresee is the problem of the expected large 
background with $4W$ or $WWZZ$ final states\cite{Jikiainprog}, 
all vectors being transverse. We have seen that the production of 
three vector bosons within the \sm is quite large and growing. Hence we 
might have to face the same recurring problem, {\it i.e.,} 
that of trying to get rid of 
the annoying    gauge transverse sector contribution.

\section{Conclusions}
The laser induced \gag collider at sub-TeV or TeV energies 
offers a rich and attractive physics program especially as regards 
the physics of $W$'s and the concomitant symmetry breaking 
phenomenon. Such machines are ideal 
$W$ factories and enviable laboratories for precision tests on 
the Higgs properties. 
As we have shown, one should, perhaps, add a slight undertone as concerns
$W$ physics. This  is 
because, while we have plenty of cross section 
as compared to the usual 
\epm and a host of appealing 
processes , the preliminary studies indicate that the most 
interesting part of the $W$ (that has a stronger link 
to symmetry breaking) has a somehow larger ``intrinsic self-background" 
from the ``gauge-transverse" sector than at the usual 
\epm mode. \\
\noi Of course, more physics can be done than the aspects 
we have covered here. For instance, one could mention 
the search for new particles 
like those of 
susy \cite{Goto,Franksusy} or those that signal a new layer of matter 
\cite{excitedIlya,excitedme} 
and how these searches could complement those  in the \epm 
mode. 
The ability to access the $J_Z=0$ quite naturally 
is a strong asset, as is the possibility to produce in an almost 
democratic way any charged particle. 

There is, though, still much work to  be done in this nascent domain. 
Leaving aside the technical feasibility, thorough investigations and 
simulations of the interaction region should be performed. One will 
have an unusual environment where the backgrounds could be more severe 
than those we encounter in the usual \epm mode. For instance, we have not 
addressed the subject of the hadronic cross-section and the minijet 
problem\cite{Manueljets}. Although, at the present time, 
this problem does not look as dramatic\cite{Forshaw,Peskinhadronic}
as it first appeared three years ago, it is nonetheless 
more severe in a \gag mode than in the \epm mode 
of the next linear collider 
especially at $TeV$ energies\cite{Peskinhadronic}. At a \gag version 
of a $500$GeV linear collider the situation looks quite ``bright"\cite{Peskinhadronic}. 
In any case, one should strive 
to further reduce any of these backgrounds even at $TeV$ energies or be 
able to efficiently simulate them. Another potential problem that we have 
alluded to a few times, is the purity of the \gag state and the 
rejection of the soft electrons which could also, through multiple 
interactions with the laser beam, lead to further disruption. 
The incidence, on the design of a dedicated detector, 
of a large magnetic field to sweep these unwanted 
relics deserves an urgent and critical clarification.  

Although, at present, one does not 
have the ideal laser that embodies all the requirements (high power, 
high repetition, short pulse length at the appropriate frequencies)
\cite{Richard,Borden} 
 for the 
conversion of the \epm NLC into the \gag mode, there is all reason to 
believe 
that such a device will be available. Free Electron Lasers 
(FEL)\cite{Borden} seem to be a hopeful prospect. 
Because of all these critical machine/detector issues it would very 
gratifying to have an unexpensive test-machine to convince us of the 
feasibility of the scheme. It has been suggested\cite{Borden} to 
build a low energy prototype.\\ 

Once the ``shadows" associated with the technical 
feasibility have been dispelled, the physics program that such a 
collider offers are so enthralling that every effort should be 
made to turn this ``bright" idea into practice. Also, 
one can not help over-emphasizing that such a scheme will be most 
beneficial as part of a package with a normal mode of the \epm and 
that it is the combination and the complementarity 
of the two modes which will make the NLC 
a dream machine. 

\vspace*{1.5cm}

\noi{\bf Acknowledgments}

We have greatly enjoyed our lengthy and valuable 
discussions with Ilya Ginzburg, George 
Jikia and Valeri Serbo. 
On QCD issues, we were lucky to have 
Patrick Aurenche and Jean-Philippe Guillet a few doors away from our offices. 
We acknowledge helpful discussions with Michel Fontannaz and Peter Zerwas
concerning the ``resolved" photon problem. 
We also wish to thank Edward Boos, Misha Dubinin 
and Slava Ilyin for confirming our results on the three body processes and 
for discussions. 
We also benefited from discussions with 
Andr\'e Courau, Abdel Djouadi, George Gounaris, David Miller and Tran Truong. 
We are indebted to Abdel Djouadi for sending us the Fortran 
code for the Higgs branching ratios, George Jikia for making available the 
figures for $\gam \ra ZZ$ and Misha Bilenky for adapting the 
$e^+ e^- \ra W^+W^-$ BMT program to the case of the chiral Lagrangian. 
MB and GB wish to express their gratitude for the hospitality LAPP extended 
to them. 

 \setcounter{section}{0} 
\setcounter{subsection}{0}
\def\thesection{\Alph{section}} 
\def\thesubsection{\thesection.\arabic{subsection}} 

\section{Appendix} 
\setcounter{secnumdepth}{2} 
\setcounter{equation}{0}
\def\thesubsection{\thesection.\arabic{subsection}} 
\def\theequation{\thesection.\arabic{equation}} 

The formulae for the   helicity amplitudes of the anomalous 
trilinear couplings, $\Delta\kappa$ and $\lambda_\gamma$ are 
collected here.
The reduced amplitude ${\cal N}^{ano}_{\lambda_1 \lambda_2 \lambda_3 
\lambda_4}$  
is defined in the same way as the    \sm 
ones, {\it i.e.} in ~\ref{eq:mn}, ${\cal N}^{sm}\ra{\cal N}^{ano}$
 
 \begin{eqnarray}
 {\cal N}^{ano}_{\hppll} &=&
  \dk(\gamma\sinsq+4\cossq) +4\lag\sin^2\theta+\klam(1-3\cos^2\theta)\nonumber\\
& &\frac{\dk^2}{2 }(\gamma\sin^2\theta-1+3\cos^2\theta)+
\frac{\lag^2}
{4}(1+\cos^2\theta)( \gamma\sinsq+4\cossq-2) \nonumber\\
{\cal N}^{ano}_{\hpmll} &=& -\biggl\{
 4 \dk+\frac{\kac }{4}(\gamma+2)+
 \frac{\klam}{2}(\gamma-2)+ 
 \frac{\lac}{4} [(\gamma-4)\cos^2\theta+2] \biggr\} \sin^2\theta\nonumber
\end{eqnarray}

\begin{eqnarray} 
 {\cal N}^{ano}_{\hpplt} &=& \frac{\cos\theta\sin\theta}{\sqrt{2\gamma}}\biggl\{ 
-\dk\left[ \gamma\beta+\lambda_3(\gamma-4)\right]-\lag\left[
\gamma\beta+\lambda_3(\gamma+4)\right]-\nonumber\\
& &\frac{\dk^2}{2}\left[ \gamma\beta+\lambda_3(\gamma-2)\right]
+\frac{\lag^2}{8}\left[\gamma(\lambda_3-\beta)(\gamma-4)\sin^2\theta+8\lambda_3  \right]
-\nonumber\\ & &
\klam \lambda_3(\gamma+2)\biggr\}\nonumber\\
 {\cal N}^{ano}_{\hpmlt} &=& \frac{\sin\theta}{\sqrt{2\gamma}}
(1+\lambda_3\cos\theta)\biggl\{
  \dk (\gamma+4)+ \lag(\gamma-4) +\nonumber\\
& &\frac{\klam}{4} (\gamma-2)
(\gamma-(\gamma-4) \cos\theta\lambda_3) + 
 \frac{\kac}{4}\biggl\{3\gamma-(\gamma-4) \lambda_3\cos\theta 
\biggr\}+\nonumber\\
& &\frac{\lac}{8}\bigl[ \gamma(\gamma-2)(1-\lambda_3\cos\theta)^2+
 2\lambda_3\gamma\cos\theta(1-\lambda_3\cos\theta)+8\cossq\bigr]
\biggr\} \nonumber\\
 \end{eqnarray} 
 \begin{eqnarray} 
 {\cal N}^{ano}_{\hpptt} &=& 
2\dk  \ppout\left[2+\beta\lambda_3(1+\cos^2\theta)\right]+
\nonumber\\
& &+ \lag \biggl\{ \sin^2\theta(\gamma-2)-\ppout
\left[ \beta\lambda_3\left(\gamma\sin^2\theta+
2(1+\cos^2\theta)\right) 
 -4\cos^2\theta\right]\biggr\}+\nonumber\\
 & &\frac{\dk^2}{8}\biggl\{\ppout\left[\gamma\sin^2\theta-
 \beta\lambda_3(
\gamma\sin^2\theta-4\cos^2\theta)+6(1+\cos^2\theta)\right]+ 
 2\pmout\sin^2\theta\biggr\}+\nonumber\\
& &+\frac{\klam}{4} \biggl\{\ppout\left[
 (\gamma \sinsq+4\cossq)(1-\beta\lambda_3)-6\sinsq \right]+
2\pmout(\gamma-1)\sinsq\biggr\}+\nonumber\\
& &\frac{\lag^2}{16 }
\Biggl(\ppout(1-\beta\lambda_3)
\left[\gamma(\gamma-2)(3-\cos^2\theta)\sin^2\theta+ 
 2\cos^2\theta(5\gamma-\gamma\cos^2\theta-4)\right]\nonumber\\
&&-2\sin^2\theta(\gamma-2)(\sin^2\theta+2\ppout)-4\cos^2\theta
(\sin^2\theta+4\ppout)\Biggr) \nonumber\\
  \end{eqnarray}

 \begin{eqnarray}
 {\cal N}_{\hpmtt} &=&
 \frac{\dk}{2}\left[\ppout\sinsq+\pmout(1+\lambda_3\cos\theta)^2\right]
  +\lag\ppout(\gamma-4)\sin^2\theta+\nonumber\\ & &
\frac{\kac}{8}\Bigl\{6\ppout\sinsq+\pmout(1+\lambda_3\cos\theta)^2
(\gamma+2-(\gamma-4)\lambda_3\cos\theta)\Bigr\}  
 +\nonumber\\
& &\frac{\klam}{ 16}\Bigl\{2\ppout\sinsq(\gamma-3)+\pmout(1+\lambda_3\cos\theta)^2
(\gamma-2-(\gamma-4)\lambda_3\cos\theta)\Bigr\}  
 +\nonumber\\
 & &\frac{\lac}{16}\Biggl\{ 2\ppout\sinsq[2-\sinsq(\gamma-4)]+
 \pmout(1+\lambda_3\cos\theta)^2 \times \nonumber\\& &\biggl[
\gamma^2(1-\lambda_3\cos\theta)(3-\lambda_3\cos\theta)
-4(1-6\lambda_3\cos\theta+2\cos^2\theta)- \nonumber\\&&
2\gamma
(6-11\lambda_3\cos\theta+3\cos^2\theta)\biggr]
   \Biggr\} 
 \end{eqnarray} 

\noi
where  $P^{\pm}_{34}= 
(1\pm\lambda_3\lambda_4)/2 $ are operators projecting 
onto the W   states with same (opposite) helicities respectively.
The amplitudes not written explicitly above are simply obtained
with the relation,
\beq
{\cal M}_{\lambda_1\lambda_2;\lambda_3\lambda_4} =
{\cal M}^*_{-\lambda_1-\lambda_2;-\lambda_3-\lambda_4} \;\;\;\;
\theta\ra -\theta 
\eeq

\noi
Apart for various signs due to a different
labelling of the polarisation vectors,  these amplitudes
agree with those of Yehudai{\cite{Yehudai}} save for the
dominant term (in $s_{\gamma\gamma}$) in the $\lambda_\gamma^2$ term 
of the $J_Z=0$ amplitude for transverse W's, 
{\it i.e.} 
$(1-3\cos^2\theta)\ra(3-\cos^2\theta)$. This is probably just a misprint.

\bibliography{bank}
\end{document}